\documentclass[fleqn,usenatbib]{mnras}

\usepackage{newtxtext,newtxmath}
\usepackage{cancel}

\usepackage[T1]{fontenc}

\DeclareRobustCommand{\VAN}[3]{#2}
\let\VANthebibliography\thebibliography
\def\thebibliography{\DeclareRobustCommand{\VAN}[3]{##3}\VANthebibliography}

\usepackage{graphicx}	
\usepackage{amsmath}	
\usepackage{caption}
\usepackage{xcolor}
\usepackage{xspace}
\def\hide#1{}
\newcommand{\eg}{e.g.}
\newcommand{\ie}{i.e.}
\newcommand{\msun}{{\mathrm{M_{\odot}}}}
\newcommand{\zend}{$z = 8.0$\xspace}
\newcommand{\mstartotal}{$M_*$}
\newcommand{\mrm}[1]{\mathrm{#1}}

\newcommand{\ramsesrt}{\textsc{ramses-rt}\xspace}           
\newcommand{\HI}{\mbox{H\,{\sc i}}\xspace}
\newcommand{\HII}{H\,{\sc ii}\xspace}
\newcommand{\HeI}{\mbox{He\,{\sc i}}\xspace}
\newcommand{\HeII}{He\,{\sc ii}\xspace}
\newcommand{\HeIII}{He\,{\sc ii}\xspace}

\begingroup
\catcode`\_=\active
\gdef_#1{\ensuremath{\sb{\mathrm{#1}}}}
\global\newcommand{_}[1][]{\ensuremath{\sb{\mathrm{#1}}}}
\endgroup
\mathcode`\_=\string"8000
\catcode`\_=12

\title[Proto-GCs at High-z]{Star Cluster Formation and Survival in the First Galaxies}

\author[F. Garcia et al.]{Fred Angelo Batan Garcia,$^{1,2}$\thanks{E-mail: fgarcia4@umd.edu}
Massimo Ricotti,$^{1}$\thanks{E-mail: ricotti@umd.edu}
Kazuyuki Sugimura,$^{3,4}$
and Jongwon Park$^{1}$
\\
$^{1}$Department of Astronomy, University of Maryland, College Park, MD 20742, USA\\
$^{2}$Department of Physics, University of Maryland, College Park, MD 20742, USA\\
$^{3}$The Hakubi Center for Advanced Research, Kyoto University, Sakyo, Kyoto 606-8501, Japan\\
$^{4}$Department of Physics, Graduate School of Science, Kyoto University, Sakyo, Kyoto 606-8502, Japan\\
}

\date{Accepted XXX. Received YYY; in original form ZZZ}

\pubyear{2022}

\begin{document}
\label{firstpage}
\pagerange{\pageref{firstpage}--\pageref{lastpage}}
\maketitle

\begin{abstract}
Using radiation-hydrodynamic cosmological simulations, we present a detailed ($0.1$ pc resolution), physically motivated portrait of a typical-mass dwarf galaxy before the epoch of reionization, resolving the formation and evolution of star clusters into individual $10\:\mathrm{M_{\odot}}$ star particles. In the rest-frame UV, the galaxy has an irregular morphology with no bulge or galactic disk, dominated by light emitted from numerous, compact, and gravitationally-bound star clusters. This is especially interesting in light of recent HST and JWST observations that -- aided by the magnifying power of gravitational lenses -- have imaged, at parsec-scale resolution, individual young star clusters in the process of forming in similar galaxies at $z>6$. Because of their low metallicities and high temperatures, star-forming gas clouds in this galaxy have densities $\sim 100$ times higher than typical giant molecular clouds; hence, their star formation efficiencies are high enough ($f_*\sim10-70$ per cent) to produce a sizeable population of potential globular cluster progenitors but typically smaller (between a few $100\:-\: 2\times10^4\:\mathrm{M_{\odot}}$, sizes of $0.1-3$ pc) and of lower metallicities ($10^{-3.5}-10^{-2.5}\:\mathrm{Z_{\odot}}$). The initial mass function of the star-forming clouds is log-normal while the bound star cluster mass function is a power-law with a slope that depends mainly on $f_*$ but also on the temporal proximity to a major starburst. We find slopes between $-0.5$ and $-2.5$ depending on the assumed sub-grid $f_*$. Star formation is self-regulated on galactic scales; however, the multi-modal metallicity distribution of the star clusters and the fraction of stars locked into surviving bound star clusters depends on $f_*$.
\end{abstract}


\section{Introduction}
Metal-poor globular clusters (GCs) and some of the smallest ultra-faint dwarf (UFD) galaxies found in the Local Group are among the oldest objects in the Universe \citep[][]{RicottiG:2005, BovillR:2009, Brown:2012, 2013ApJ...775..134V, Brown:2014, Simon:2019UFDReview}, but their cosmological origin is still unknown. In \cite{Ricotti:2002}, it was first pointed out that even if a relatively small fraction of the progenitors of today's old/metal-poor GCs were formed before the epoch of reionization, it would still imply that star formation in bound star clusters (BSCs) was the primary mode of star formation in the first galaxies. As such, these GC progenitors (proto-GCs) may have played an important or dominant role as sources of cosmic reionization. Since this early proposal, the role of proto-GCs in the cosmological context and as reionization sources has been more widely recognized and investigated \citep[][]{Schaerer:2011MNRAS.413.2297S, KatzR:2013, KatzR:2014, Renzini:2017MNRAS.469L..63R, Boylan-Kolchin:2018}. From an observational point of view, recent discoveries by the \textit{Hubble Space Telescope} (HST) and the \textit{James Webb Space Telescope} (JWST) of strongly lensed galaxies have confirmed that indeed high-z galaxies are characterized by compact regions (<$10$~pc) of star formation \citep[][]{Welch:2022, Vanzella:2022arXiv221109839V}, possible sites of the formation of proto-GCs.

Simulations of star formation in the first galaxies have also revealed a possible connection between the smallest UFD galaxies and small mass analogs of today's GCs. The idea, suggested in \citep{RicottiPG:2016}, is that the sizes and masses of the faintest UFD galaxies can be explained as produced by the dissolution of small-mass compact star clusters that remain confined by the gravitation potential of their dark matter (DM) haloes. The remnants of these compact BSCs produce low-surface brightness and DM-dominated dwarf galaxies. Hence, UFDs and some metal-poor GCs may have had a similar cosmological origin. Star formation efficiency may play a critical role in understanding this link \citep{Salvadori:2009MNRAS.395L...6S, Tassis:2012ApJ...745...68T, Madau:2014ApJ...790L..17M}. It is, however, unclear what is the efficiency of formation and the survival rate of compact star clusters in high-z galaxies \citep{Bromm:2009Natur.459...49B, Behroozi:2013ApJ...762L..31B, Sun:2016MNRAS.460..417S}, and whether UFDs can be produced from dissolving star clusters over timescales of billions of years due to two-body relaxation, or on short timescales due to natal mortality in molecular clouds with low star formation efficiency (SFE).  

Another open question regards the initial BSC mass function at high-z. Due to secular evolution, mainly produced by two-body relaxation and tides, only GCs more massive than about $10^4~\msun$ can survive for a Hubble time. Therefore, even if GC progenitors at high-z were forming copiously with masses smaller than this threshold, such objects would not exist today. But small-mass BSCs can be important at high-z for understanding galaxy formation/feedback and as reionization sources. In addition, old GCs are metal-poor, but their metallicity is typically $\sim 10^{-1}-10^{-2}$~Z$_\odot$. So far, no GC has been observed with a metallicity below $10^{-2.5}$~Z$_\odot$ \citep{Fernandex:2021ApJ...908L..42F}, while stars in UFD galaxies with masses $<10^4~\msun$ can have metallicities as low as $10^{-3}$~Z$_\odot$ \citep{McConnachie:2012, Fu:2022ApJ...925....6F}.

In the past, resolving GC formation with high-fidelity cosmological simulations has been impossible due to limits in both spatial and mass resolutions. Furthermore, the methods in which star formation is typically modeled in low-res simulations do not allow such a detailed view, with earlier efforts to model this process often treating the whole star cluster as a single particle \citep[\eg,][]{Bournaud:2008MNRAS.389L...8B, Griffen:2010MNRAS.405..375G, 2019MNRAS.486.3134K, Li:2019MNRAS.486.4030L, Phipps:2020A&A...641A.132P}. However, gradual improvements in hardware capabilities as well as the development of more robust sub-grid models have allowed the study of star clusters in realistic galactic environments. Recent high-resolution simulations of star cluster formation can roughly be categorized into two general areas: (i) the study of the formation of contemporary young massive clusters (YMCs) at low-$z$ \citep[\eg,][]{Lahen:2020, Li:2022MNRAS.514..265L, Hislop:2022} and (ii) those dealing with the formation of the earliest BSCs at high-z through cosmic time scales \citep[\eg,][]{RicottiPG:2016, Kim:2018, Ma:2020, Sameie:2022, Phipps:2022}. Simulations of galaxies and galaxy mergers in the local universe have been able to form young star clusters with masses on the order of $10^5$ to $10^6~\msun$ \citep[][]{Li:2019MNRAS.486.4030L, Lahen:2020}. On the low mass end, some have resolved star clusters with masses of a few hundred solar masses, with stellar mass resolution as small as $4~\msun$ \citep{Hislop:2022}. Furthermore, cosmological simulations at high-z are able to form star clusters as massive as $\sim10^7~\msun$ \citep{Grudic:2022arXiv220305732G}. Though, in several cases \citep[\eg,][]{Ma:2020, Phipps:2022, Sameie:2022} the relatively large masses of stellar particles (500 to $10^3~\msun$) prevent the identification and formation of low-mass BSCs that may play a key role in the early evolution of the first galaxies.

In this work, we present cosmological simulations with $0.1$~pc resolution and star masses of $10~\msun$, of a high-z dwarf galaxy ($M_{{halo}} \sim 2 \times 10^8~\msun$ at $z=8$). We resolve the formation of compact star clusters with masses between few 100 and $10^4~\msun$ and study the conditions for their formation, their dynamical evolution, and their properties. Our goal is to understand what determines the formation of BSCs and their mass function in the first galaxies, providing predictions for existing and future observations of lensed galaxies at $z>6$, in which compact star-forming regions have started to be resolved on scales ranging from few tens of parsec to even 0.1~pc in the case of one lens \citep{Welch:2022, Welch:2022ApJ...940L...1W}.

This paper is organized as follows. In \S~\ref{sec:sim}, we briefly present the simulation setup and methods. In \S~\ref{sec:res} we show the results of our simulations, discussing how the properties of the star-forming gas clouds and the population of BSCs are shaped by the assumptions on the sub-grid star formation efficiency in our simulations. In \S~\ref{sec:dis}, we present a discussion of our result in light of previous work, and in \S~\ref{sec:sum} we summarize our main results.
\section{Methods and Simulations}\label{sec:sim}
The simulations in this paper are run with the adaptive mesh refinement (AMR) radiative hydrodynamics (RHD) code \ramsesrt\ \citep{teyssier2002,rosdahl2013}. Our version of the code includes several physics modules that have been described in previously published work by our group for RHD simulations of galaxy formation \citep[\eg,][]{kimm2017,katz2017}, metal-rich star formation in molecular clouds (\citealp*[\eg,][]{HeRG:2019,HeRG:2020,HeR2022}) and Pop~III star formation \citep{parkR2021a,parkR2021b, parkR2022}. 

The physics for chemistry and cooling due to metal-enriched gas and dust is described in \cite{katz2017}, while primordial gas physics is described in \cite{parkR2021a,parkR2022}. However, in the present paper, we have made significant modifications to the recipe for star formation and SN feedback with respect to previously run simulations. Details of the modifications and test runs that led to the adoption of the model parameters used in this paper will be discussed in a separate paper (Sugimura, Ricotti, \& Park, in preparation). Here, we summarize the main physics included in the code and emphasize the new features mainly with regard to the sub-grid star formation recipe.

\textbf{\em Cosmological initial conditions}: We run a cosmological zoom simulation of a DM halo that at $z=0$ is relatively isolated and has a mass of $\sim 10^{10} \: \msun$ (an analog of the dwarf galaxy WLM (DDO 221) in the Local Group). This is Halo~D in \cite{RicottiPC:2022}; details on its accretion history and mass profile can be found in that reference. In this simulation, the refined region consists of a cubic zoom region of 300~$h^{-1}$~kpc comoving on each side with DM mass resolution of $\sim$ 800 $\msun$. This region is within a cubic simulation volume of 35~$h^{-1}$~Mpc (49.3 Mpc) on a side. The zoom region was selected such that the halo at $z=0$ appeared isolated, had identifiable progenitors for at least $z<6$, and had not fallen into larger structures or merged with a halo $> 50$ per cent its mass after $z = 6$.

\textbf{\em Refinement criteria}:
We use a Lagrangian refinement criterion for the DM and stars; we refine a cell if it contains more than 8 particles, including both DM and star particles.

For the gas component, in addition to the Lagrangian criterion that refines cells with gas mass exceeding eight times the initial mean value in the zoom region ($\sim 160~\msun$) we resolve the Jeans length with at least $N_J$ cells, where $N_J=8$ if $1.2[(1+z)/10]^{-1}<\Delta x \leq 76.8~[(1+z)/10]^{-1}$~pc, and $N_J=4$ if $\Delta x \leq 1.2~[(1+z)/10]^{-1}$~pc. The highest resolution cell (level $l=25$) has a physical size of $\Delta x_{min}=0.15[(1+z)/10]^{-1}$~pc.

\textbf{\em Gas physics}: The primordial gas physics is treated using non-equilibrium chemistry and cooling/heating of species (\HI, \HII, \HeI, \HeII, \HeIII, H$_2$, and H$^-$) as in \cite{parkR2021a}, with H$^-$ assumed to be in the chemical equilibrium with other species. \hide{(KS: HD and H$_2^+$ are not considered in the current runs. H$^-$ is considered but assumed to be in chemical equilibrium with other species. Also, the chemical heating due to H2 formation, which becomes important at $n\gtrsim 10^{9}\mathrm{cm^{-3}}$, is neglected in the current runs, so I have changed the number in the following sentence.)} The primordial processes included are accurate to number densities up to $n \sim 10^{9}$~cm$^{-3}$. Metal and dust cooling/heating are treated as in \cite{katz2017}, using Grackle tables \citep{Grackle:2017}, with the assumption of a constant dust/metal ratio. Unlike in \cite{katz2017}, we do not introduce a sub-grid clumping factor leading to enhancement of H$_2$ formation on dust, presuming that the clumping is directly followed in our higher resolution simulations. 

\subsection{Star formation and SN feedback}
Star formation is implemented as follows. At each time step we identify cells at the maximum refinement level with density above the density threshold:
\begin{equation}
    n_{{crit}}=\left(5.0\times 10^4 {\rm cm}^{-3}\right) 
    \left( \frac{T}{100~{\mrm{K}}} \right)\left(\frac{1+z}{10}\right)^2
    \left(\frac{N_{cr}}{4}\right)^{-2},
    \label{eq:n_cr}
\end{equation} 
where $T$ is temperature, $z$ is the redshift, and $N_{cr}=4$ is a fiducial parameter assumed in our simulations. This threshold is derived by imposing that the Jeans length is resolved with $N_{cr}$ cells at the maximum refinement level: $\lambda_J=N_{cr} \Delta x_{min}$, where $\lambda_J$ is the Jeans length in physical units, $\Delta x_{min}$ is the cell size at the maximum refinement level ($l=25$). If at any given time, at the max refinement level, the gas number density is $n>n_{crit}$, instead of refining the cell and introducing a higher level, we switch on one of the following "sub-grid" recipes for star formation, depending on the gas metallicity.

\textbf{\em{Population~III stars}}: If $Z<Z_{crit}=10^{-5}$~Z$_\odot$ we form a single Pop~III star particle, treated as one binary system of Pop~III stars with 40 and 80~$\msun$ (in total 120~$\msun$). The 40 $\msun$ star explodes as a Hypernova after 4~Myr, injecting a thermal energy $E_{\rm SN, PopIII}=3\times 10^{52}$~ergs in the gas, while the other 80~$\msun$ star directly collapses into a BH without explosion. After their lifetime, they leave a BH particle composed of binary BHs of 20 and 80~$\msun$, respectively (or a single 100~$\msun$ BH if the binary is assumed to merge), which emits no radiation. Note that in this paper, there is no difference in our numerical treatment between BHs in a binary and a single BH. Of course, the recipe for Pop~III star formation is too simplified considering the current advances in our understanding of Pop~III stars' mass function and multiplicity found in AU-scale simulations \citep[\eg][]{sugimura:2020, parkR2021a, parkR2021b, parkR2022}, where the formation of the first galaxies is likely affected by the assumed Pop~III IMF \citep{Abe:2021}. We will study the dependence of the formation of the first galaxies on different Pop~III star formation recipes using our simulation suite in future work.

\textbf{\em Population~II star clusters}:
If $Z>Z_{crit}$, we form star clusters as follows. We calculate, on the fly, the gas density profile around the peak density and define the cloud mass and the cloud radius where the mean density reaches $1/N_{{cut}}$ of the peak density. Here we assume $N_{{cut}}=10$. Within this cloud, the stars are formed with an efficiency $f_*=m_{SC}/M_{MC}$, where $m_{SC}$ is the total mass in stars produced in a star-forming cloud with a total mass of $M_{MC}$. We run two identical simulations in which $f_*$ is the only parameter that is changed: $f_*=0.70$ and $f_*=0.35$. The stars are distributed randomly in a density-weighted way within the gas inside the cloud, but only in regions where the density is above the mean density. For instance, if the star-forming cloud is an elongated filament rather than a spheroid, the stars can only form within the filament. The star particles have equal masses  $m_*=10 \: \msun$. SN explosions with thermal energy ($E_{{SN}}=10^{51}$~ergs) are produced stochastically between times 4~Myr and 40 Myr, with about one SN every 100 $\msun$ in stars. The number of SNe, as well as the metal yield, H and He ionizing radiation, and Lyman-Werner (LW) radiation from star clusters are calculated assuming a Salpeter IMF (1 -- 100 $\msun$) \citep{Salpeter:1955}.

\textbf{\em Radiation transfer}: The code \ramsesrt\ includes an approximate treatment of 3D radiation transfer based on solving the moments of the radiation field with M1 closure \citep{rosdahl2013}. We include photoionization of hydrogen and helium ions and photoionization heating. The code also includes a treatment of secondary ionization/heating from fast photo-electrons produced by X-ray ionization as in \citep{RicottiGS:2002a, Ricotti:2016, parkR2021a}, but in this work, we do not include sources of X-rays. We consider the self-shielding of LW photons in a local cell using a fitting formula from \cite{Draine:1996ApJ...468..269D}. We adopt the metallicity-dependent 
dust attenuation cross-section of $\sigma_{\mathrm{d,eff}}=4\times10^{-21}\mathrm{cm^2}\,(Z/Z_\odot)$ assuming a constant dust-to-metal ratio \citep{katz2017}. We include radiation bins for the LW band, \HI, \HeI, and \HeII ionizing radiation.

\section{Results}\label{sec:res}
\begin{figure*}
	\includegraphics[width=\textwidth]{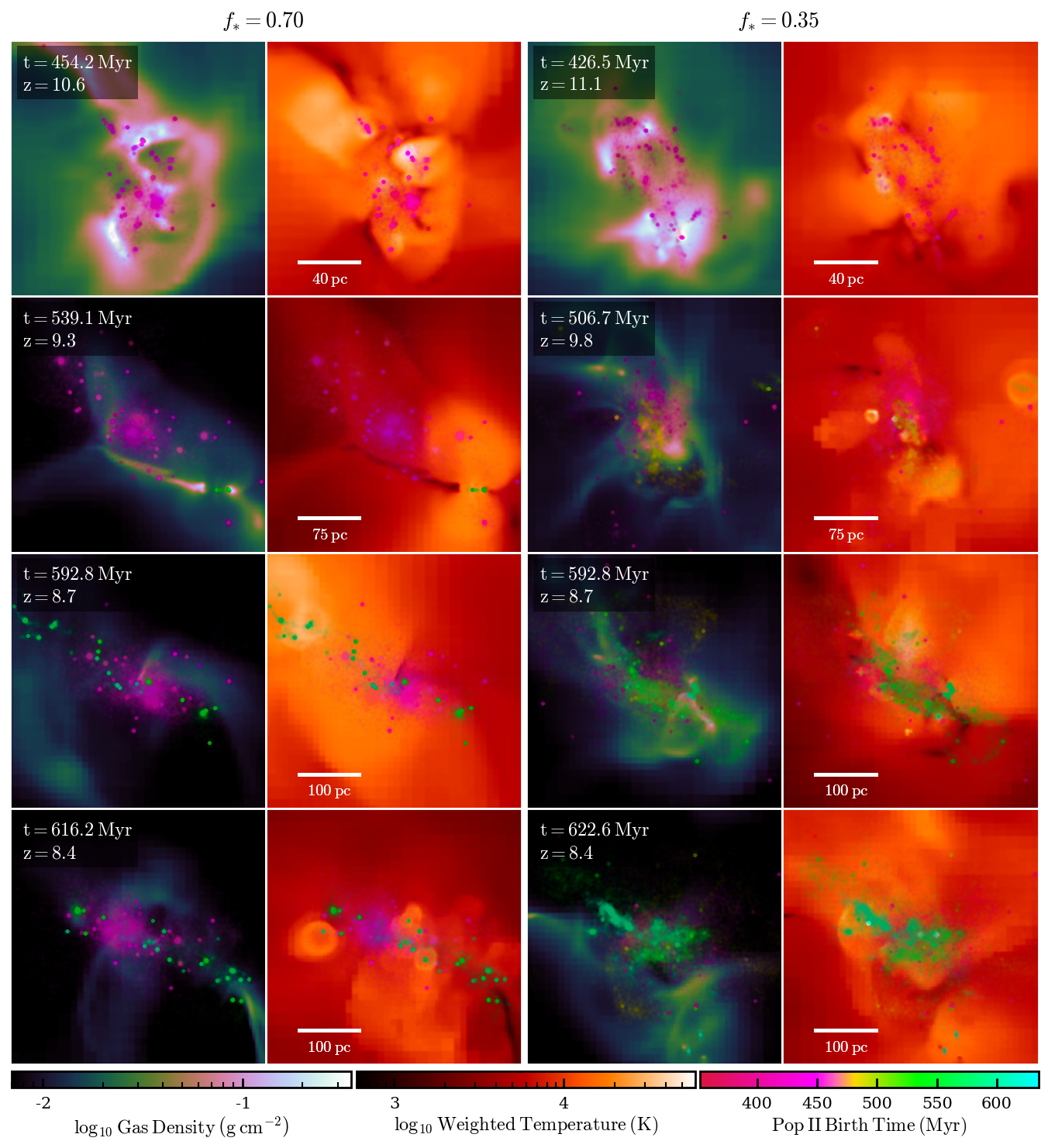}
    \caption{Line-of-sight projected gas density and density-weighted temperature in our simulated dwarf galaxy as a function of time, for two choices of the sub-grid SFE: $f_{*} = 0.70$ (\textit{first two columns}) and $f_{*} = 0.35$ (\textit{last two columns}). The panels, from \textit{top to bottom}, show the time evolution of each galaxy. The images in each row (some of which differ in spatial scale, shown by the annotated bar) compare the galaxies at similar evolutionary stages of star formation. In each panel, the stars are colored (see \textit{bottom right} color bar) based on the time of birth. The stars themselves are too small to be seen individually; the spheroidal objects are bound star clusters (BSCs). A movie rendering the time evolution of these two simulations is available as supplementary material in the electronic version of this paper.}   
   
    \label{fig:Gas Projection}
\end{figure*}

\begin{figure*}
	\includegraphics[width=\columnwidth]{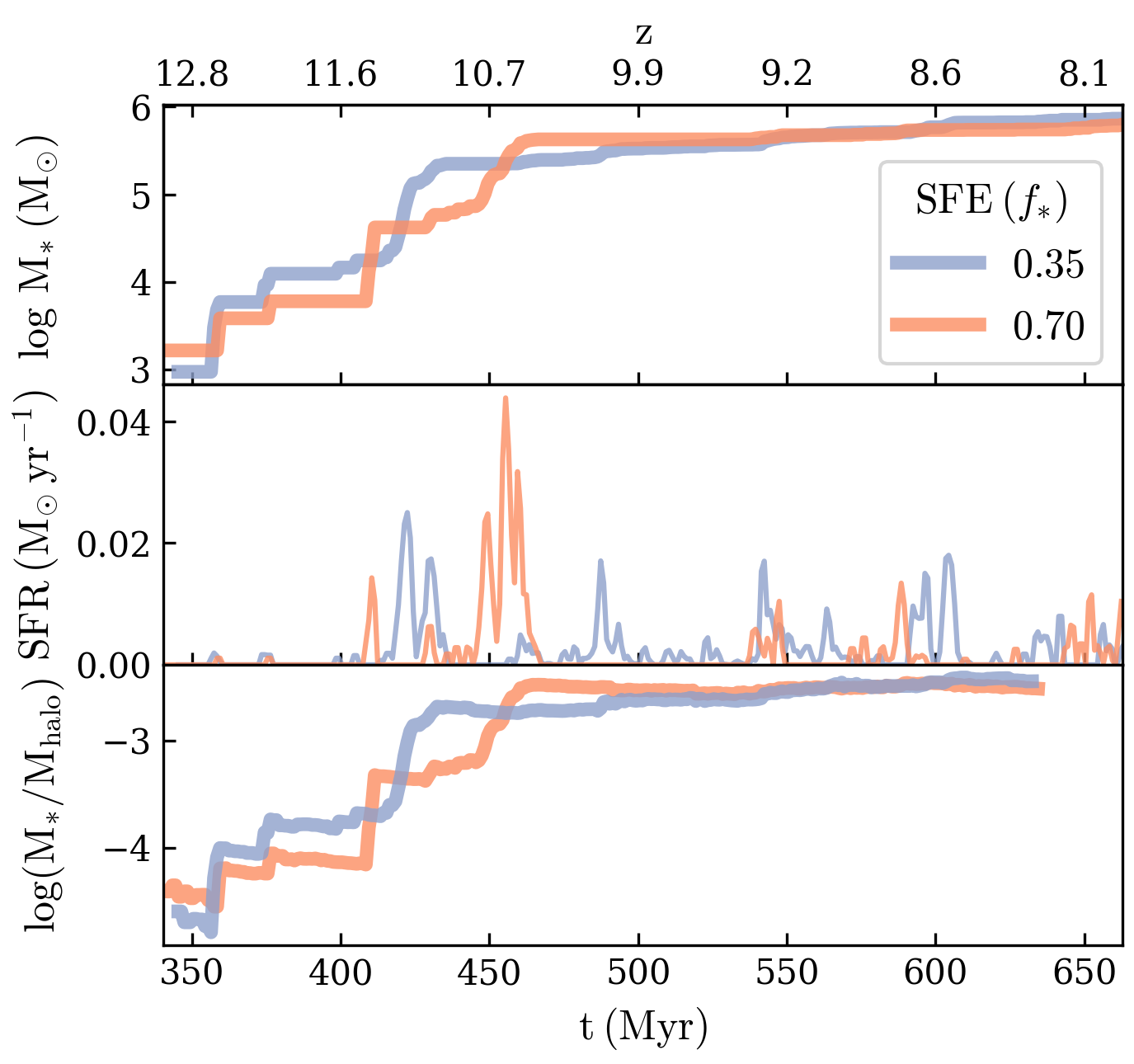}
	\includegraphics[width=\columnwidth]{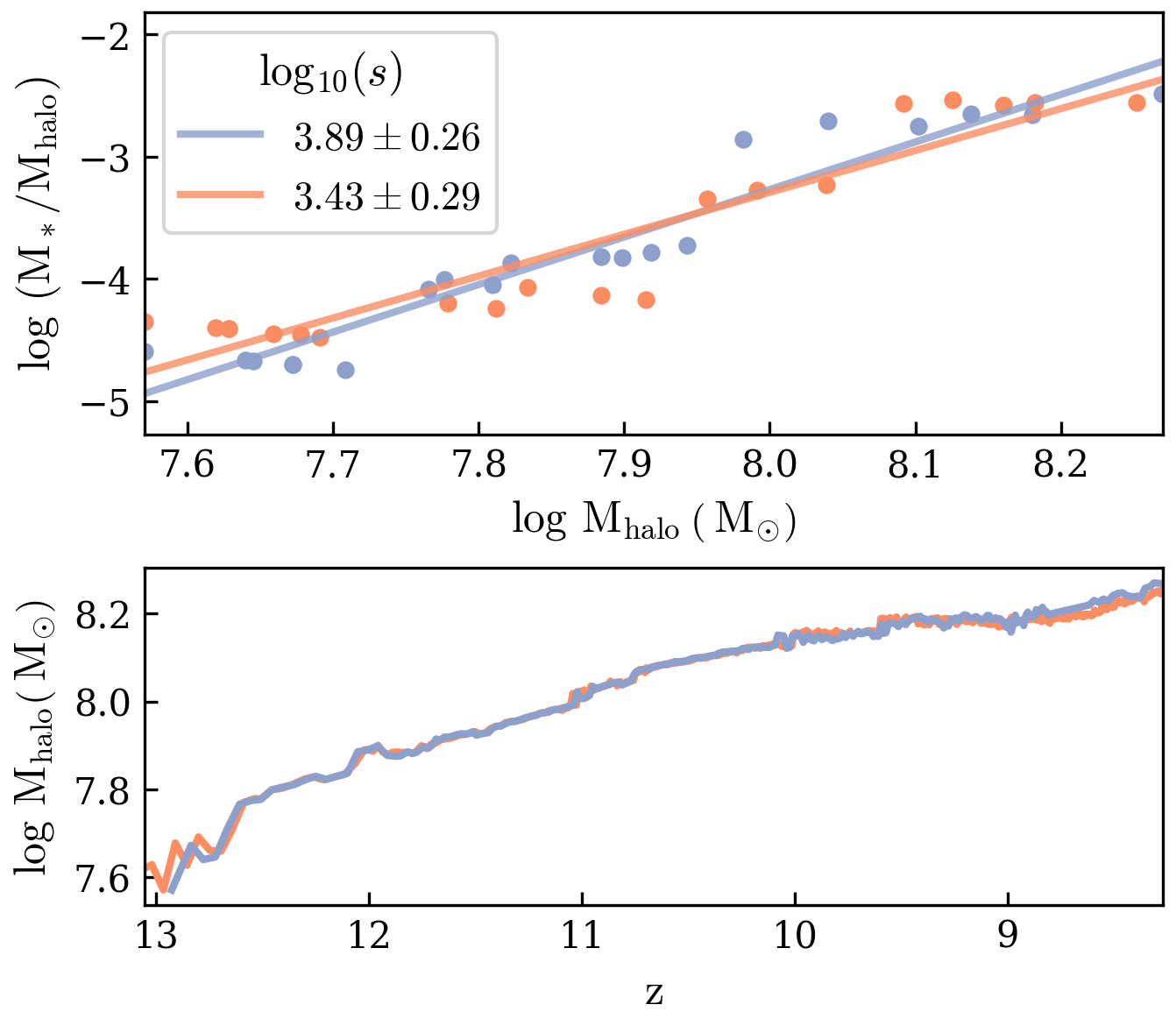}
    \caption{ (\textit{Left column.}) Top panel: total mass in Pop II stars for both low-SFE and high-SFE runs as a function of time and redshift, see left column legend. Middle panel: the corresponding SFR in solar mass formed per year for each efficiency. Bottom panel: galaxy-scale star formation efficiency ($M_*/M_{halo}$) as a function of time. (\textit{Right column.}) Top panel: galaxy-scale star formation efficiency as a function of halo mass for the same halo for the two different $f_{*}$ runs (colored symbols). The solid lines show the best power-law fits with values of the best-fit slopes $s$ and 1$\sigma$ errors shown in the right column legend. Note, the halo efficiencies were found by sampling the same halo over time. We point out that the observational bias is to see the galaxy during or shortly after a burst of SF, when $M_*/M_{{halo}}$ is at a relative maximum. Bottom panel: DM halo mass as a function of redshift for both simulations. }
    \label{fig:SFC Mass Over Time} 
\end{figure*}

In this work, we run two identical simulations in which the only parameter that we changed is the sub-grid SFE, $f_*$. We will refer to the run with $f_*=0.70$ as the high-SFE run, and the one with $f_*=0.35$ as the low-SFE run (see \S~\ref{ssec:Are the assumed values of the sub-grid SFEs realistic?} for an explanation of this choice). In future work, we will present the results of simulations in which $f_*$ depends on the properties of the gas cloud wherein stars are formed, such as its mass, mean density, and metallicity. Also, we will refer to the star-forming gas clouds as molecular clouds (MCs) despite the fact that the gas in them may be mostly atomic rather than fully molecular (\ie, the molecular fraction can be less than unity in a gas cloud with very low metallicity irradiated by a strong LW background).

Fig.~\ref{fig:Gas Projection} shows representative evolutionary snapshots of the high-SFE simulation ($f_*=0.70$, left columns) and low-SFE simulation ($f_* = 0.35$, right columns). A movie is available in the electronic version of this paper. Each panel shows projected quantities -- either gas density or density-weighted temperature -- along the same line-of-sight for each simulation. As the simulations progress, from top to bottom, the interaction between the gas and Pop~II stars (color-coded by age), in the form of heating from radiative feedback and SN explosions, is clearly displayed by the density-weighted temperature (shown side-by-side with the gas density). In these plots, Pop~II stars, although too clumped and small to be seen individually, appear clustered in compact spheroids. Each row depicts the galaxies at a similar star-forming stage. The first row depicts each galaxy during its first (and largest) starburst (see Fig.~\ref{fig:SFC Mass Over Time}). The second and third rows aim to show corresponding gas heating and the result of a second burst of star formation, respectively. The final row depicts each galaxy near the latest stage of evolution reached by the simulations (\zend). For both choices of the sub-grid $f_*$, the galaxy does not form a disk or a central bulge. The morphology is highly irregular, with star formation happening nearly all in compact star clusters. By the end of the simulation at \zend, both galaxies have a surviving population of compact, BSCs with masses between $\sim 300 \: \msun$ and a few $10^4 \: \msun$ for the high-SFE, and between $\sim 300 ~\mathrm{M}_\odot$ and  $10^3 \: \msun$ for the low-SFE. 

In this work, the unprecedented spatial resolution ($\sim 0.1$~pc, physical at $z=10$) and star mass resolution (10 $\msun$), allow us to reliably simulate the formation of individual compact star clusters and their dynamical evolution, even for very small/compact BSCs of mass $>10^3 \: \msun$ and core radii $>0.1$~pc. We can therefore answer questions on the sites and efficiency of the formation of star clusters in the galactic environment, and whether star formation in primordial MCs produces BSCs or unbound "open" star clusters, that eventually become field stars.

Although an irregular morphology has been observed in other simulations of high-z dwarf galaxies \citep[\eg,][]{Besla:2012MNRAS.421.2109B, Mastropietro:2021MNRAS.504.3387M}, it is important to note that in other simulations of high-z dwarf galaxies \citep[\eg,][]{Duc:2004A&A...427..803D, Pawlik:2013ApJ...767...59P, RicottiPG:2016, Arata:2018MNRAS.475.4252A}, the formation of thick and highly Toomre-unstable disks, have been observed. It is unclear if the lack of a disk can be attributed to differences in numerical methods (\eg, differences in resolution, details of feedback processes, and/or star formation recipes) or the choice of halo we zoom-on, specifically its accretion history. A simulation with similar physics and resolution of a representative volume at $z=10$, rather than a zoom on a single halo, should produce enough statistics in terms of the number of galaxies with different morphologies to discern between these two interpretations. We will address this question in future work.

In the remainder of this section, we will start by analyzing the effects of the assumed sub-grid SFE on the global star formation within the galaxy through cosmic timescales (\S~\ref{ssec:sfe}). We then move on to study the properties of star-forming gas clouds and the formation of BSCs in \S~\ref{ssec:bsc}, focusing on their natal cradles along with the cluster mass functions for both high- and low-SFE simulations. Afterwards, we take a look at the demographics of the BSCs and their evolution through cosmic timescales (\S~\ref{ssec:dynamics}). We then offer insights into how this evolution influences the overall morphology of the host galaxies in \S~\ref{ssec:morpho}, contextualizing them with respect to current and forthcoming observations.

\begin{figure*}
    \includegraphics[width=\columnwidth]{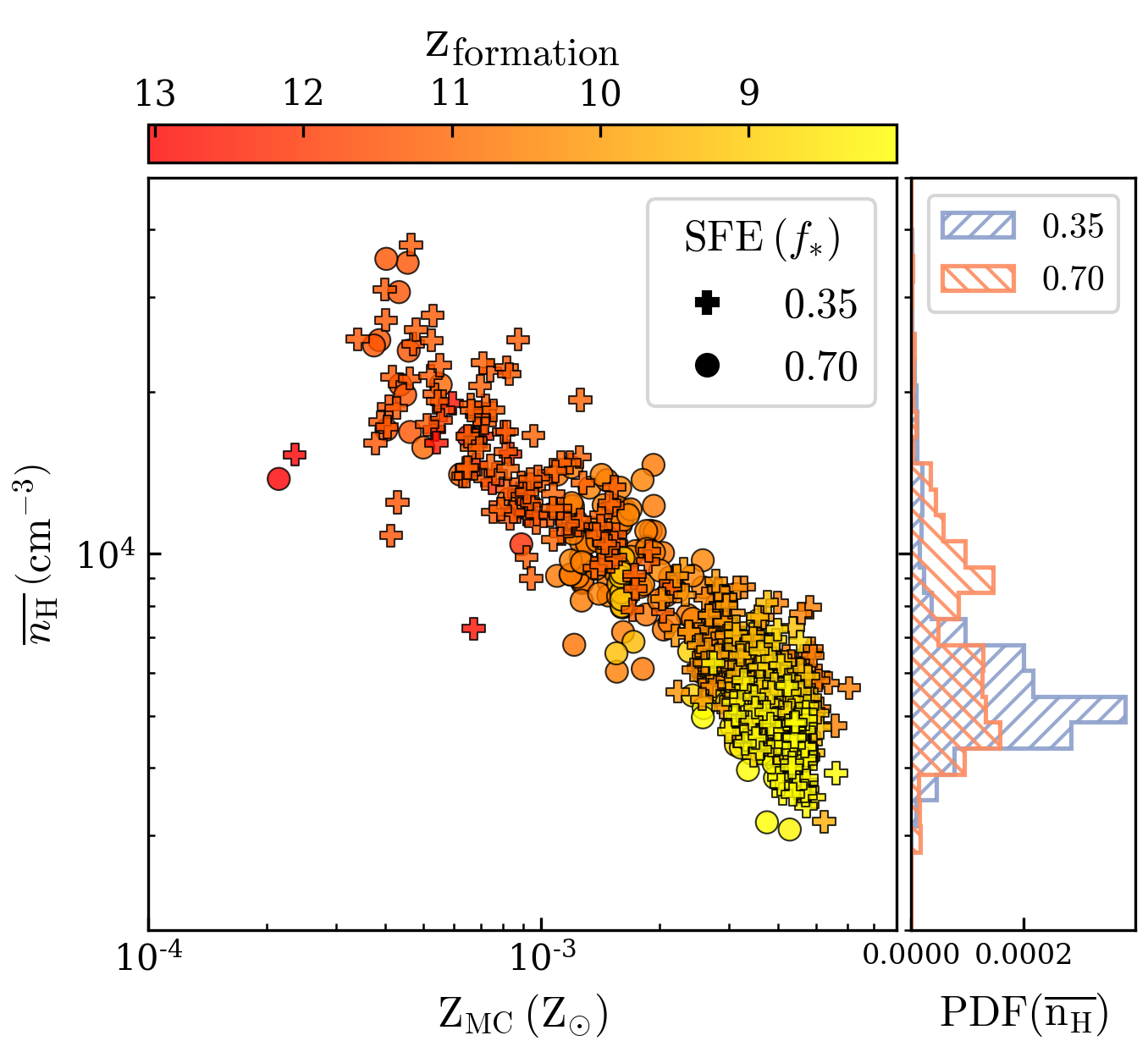}
    \includegraphics[width=\columnwidth]{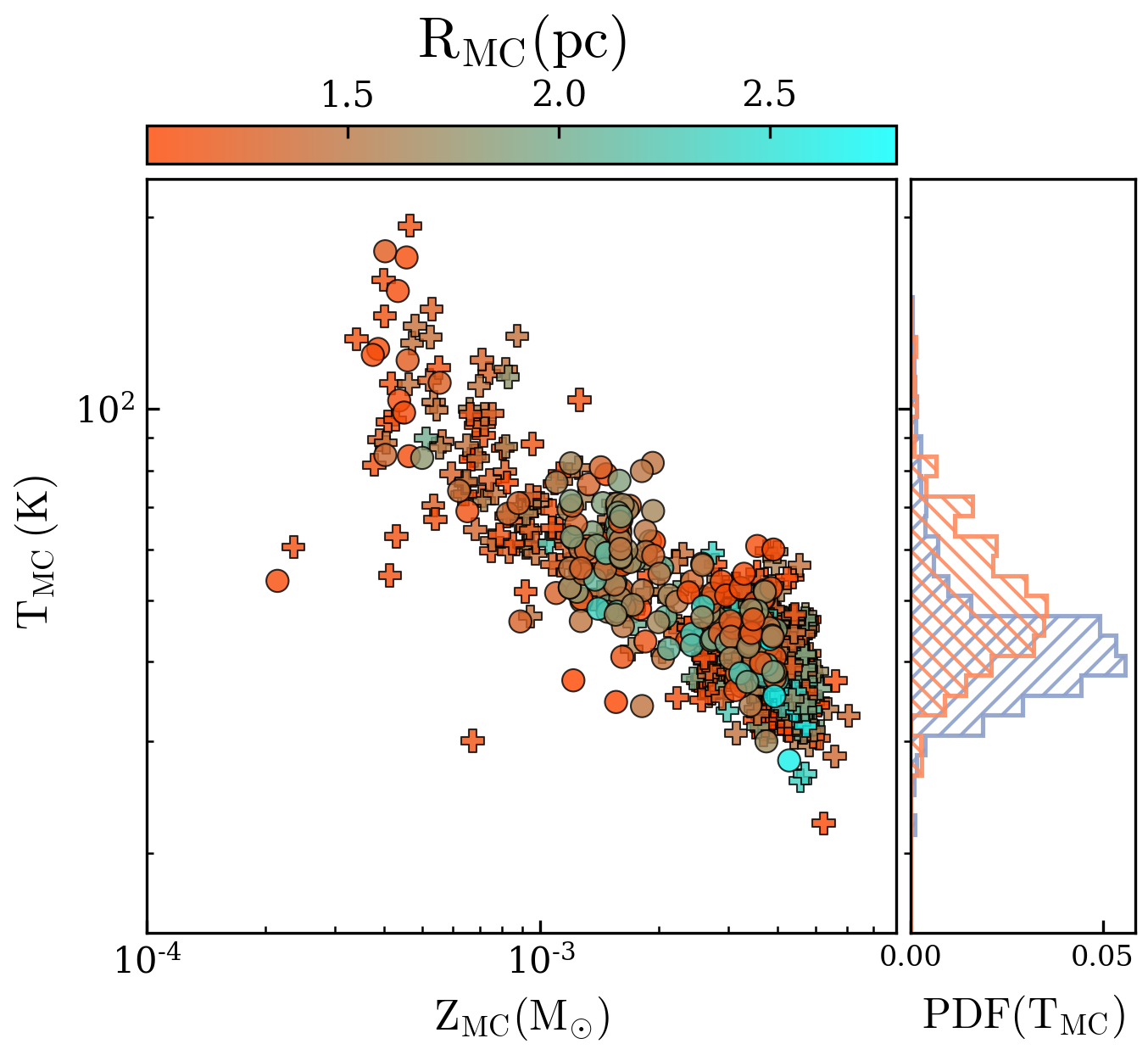}
    \caption{({\it Left.}) Mean hydrogen number density ($\overline{n_{{H}}}$) in star-forming clouds (MCs) as a function of their gas metallicity in the $f_{*} = 0.35$ and $f_{*} = 0.70$ runs (see legend). The clusters are colored by star-formation redshift, with older MCs depicted in red and younger in yellow. ({\it Right.}) Same as the left panel, but shows the gas temperature of the MCs as a function of their metallicity. The symbols are colored according to the cloud radii in parsec. To derive the temperature, we assumed that the cloud is approximated by an isothermal sphere supported against gravity by both thermal and turbulent pressures. On the right side of each scatter plot, a PDF is shown for the likelihood that a MC has the given $\overline{n_{{H}}}$ (\textit{left}) and $T_{{MC}}$ (\textit{right}) for each sub-grid SFE (see \textit{right-most} legend).} 
    \label{fig:SFC Mean Density vs Metallicity} 
\end{figure*} 

\subsection{Self-regulation of Star Formation on Galactic Scales}
\label{ssec:sfe} 

For now, we will focus on how star formation unfolds on galactic scales and its dependence, if any, on the assumed sub-grid SFE. 

\subsubsection{Galaxy-scale star formation histories and self-regulation} 

In both simulations, Pop~III star formation begins at $z\sim 14$ ($t=300$~Myr), when the main halo mass is $M_{\mathrm{DM}} \sim 10^6$~M$_\odot$. Pop~II star formation begins about $30$~Myr later, at z $\lesssim 13$. The top panel in Fig.~\ref{fig:SFC Mass Over Time} (left) shows the total mass in Pop~II stars, $M_*$, in the galaxy as a function of time. In this figure, we see a slight delay (< 5~Myr) in the initial starburst for the galaxy with low-SFE. We attribute this delay to the stochastic nature of Pop~III star formation -- and subsequent SNe -- in our simulations. These 120~$\mathrm{M_{\odot}}$ Pop~III stars are short-lived, however, their deaths can sterilize MC nurseries, delaying the onset of initial Pop~II star formation and the occurrence of major bursts of star formation \citep{Chiaki:2018,Abe:2021}.

In the same figure, we show the star formation rate (SFR) in solar masses per year. Here, we use equal-width 1~Myr bins (sampling the latest total stellar mass in between bursts) which we then scale to obtain the appropriate SFR. In the simulation, we form stars instantaneously within star-forming clouds; however, here the choice of the bin width is physically motivated as follows: \cite{HeRG:2019} found that the duration of the star formation burst in compact molecular clouds is $\tau_{SF} \sim 6 t_{cr}$, where $t_{cr}\equiv R_{MC} / c_s$ is the sound crossing time of the cloud, that depends on $c_s$ = 10 $\mathrm{km \: s^{-1}}$, the sound speed in a low-metallicity HII region. The average radius of molecular clouds in our simulations is $R_{MC} \sim 1.5$~pc (see Fig.~\ref{fig:MC IMF}), giving us the typical duration of a single starburst on the order of $\tau_{SF} \sim 0.9$~ Myr, that we approximate to 1~Myr bins.
We notice the following trends with regard to the magnitudes and frequency of these bursts. The low-SFE run produces more frequent -- albeit less intense -- star formation periods, while the high-SFE run experiences more sporadic but stronger starbursts, with peak SFR of $\sim$0.050 $\msun \: \mrm{yr}^{-1}$, approximately a 50 per cent increase with respect to the maximum SFR for the low-SFE run ($\sim$0.034 $\mathrm{M_{\odot} \: yr^{-1}}$). However, this relatively high star-forming episode (centered at $\sim$450~Myr) results in the galaxy having a longer quenched star formation period, lasting $\sim$100~Myr. Indeed at \zend, the two simulations have nearly identical mass in stars $M_* \sim 5 \times 10^{5} $~M$_\odot$. In a cosmological context, this confirms that when a feedback loop is in place, the total amount of stars a galaxy produces is rather insensitive to the assumed sub-grid SFE \citep[\eg,][]{RicottiGS:2002b, Hopkins:2011MNRAS.417..950H, Yajima:2017ApJ...846...30Y}. We will see in section \S~\ref{ssec:bsc} that, although star formation is self-regulated and $M_*$ is insensitive to the assumed sub-grid $f_*$, the metallicity of the gas and stars as well as the fraction of stars in bound star clusters both depend on $f_*$.

\subsubsection{Galaxy-scale star formation efficiency} 

In Fig.~\ref{fig:SFC Mass Over Time} we show the overall star-formation efficiency in the DM halo ($f_{\mathrm{*,halo}}\equiv \mathrm{M_* / M_{halo}}$) as a function of time (bottom left panel) and as a function of halo mass (top right panel). Note that $M_* / M_{halo}$ as a function of halo mass refers to only {\it one halo} during its growth. We expect that haloes with similar mass but with different growth histories will show different relationships between $M_*$ and $M_{halo}$. Because star formation in the high-SFE run begins slightly earlier than in the low-SFE run, this results in a marginally higher $f_{\mathrm{*,halo}}$ when the halo mass was small, and produces the trend seen in the top right panel whereby $f_{\mathrm{*,halo}}\propto M_{halo}^{s}$ has a slightly flatter slope with halo mass than the case with low-SFE. Though given the large scatter around the mean of about one dex, both fits are in agreement within the uncertainties of the fitting parameters. We also point out that the observational bias is to observe the galaxy during or shortly after a burst of SF, when $f_{\mathrm{*,halo}}$ is at a relative maximum.

The DM halo growth histories for different sub-grid SFEs are also quite similar, as expected if baryonic feedback effects are negligible (bottom right panel). However, the DM halo in the high-SFE run has a slightly smaller halo mass at later times (4 per cent smaller $M_{\mathrm{DM}}$ at $z=8.4$). Simulations have shown that baryonic effects in the form of SNe-driven outflows can introduce cores in DM profiles \citep{Governato:2010}. \cite{Chan:2015} suggest that in dwarf galaxies, halo cores originate from early bursts of SF, but they are not fully established until later times. Larger bursts of star formation in the high-SFE simulation can seed this effect, only noticeable at later times. However, in our simulations, this rather small discrepancy is most likely a numerical artifact caused by slightly different halo histories; \eg, a greater $\mathrm{M_{200}}$ for the low-SFE caused by an in-falling sub-halo, unaccounted for in the other run. This slight over-counting of mass may manifest more strongly at later times when the mass evolution of the two haloes sufficiently diverges.

\subsection{Formation of Bound Star Clusters }
\label{ssec:bsc}
After discussing the SFE in the context of the entire galaxies' time evolution, we will now shift our focus to the properties of the BSC population in each galaxy, starting with the properties of the gas MCs in which the star clusters are born.
\subsubsection{Properties of molecular clouds in the galaxy} 
\begin{figure*}
	\includegraphics[width=\textwidth]{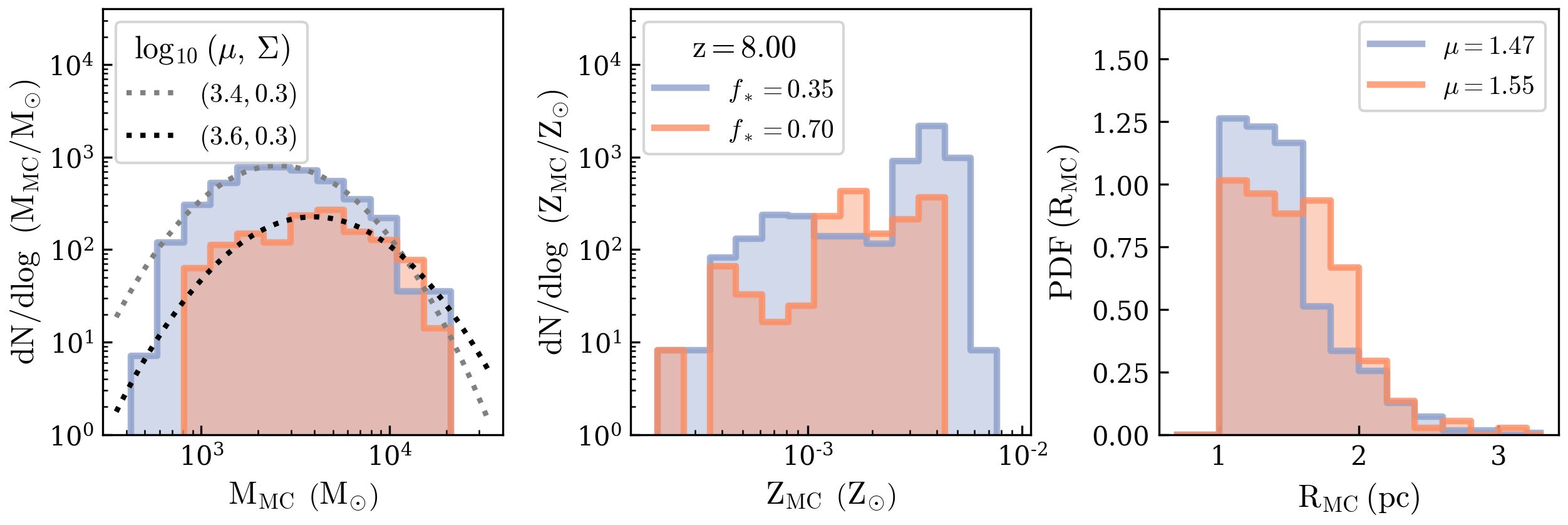}
    \caption{(\textit{Left}.) Latest (at \zend) initial mass function of MCs along with a log-normal fit (dotted lines) for both $f_*$ runs (see legend). The legend shows the mean and standard deviation of the fitted distributions. (\textit{Center.}) The initial metallicity function of the MCs for the same two runs. (\textit{Right.}) PDF of the MC radii, ${R_{MC}}$, for the two runs (the legend shows the average $R_{MC}$ in the two runs). Although the distributions change throughout the simulations as more MCs are formed, the MC properties remain constant as they are those at the time when star formation begins ($n_{H} > n_{crit}$ and $Z > Z_{crit}$). 
    }
    \label{fig:MC IMF} 
\end{figure*}
In our simulations, upon the satisfaction of the star-formation criteria, a test particle is formed storing -- among other things -- the mass of the star-forming MC, the total mass in stars generated in the MC itself, the metallicity of the cloud at formation, its formation redshift, and the mean gas density in the cloud. The test particle properties, apart from their positions, remain fixed over time. Fig.~\ref{fig:SFC Mean Density vs Metallicity} (left) shows the mean number density of MCs as a function of metallicity for the two simulation runs. The symbols are color-coded according to the formation redshift (see color bar). We find that nearly independently of the assumed sub-grid $f_*$, the mean density of MCs is between $2 \times 10^3$ and $5 \times 10^4$~$\mathrm{cm^{-3}}$, about 10 to 100 times higher than the mean density of MCs in the Milky Way. The MC density increases with decreasing gas metallicity of the cloud and with increasing redshift. Clearly, the gas metallicity in the galaxy and in MCs increases with time as more star clusters are formed and the gas within the galaxy is enriched with metals. 

But why do MCs tend to be denser if the gas has lower metallicity? Wouldn't the reduced cooling rate at low metallicity produce hotter and therefore less compressed/dense clouds? The answer is the opposite: a gravitationally unstable hotter cloud is denser (see the PDFs in Fig.~\ref{fig:SFC Mean Density vs Metallicity}) as clearly shown if we approximate the cloud profiles with an idealized self-gravitating singular isothermal sphere:
\begin{equation}
    n_{H}(r) = \frac{n_{crit}}{N_{{cut}}}\left(\frac{R_{MC}}{r}\right)^2=\frac{c_s^2(1+{\cal M}^2_{tur})}{2\pi G \mu_{\mathrm{mol}}\,m_{\mathrm{H}}r^2},
    \label{eq:iso}
\end{equation} where $N_{{cut}}=10$, $R_{\rm MC}$ is the cloud radius (see \S~\ref{sec:sim}), $\mu_{\mathrm{mol}}$ is the mean molecular weight, and $m_{\mathrm{H}}$ is the mass of hydrogen atoms. Here, we assumed that the cloud is supported by an effective pressure contributed by thermal pressure and turbulence, where ${\cal M_{{tur}}}\equiv \sigma_{{tur}}/c_{s}$ is the turbulent Mach number, $\sigma_{{tur}}$ is the rms of the turbulent velocity, and $c_s \approx 1~{\rm km \: s^{-1}}(T/100~{\rm K})^{1/2}$ is the sound speed. Clearly in the limit of subsonic turbulence ${\cal M}_{{tur}}\ll 1$, but in general we will not neglect this term in Eq.~(\ref{eq:iso}). So, at a given radius $r$ and for a fixed $\cal M_{tur}$, the density increases linearly with the gas temperature $T$.

Recall that the general non-singular solution \citep[][]{Ebert:1955ZA.....37..217E, Bonnor:1956MNRAS.116..351B} has a core that becomes denser and smaller as a function of time for a gravitationally unstable sphere. Since we identify our MCs when the cloud core density reaches $n_{crit}$ and the cloud radius $R_{MC}$ is identified where the spherically averaged density in the profile is $n_{crit}/N_{{cut}}$, it is easy to show that $R_{MC}=R_{core} N_{{cut}}^{1/2}\sim 3 R_{core}$, $M_{MC} \approx 7 M_{\rm core}$ and the mean density is $\overline{n_{H}} \approx 0.26 n_{crit}$\footnote{More generally, in terms of the free parameter $N_{{cut}}$ we have $M_{MC}=(3 N_{{cut}}^{1/2}-2)M_{core}$ and $\overline{n_{H}} = (3 N_{{cut}}^{1/2}-2)N_{{cut}}^{-3/2} n_{{crit}}$.}. Using Eq.~(\ref{eq:n_cr}) we then find the relation between the cloud temperature and the mean density:
\begin{equation}
    \frac{T_{MC}}{100~K} = \frac{\overline{n_H}}{1.3 \times 10^4~{\rm cm}^{-3}} \left(\frac{1+z}{10}\right)^{-2}.
    \label{eq:temp}
\end{equation}
Note that, in this model, $\overline{n_H}$ (and $T_{MC}$) do not depend on the turbulent support, while the core radius and mass depend on ${\cal M}_{{tur}}$:
\begin{equation}
    \begin{split}
    R_{\rm core} &\approx 0.18~{\rm pc} \left(\frac{1+z}{10}\right)^{-1}(1+{\cal M}_{{tur}}^2)^{1/2},
    \label{eq:rcore}\\
    M_{core} &\approx 26~{\rm M}_{\odot}\left(\frac{1+z}{10}\right)^{-1}
    \left(\frac{T}{100~{\rm K}}\right)(1+{\cal M}_{{tur}}^2)^{3/2}.
    \end{split}
\end{equation}
In Fig.~\ref{fig:SFC Mean Density vs Metallicity} (right) we show the MC temperature in Eq.~(\ref{eq:temp}) as a function of gas metallicity. This shows that lower-metallicity MCs are hotter and therefore denser than higher-metallicity MCs. These results are also confirmed by inspecting the density and temperature profiles of each MC, although the profiles are not always perfectly isothermal or spherical, as the collapsing cloud is often elongated/filamentary. 

In Fig.~\ref{fig:MC IMF} we compare the mass functions (left panel), the metallicity distribution (middle panel), and the radii PDF (right panel) of MCs in our two simulations at \zend. The distribution functions evolve as a function of time as new MCs are formed throughout the course of the simulations; however, the properties of the MCs in our plots are at the time of formation. We also show best-fit models with uni-modal or multi-modal log-normal distributions:
\begin{equation} 
    f(x) = \sum_{\textit{i}=1}^n a_{\textit{i}} \exp\left({-\frac{(x-\mu_{\textit{i}})^2}{2\Sigma^2_{\textit{i}}}}\right),
    \label{Eq Gaussian}
\end{equation}
where $a$ is the amplitude, $\mu$ is the mean, $\Sigma$ is the standard deviation of each peak, and $n$ is the number of modes. We find that the MCs are quite compact and low-mass: their radii range between 1~pc and 3~pc and masses between $500~\msun$ and $2\times 10^4~\msun$.

We observe a unimodal distribution ($n = 1$) for the initial mass function of the MCs (Fig.~\ref{fig:MC IMF}, left panel), well fitted by a log-normal distribution. For our two simulations, the distributions have the same variance but different means, with a larger $\mu=\log_{10}{\overline{M_{MC}} }$ for the run with higher sub-grid SFE ($\overline{M_{MC}} \sim 4 \times 10^3~\msun$ vs $\overline{M_{MC}} \sim 2.5\times 10^3~\msun$).  The total mass in MCs is roughly twice as large in the low-SFE run with respect to high-SFE run: $2\times 10^6~\msun$ vs $0.9 \times 10^6~\msun$, respectively. Since the sub-grid SFE is half in the low- vs high-SFE runs, the total mass in stars in the two runs is nearly identical, highlighting the self-regulation produced by the feedback loop mentioned before. This also means that the MCs in the high-SFE run are fewer but on average, more massive than in the low-SFE run, as shown in the left panel of Fig.~\ref{fig:MC IMF}. 

If we look back at Fig.~\ref{fig:SFC Mean Density vs Metallicity} (left), we notice that MCs in the high-SFE run are on average more dense and more metal-poor. Since the radii of the MCs are similar between the two simulations -- Fig.~\ref{fig:MC IMF}, right panel shows a similar PDF with a peak value at $R_{MC} \sim 1.5$~pc and a range between $1$ and $3$~pc -- we interpret the larger masses as due to the lower metallicities (and in turn the higher densities) of the MCs in the high-SFE run. 

According to Eq.~(\ref{eq:rcore}), when turbulence is negligible, we expect $R_{MC} \approx 0.57[(1+z)/10]$~pc, independent of the gas temperature. Hence, the observed range of radii suggests values of the turbulent Mach number in the range $1\lesssim {\cal M}_{{tur}}\lesssim 5$. In addition, Fig.~\ref{fig:SFC Mean Density vs Metallicity} (right) shows that the MC radii are larger for higher metallicity clouds, suggesting that these clouds have a larger turbulent Mach number. This is not surprising given that these clouds are colder, hence the lower sound speed increases the Mach number even at fixed $\sigma_{{tur}}$. Another interpretation that would produce a similar trend for the radii is that in the colder (higher-metallicity) clouds, the temperature profiles are not perfectly isothermal: rather, the gas gets hotter in the outer parts decreasing the slope of the density profile, increasing the cloud's radius.

We also observe a multi-modal metallicity distribution consistent with episodic star formation (middle panel in Fig.~\ref{fig:MC IMF}). At $z\sim 8.4$, we see two modes for the low-SFE run. In the high-SFE run, there appears to be a third low-metallicity mode, reflecting the three major episodes of star formation in that run. We expect the number of modes to increase as the simulation further progresses for both simulations.
Bimodal metallicities in the GC stellar populations is a well-documented phenomenon in the local universe at $z = 0$ \citep{Peng:2006}, with populations differing in metallicity by around a factor of 10 and peaks at around $1-0.1$~Z$_\odot$  for the red-population and $0.1-0.01$~Z$_\odot$ for the blue population. However, it is worth noting that {\em GCs with metallicities below} $10^{-3}$~Z$_\odot$ {\em have not been observed so far} \citep{Forbes:2018RSPSA.47470616F, Beasley:2019MNRAS.487.1986B, Wan:2020Natur.583..768W}, though some have observed GC remnants with metallicities bellow this metal floor \citep[\eg,][]{Martin:2022Natur.601...45M}. The bimodal population observed in many galaxies with masses similar to the Milky Way is typically consistent with models in which the red populations forms in-situ in the galaxy and the blue population is dominated by accreted GCs from dwarf galaxies \citep{Muratov:2010, KatzR:2014}.

We observe a population of BSCs with metallicities of about $10^{-3} ~{Z_{\odot}}$ in both runs. However, there are no Pop~II star clusters with metallicities below $2\times 10^{-4}$~Z$_\odot$, even though the critical density for Pop~II formation in our simulations is $Z_{crit}=10^{-5}$~Z$_\odot$. In particular, in the low-SFE run, the blue population has metallicity $Z\sim 10^{-3}$~Z$_\odot$ and the red population  $Z\sim 5\times 10^{-3}$~Z$_\odot$. In addition, in the low-SFE run the metallicity distribution of MCs (and of BSCs) is broader (\ie, more smooth) and reaches higher values ($Z\sim 10^{-2}$~Z$_\odot$) than in the high-SFE run. This is expected because the weaker and more numerous star formation episodes in the low-SFE run produce a more gradual enrichment of the gas, with fewer of the metals ejected from the galaxy through galactic winds during each burst.

We have already seen that the galaxies in the two runs form a nearly identical mass in stars \mstartotal, regardless of the sub-grid SFE. This also means that the total metal production in the two simulations is comparable. Most of the metals should be in the interstellar medium and a fraction expelled from the galaxy. However, the mass in metals locked in MCs is nearly 4 times higher in the low-SFE run than in the high-SFE run: $88~\msun$ compared to $23~\msun$, assuming $Z_\odot=0.0134$. Given that the mass in MC is roughly twice in the low-SFE run, we find that the mean metallicities of the gas in MCs (and of the stars) are $3\times 10^{-3}$~Z$_\odot$ and $1.9\times 10^{-3}$~Z$_\odot$, respectively. The lower metal enrichment of the stars in the high-SFE run can be due to a combination of stronger galactic winds and the more bursty nature of star formation, with longer quiescent periods in which the gas is gradually enriched by SNe but the stars are not (since they are not forming).

Furthermore, in the high-SFE run, the most massive MC produces a star cluster with a mass of $1.2 \times 10^{4} ~\mathrm{M_{\odot}}$ and metallicity of $6 \times 10^{-4}~\mathrm{Z_{\odot}}$. In the low-SFE run, the most massive MC produces a  $6.5 \times 10^{3} ~\mathrm{M_{\odot}}$ star cluster with a metallicity of about $4 \times 10^{-3}~\mathrm{Z_{\odot}}$, suggesting that the most massive star-forming MCs, in both runs,  form during the first burst of star formation and are in lowest metallicity mode of the distribution. This is consistent with the idea that lower metal enrichment requires a higher cloud core density -- and therefore on average, a higher mass cloud -- to reach the critical star-forming density.

In summary, the sub-grid SFE does not determine the total stellar mass in a galaxy and is instead regulated by the implementation of feedback processes in the simulation. However, it can determine the masses of MCs and star clusters, and the metallicity distribution of the different stellar populations within the galaxy. Even though the stellar mass in the two runs is basically the same, the lower sub-grid SFE run forms star clusters with roughly 40 per cent higher metallicity than in the high-SFE run, just during the first $\sim$~300 Myr of star formation. A consequence of the different metallicity distribution of the MCs is that the density and masses of MCs are on average higher for lower-metallicity clouds (\eg, in the high-SFE run). 

\begin{figure}
    \includegraphics[width=0.95\columnwidth]{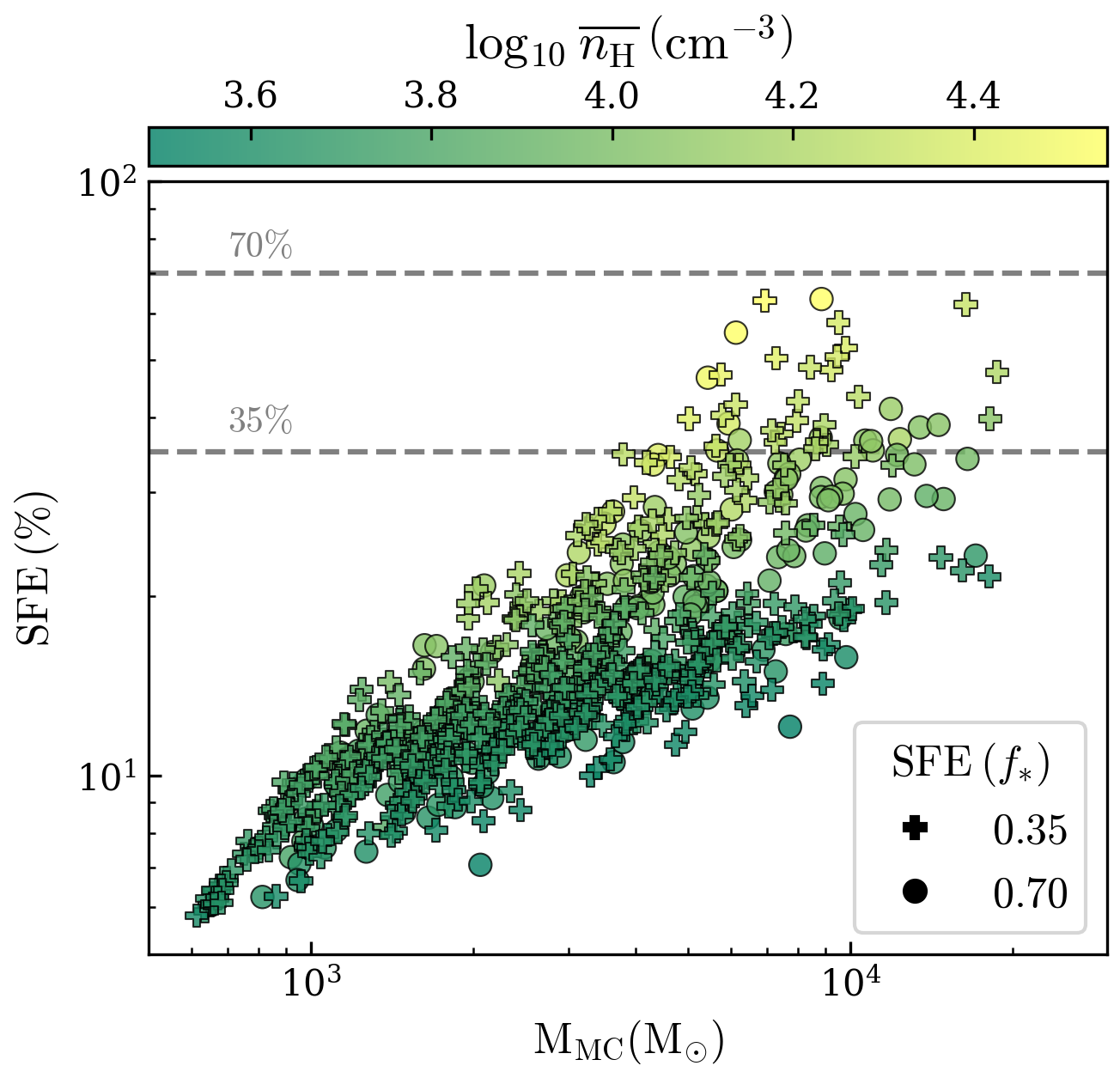}
    \caption{Theoretical estimate of the true sub-grid star formation efficiency as a function of MC mass, $M_{MC}$, in the two runs in which we instead assumed a constant sub-grid $f_*$ (see legend). Note that the derived SFEs, as well as the MC properties, are relatively insensitive to the assumed $f_*$. We use Eq.~(\ref{Eq SFE}), derived in \citet{HeRG:2019} based on AU-scale radiation-MHD simulations of star formation in MCs, to estimate the SFE in each MC, based on the MC mass, mean density, and metallicity. The markers (see legend) are colored according to the mean hydrogen number density in each MC.} 
    \label{fig:SFE} 
\end{figure}
\subsubsection{Are the assumed values of the sub-grid SFEs realistic?}
\label{ssec:Are the assumed values of the sub-grid SFEs realistic?}
In our simulations, we run two identical simulations assuming two different constant values for the sub-grid SFE (35  per cent  or 70  per cent ) to study how the properties of the star cluster population depend on that assumption. The two values we choose are sufficiently large to produce bound star clusters but are these values realistic; \ie, within the range we expect based on properties of the star-forming MCs in the simulations? Simulations of star formation in MCs at few tens or hundreds of AU scale can help us answer this question. Most of these simulations show that the SFE in MCs increases with increasing mean gas density and mass  \citep{HeRG:2019, Fukushima2019, FukushimaHajime:2020}. If we adopt the fitting function for the SFE proposed in \cite{HeRG:2019}:
\begin{equation}
   f_{*} =
   \min\left[0.9,
     0.004
     \left(\frac{Z_{MC}}{10^{-3}~{\rm Z}_\odot}\right)^{0.25} 
     \left(\frac{M_{MC}}{10^{4}~{\rm M}_\odot}\right)^{0.4} 
     \left(1+\frac{\overline{n_{{H}}}}{n_{0}}\right)^{0.91}
     \right],
     \label{Eq SFE}
\end{equation}

\begin{figure*}
	\includegraphics[width=0.95\textwidth]{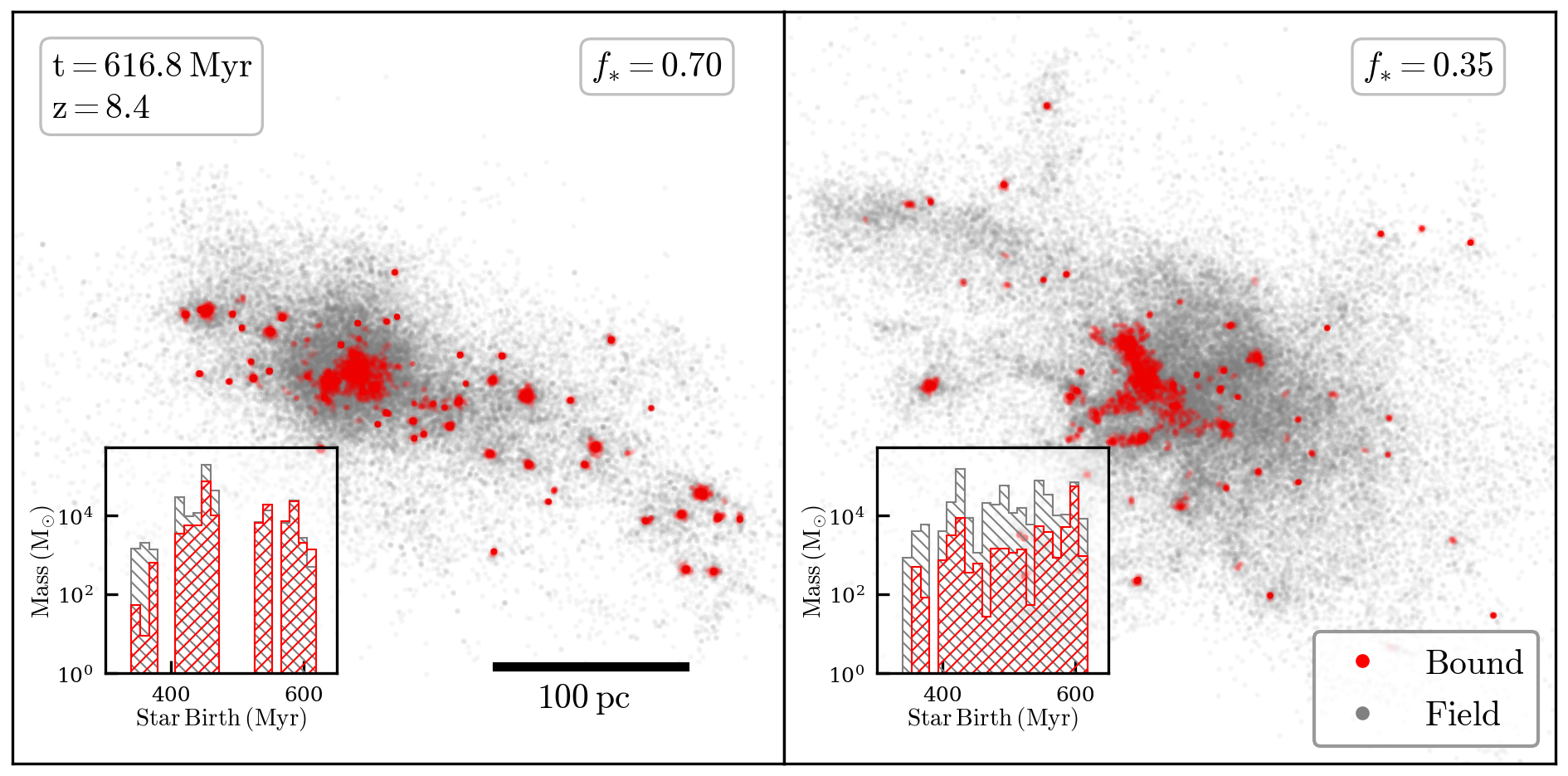}
	\centering
    \caption{A spatial plot of Pop II stars within the galaxy for the two sub-grid SFE runs (\textit{left} and \textit{right} are for $f_{*} = 0.70$ and $f_{*} = 0.35$, respectively). Star populations belonging to BSCs are colored in red, while field stars are in grey. Panel insets show overlapping histograms (use the histogram hatch fill as a guide and note that the y-axis is in log-scale) of the Pop II stellar mass contribution as a function of star birth time in Myrs for both bound and unbound star systems. We emphasize that the colors specify the status of stars at the time of this snapshot, indicated at the top left corner.}
    \label{fig:Bound Unbound Scatter} 
\end{figure*} 
where $n_{0} = 100 \: {\rm cm}^{-3}$, we can estimate the expected sub-grid SFE in each of the MC in our simulations as a function of MC gas mass ($M_{MC}$), metallicity ($Z_{MC}$), and mean hydrogen number density $\left(\overline{n_{{H}}}\right) $. The sub-grid SFE using Eq.~(\ref{Eq SFE}) is shown in Fig.~\ref{fig:SFE} for both simulations as a function of MC mass. The metallicity dependence is poorly known. Simulations show that decreasing the metallicity from solar to $\sim 1/100$~Z$_\odot$ reduces the star formation efficiency roughly by a factor of 5. This is due to stronger feedback produced by \HII regions at low metallicity, which are hotter and therefore more pressurized. However, the functional form of the dependence on $Z$, as well as the efficiency at $Z<10^{-3}$~Z$_\odot$ typically found in the compact mini star-cluster in our simulations, is unknown.
Note that, according to Eq.~(\ref{Eq SFE}), in the regime when $f_*$ is proportional to $\overline{n_H}^{0.91}$ (\ie, $n \gg n_0$), the value of $m_{SC}=f_*M_{MC}$ is not sensitive to our sub-grid assumption of MC size $R_{MC}\propto N_{{cut}}^{1/2}$, defined as the radius where the density of the isothermal sphere becomes less than $n_{crit}/N_{{cut}}$. This is because, assuming an isothermal-sphere profile, $M_{MC} \propto R_{MC}$ and $\overline{n_H} \propto R_{MC}^{-2}$, hence $M_* \propto R_{MC}^{-0.42} \propto N_{{cut}}^{-0.21}$. However, for the reasons explained in the next paragraph, the fraction of star formation in bound star cluster would likely become smaller if we increase $N_{{cut}}$, because $f_* \propto R_{MC}^{-1.42} \propto N_{{cut}}^{-0.71}$ would decrease.

Simple analytic arguments \citep{1980ApJ...235..986H} and more sophisticated simulations \citep{Shukirgaliyev:2017A&A...605A.119S,HeRG:2019} show that if the SFE in a star-forming MC is $>20 - 30$ per cent, the resulting star cluster should remain bound even after it has lost all its natal gas. We find that the expected sub-grid SFE in the majority of MCs in both simulations, using Eq.~(\ref{Eq SFE}), range between 10 per cent for the smallest mass clouds ($10^3$~M$_\odot$) and 70 per cent for the most massive MCs (few $10^4$~M$_\odot$). So, we expect that all but the least massive star clusters should form self-gravitating bound star clusters, despite their small ($\lesssim 2\times 10^4$~M$_\odot$) masses when compared to today's GCs. As shown later in \S~\ref{ssec:dynamics}, the short-term survival rate of these star clusters changes between our two simulations and it is typically low for the smallest mass star clusters.

Clearly, the relationship in Eq.~(\ref{Eq SFE}) is a more realistic choice for the sub-grid SFE than our adopted constant values. However, the emphasis of this paper is to understand the physics of star cluster formation rather than produce the most realistic galaxy at high-z we can at this time: we will publish results from such a simulation in a follow-up paper. Though, we stress that Eq.~(\ref{Eq SFE}) is rather uncertain for several reasons: (i) most MHD simulations of star formation in turbulent MCs adopt idealized initial conditions (ICs) instead of realistic ones extracted from larger scale simulations; (ii) the results at low-metallicities ( $Z <10^{-2}$ ~Z$_{\odot}$) are not well known/understood; (iii) the value of the initial magnetic field in the cloud (unknown in our simulations) can affect the SFE, the duration of the starburst and the IMF of the stars \citep{Vazquez:2011, Kim:2021, Hix:preprint}.

\subsubsection{Fraction of stars in BSCs}
\label{ssec:Fraction of stars in BSCs and their mass functions}
We use a basic friends-of-friends (FOF) algorithm \citep{Efstathiou:1985} operating exclusively on the Pop II particles to extract the boundaries of each bound star cluster (BSC). The halo-finding algorithm returns the centers of the BSCs, their constituent star particles (with their IDs and positions), as well as their virial radii $\left( r_{\mathrm{200}} \right)$. However, we do not attach any meaning to $r_{\mathrm{200}}$; instead, we use physically motivated quantities such as the half-mass radius to quantify the extent of the BSCs. Fig. \ref{fig:Bound Unbound Scatter} shows the results of the FOF algorithm on both runs at a redshift near \zend, when ran with a linking length $l_{\mathrm{link}} = 10^{-4}$, which groups particles whose fractional distance relative to the mean particle separation is $< l_{\mathrm{link}}$. We tested different values for this unit-less parameter until we found the smallest amount of sub-clustering within the BSCs. The insert within each figure shows a histogram of the mass contribution -- for field stars (in grey) and cluster stars (in red) -- as a function of formation times of the stars. Note that we show the histograms in log-scale. For the low-SFE run, the mass in field stars consistently exceeds that of in bound star clusters. In other words, stars formed during each star-forming episode, by redshift \zend, are more likely to end up in an unbound configuration than in a BSC for the low-SFE galaxy. The contrary is true for the high-SFE run. Also, in both simulations, the probability of stars becoming (unbound) field stars by \zend\ ($t = 617$~Myr) is higher for stars formed in earlier star formation episodes, as bound star clusters dissolve over time.

\begin{figure}
	\includegraphics[width=\columnwidth]{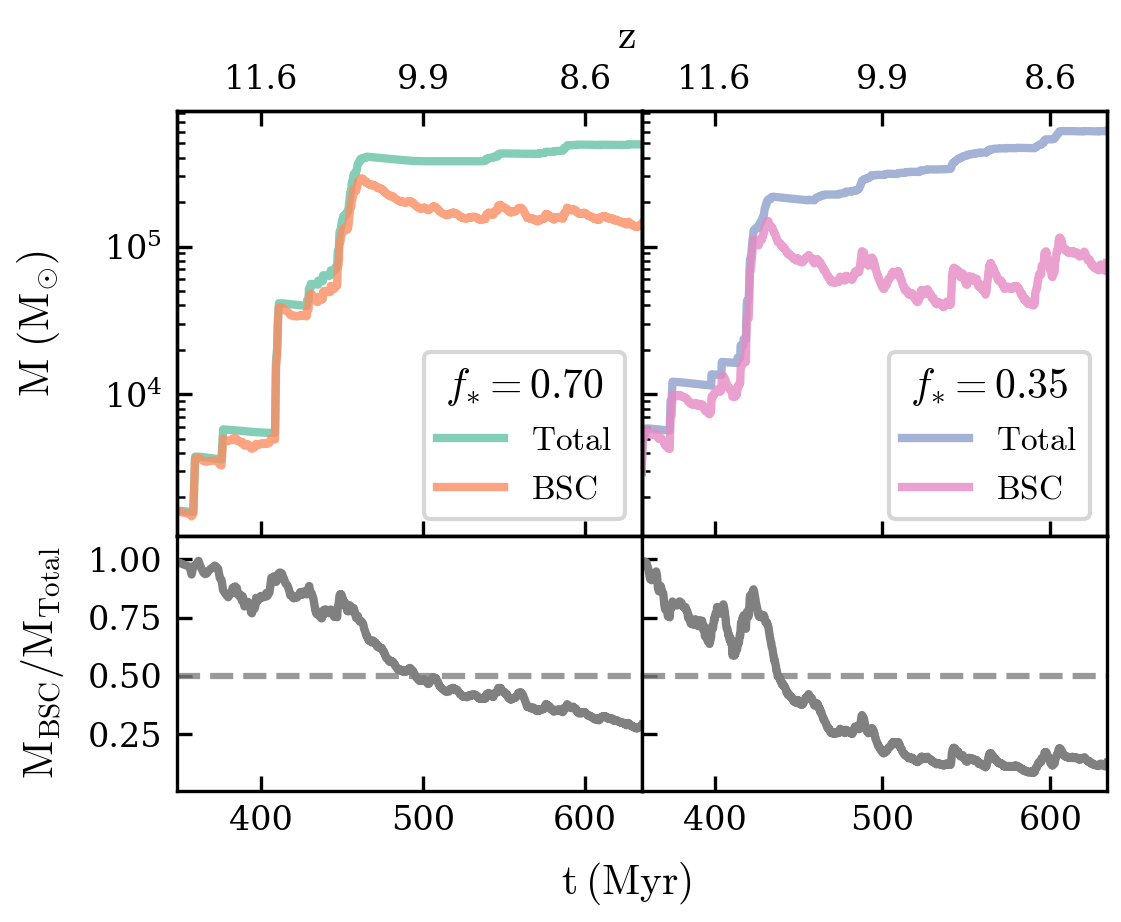}
	\centering
    \caption{The total Pop~II star mass within the simulation box along with the total mass in BSCs as a function of time (see legend). The \textit{top row} shows the mass in Pop~II stars as a function of time and redshift for $f_{*} = 0.70$ (\textit{left column}) and $f_{*} = 0.35$ (\textit{right column}) simulations. The \textit{bottom row} shows the fraction of the total mass residing in bound star clusters as a function of time for each simulation. The dashed line shows $\mathrm{M_{BSC} / M_{Total} = 0.50}$, which indicates an equal stellar mass contribution from BSCs and field stars.}
    \label{fig:Mass Bound Unbound} 
\end{figure} 

The trends discussed above are shown more quantitatively in Fig.~\ref{fig:Mass Bound Unbound}, which shows the mass in BSCs ($M_{BSC}$), the total stellar mass, and their ratio as a function of time. During the first starburst phase (from $z\sim 13$ to $z\sim 11$), stars are formed nearly exclusively in BSCs in both runs, independent of the sub-grid SFE. Indeed, the ratio of mass in BSCs to the total mass in stars is between 50 per cent and 100 per cent from $t \sim 350$~Myr to $450$~Myr. Later, during a more quiescent star formation period, this ratio starts decreasing in both runs, reflecting the partial destruction of BSCs. However, the rate at which BSCs dissolve differs between the two sub-grid SFE runs. In particular, in the high-SFE run, more than half of stars are still locked in BSCs about 150~Myr after the onset of star formation at $t \sim350$~Myr, compared to the low-SFE run in which it takes less than 100~Myrs to reach the same 50/50 fraction of field-stars/BSCs. 

Note that all stars in both simulations form in compact MCs, so there is no star formation in the "field" (\ie, in low-density gas). However, MCs with small mass and low $f_*$ form gravitationally unbound "open star clusters" that dissolve in the field on a few dynamical timescales (\ie, in a few Myrs).

Furthermore, the later evolution (after reaching 50  per cent bound mass fraction) of the BSC fraction also plays out quite differently between the two runs. In the low-SFE run, the formation rate of new BSCs eventually reaches an equilibrium with the destruction rate of previously formed BSCs, such that the BSC fraction remains fairly constant over time at 10 per cent in mass. For the high-SFE run, this equilibrium is not quite reached: the fraction of BSCs slowly decreases over time, but it is 25  per cent by the end of the simulation, much higher than in the low-SFE run.

\begin{figure}
	\includegraphics[width=\columnwidth]{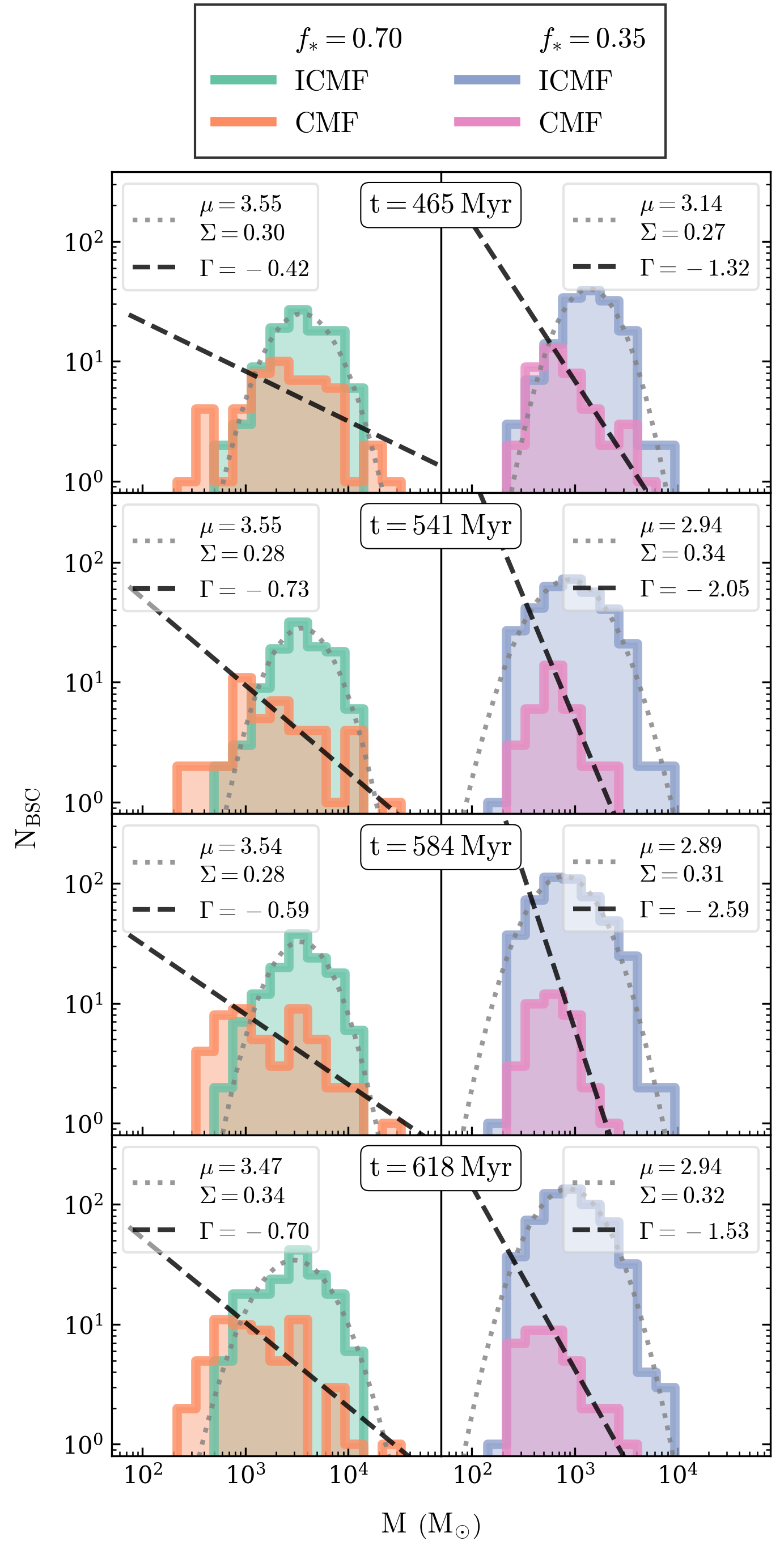}
	\centering
    \caption{Time evolution (\textit{top to bottom}) of the BSC mass function (CMF) for both sub-grid SFEs (see top legend). The \textit{first row} depicts each galaxy right after the initial, and biggest, starburst at $t = 465$ Myr (refer back to Fig.~\ref{fig:SFC Mass Over Time}). The \textit{second row} shows the CMFs during a period of stagnation in star formation. The \textit{third row} shows the CMFs at $t = 584$ Myr, during a period of star formation in both galaxies, although milder than the first burst (top row). Finally, the \textit{fourth row} samples the CMF right after this star formation period. The black dashed lines show power-law fits to the CMFs with slope $\Gamma$. In all panels, the $\mathrm{ICMF}$ is also depicted for comparison, and fitted with a log-normal distribution (grey dotted line).
    }
    \label{fig:CMF} 
\end{figure} 
\subsubsection{BSC mass functions}
The evolution of the BSC mass function, that hereafter we refer to as cluster mass function (CMF), can give us further insights into the process of BSC dissolution. The BSCs evolve over time through internal and external physical processes, the specifics of which we will delve into in the discussion section (\S~\ref{sec:dis}). For now, we will focus on trends we observe. The time evolution of the CMF is shown in Fig.~\ref{fig:CMF} for both the high and low-SFE runs. For reference, along with the CMF, we show the zero-age initial star cluster mass function (ICMF) at the same snapshot, obtained from the molecular cloud mass function ($\mathrm{m_{SC}}\equiv f_* M_{{MC}}$), which depicts the mass converted into stars for a given MC at the onset of star formation. The grey dotted lines show the non-linear least-squares fit using Eq.~(\ref{Eq Gaussian}) in linear-log space for each distribution of the ICMF. The initial log-normal distribution transitions to a power-law CMF due to the destruction processes of star clusters, which we show on each panel along with the extracted best fit power-law slope $\Gamma$. This CMF fit primarily serves to quantify in which mass bin most of the mass in BSCs resides; \ie, a shallower slope means a larger fraction of BSCs occupies the high-mass end.

We see that the CMF evolves as new stars -- and subsequent BSCs -- are formed. Each row depicts the galaxies at a specific stage of their evolution. As we see in Fig.~\ref{fig:CMF}, although the CMFs are fairly stable throughout the galaxies' evolution, starbursts replenish the BSC population. This has the effect of flattening $\Gamma$ right after each star-forming period. Then, the low mass-end gets truncated and the high-mass end shifts to lower masses. This truncation and shift are due to the dissolution of the smallest star clusters formed and the reduction of mass of the most massive BSCs. This effect can be seen when comparing the first and second rows, whereby the $\Gamma$ grows steeper. Similarly, comparing the second to the third rows, we see a flattening due to new BSC formation. The final row shows a steepening which follows the starburst in both simulations.

The main result in Fig.~\ref{fig:CMF} is that although the CMF shows some time dependence, the high-SFE run is on average much flatter ($\Gamma \sim 0.6 \pm 0.1$) than in the low-SFE run ($\Gamma \sim 2 \pm 0.5$). This means that in the high-SFE galaxy, most of the mass in BSCs is locked in the few most massive clusters, while low-mass BSCs dominate the mass budget in the low-SFE run.
\begin{figure}
	\includegraphics[width=\columnwidth]{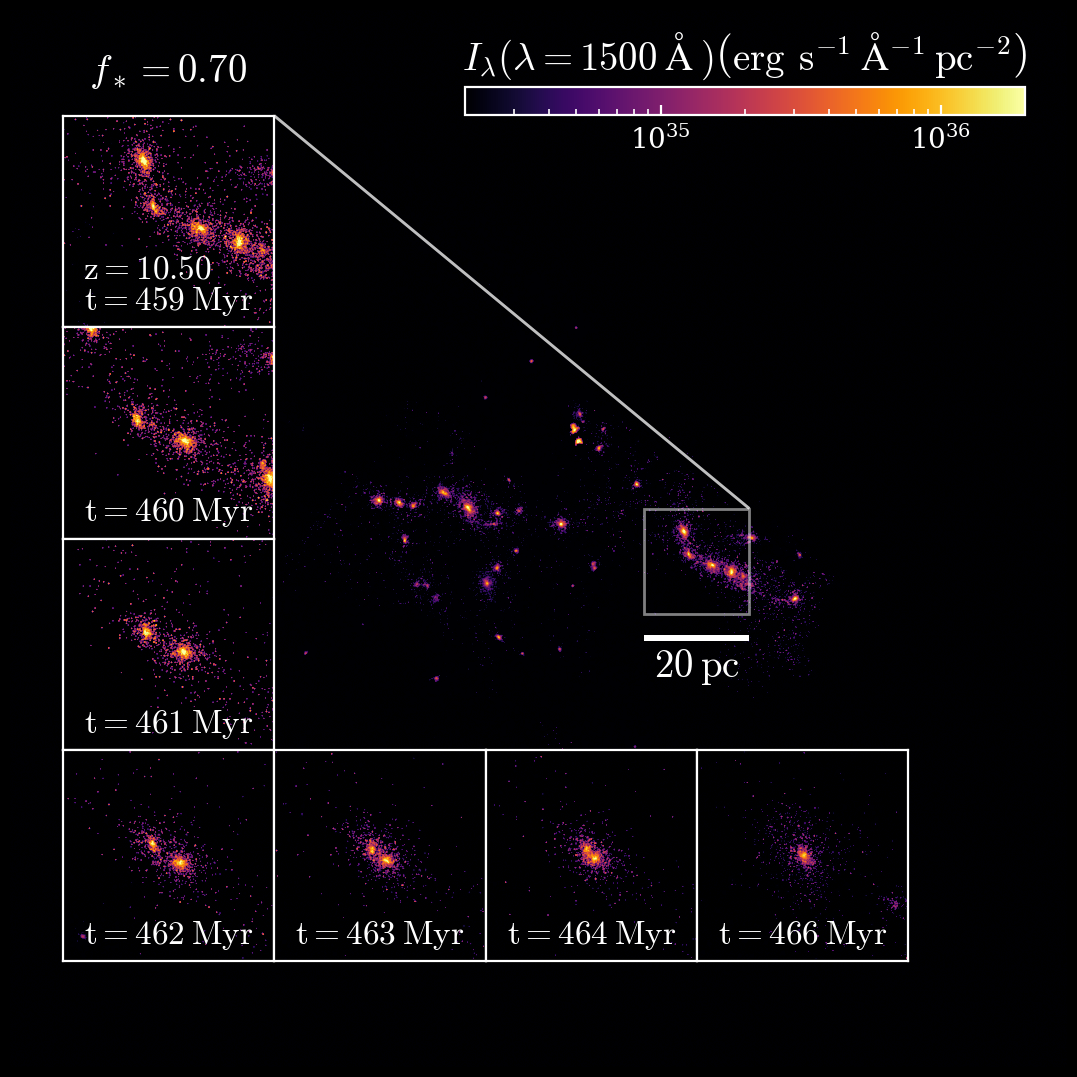}
	\centering
    \caption{Specific surface brightness $I_\lambda$ at $\lambda=1500 \text{\AA}$ (rest frame) for the galaxy with sub-grid SFE $f_{*} = 0.70$ at redshift $z=10.5$. The inset shows the time evolution (\textit{top left}, then \textit{counter-clockwise}) of a group of BSCs in which two of them merge to become a massive and compact BSC. Note, we used models which give $L_{\lambda}$ at  $\lambda= 1500 \: \text{\AA}$ as a function of star mass and age, assuming instantaneous star formation \citep{1999ApJS..123....3L}.}
   
    \label{fig:BSC Merger} 
\end{figure} 
\subsubsection{Early-stage mergers of BSCs}
The most notable difference between the populations of BSCs in the high and low-SFE runs is the high-mass end of the CMF: the former contains BSCs as massive as a few $10^4\: \msun$, while the latter has a maximum BSC mass of a few $10^3\: \msun$. Note that in the high-SFE run, the $\mathrm{ICMF}$ has a maximum value of $\lesssim 10^4 \: \msun$, less than the most massive BSC in that galaxy. This can only be explained by a merger of two or more BSCs.

Fig. \ref{fig:BSC Merger} captures such a merger of two BSCs born from separate MCs. The figure shows the surface brightness of the stars in the rest frame UV (at $\mathrm{\lambda}$ = 1500 \text{\AA}) for a snapshot in the high-SFE run. We compute the spectral flux density at this wavelength using a stellar luminosity model from Starburst99 \citep{1999ApJS..123....3L} via an age-based look-up table. We use a model with Salpeter IMF, stellar metallicity $\mathrm{Z = 10^{-3}\: Z_{\odot}}$ and instantaneous star formation of a constant stellar mass, which we scale to the mass in stars in each pixel. With the stellar luminosities attributed to a given age, we then project and sum individual luminosity contributions along the line of sight. What results is the specific surface brightness of the galaxy at a given time $I_{\lambda} \: (\lambda =1500 \text{\AA}) $.

We observe that distinct star clusters forming nearly simultaneously ($\Delta t_{\mathrm{SF}} \lesssim 0.50$~Myr) and near one another, can merge into a single bound object. These mergers are common in both simulations as the result of nearby stellar clusters forming from MC fragments that reach the star-forming threshold independently. These objects sometimes remain bound, surviving initial gravitational disruptions, and then subsequently merge over a span of $\lesssim 10$ Myr. This is not surprising as, even in idealized simulations of MCs, we observe hierarchical sub-structures within MCs \citep{HeRG:2019}.

Early-stage mergers do also occur in the low-SFE run, though, most mergers typically produce BSCs with lower masses than the high-mass end of the $\mathrm{ICMF}$. This is produced by the higher natal mass-loss in this simulation, a process we will discuss in \S~\ref{sec:dis}. In the low-SFE run, the high mass end of the CMF is shifted by almost an order of magnitude to lower masses with respect to the $\mathrm{ICMF}$ because of significant star mass loss into the field at the onset of cluster formation. Radiative feedback is the dominant culprit for unbinding young star clusters, ejecting most of the natal cloud, and taking some of the potential from the cluster along with it. The CMFs for the high-SFE galaxy are also consistently wider in comparison to the low-SFE galaxy. This suggests a more diverse BSC population for the high-SFE galaxy, an idea we will expand upon in \S~\ref{ssec:morpho}. In the next subsection, we will further analyze the diverging properties of the star cluster population by comparing the dynamical evolution of BSCs in each of the two runs.

\subsection{Dynamical Evolution and Mass Loss of Star Clusters}
\label{ssec:dynamics}
To quantify the morphology and dynamical evolution of the star clusters, we use a generalized version of the empirically-derived King's profile \citep{1962AJ.....67..471K} to fit the projected surface density $\Sigma$ of the clusters as a function of radial distance from the center r: 
\begin{equation} 
    \Sigma (\mathrm{r}) = \Sigma_{\mathrm{bg}} + \frac{\Sigma_0}{1 + (\mathrm{r} / r_{\mathrm{core}})^{\alpha}},
    \label{Eq King}
\end{equation}
where we introduce a power-law fitting parameter $\alpha$ (with $\alpha=2$ in King's model), central surface mass density $\Sigma_0$, and background surface mass density $\Sigma_{\mathrm{bg}}$, in addition to the characteristic core radius $r_{\mathrm{core}}$ (not to be confused with the MC core radius $R_{\mathrm{core}}$). See Appendix \ref{sec:Profiling BSCs} for a demonstration of how the clusters are fitted. In addition to these parameters, we also derive other characteristic properties of the BSCs; namely, the mass of the core $m_{\mathrm{core}}$ which is simply the mass integrated from the center to $r_{\mathrm{core}}$, the half-mass $m\mathrm{_{half}}$, and the half-mass radius $r\mathrm{_{half}}$ which is the radius within which contains $m\mathrm{_{half}}$.  

\begin{figure}
	\includegraphics[width=\columnwidth]{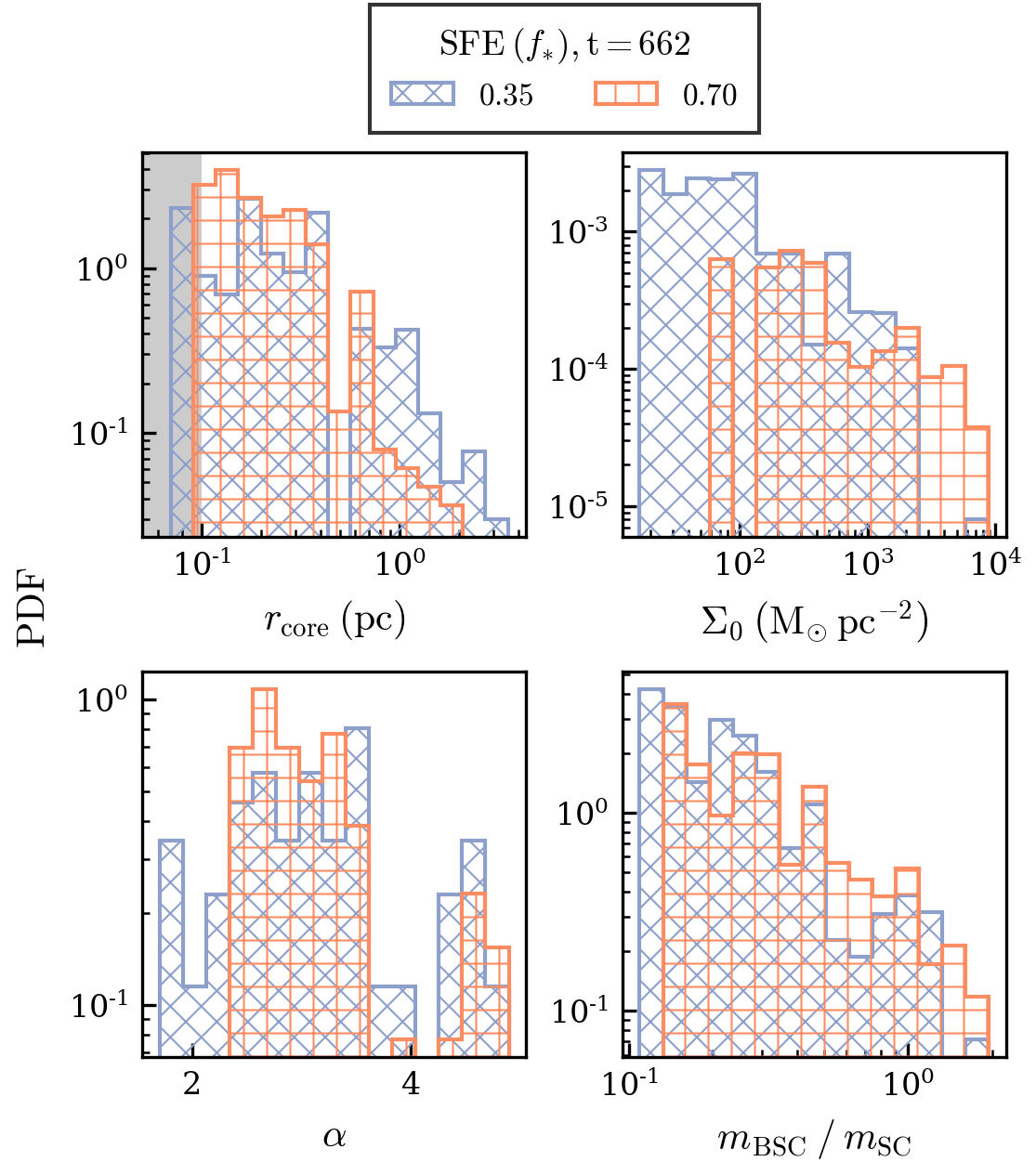}
    \caption{A collection of histograms depicting the PDFs for the following BSC properties (\textit{top left}, then \textit{clock-wise}): the core radius, $r_{\mathrm{core}}$ (the region where the radii are smaller than the softening length, $\le 0.1$~pc, is shaded in grey); the core surface mass density, $\Sigma_0$; the ratio between the mass of the BSC and the total mass in stars produced by the MC ($m_{SC} = f_* M_{MC}$) in which the BSC formed, $m_{{BSC}} / m_{SC} $; and the power-law slope $\alpha$ of the surface mass density profile. Refer to Eq.~\ref{Eq King} for the definition of these parameters. }
    \label{fig:Parameter Histogram} 
\end{figure}

\begin{figure*}
	\includegraphics[width=\textwidth]{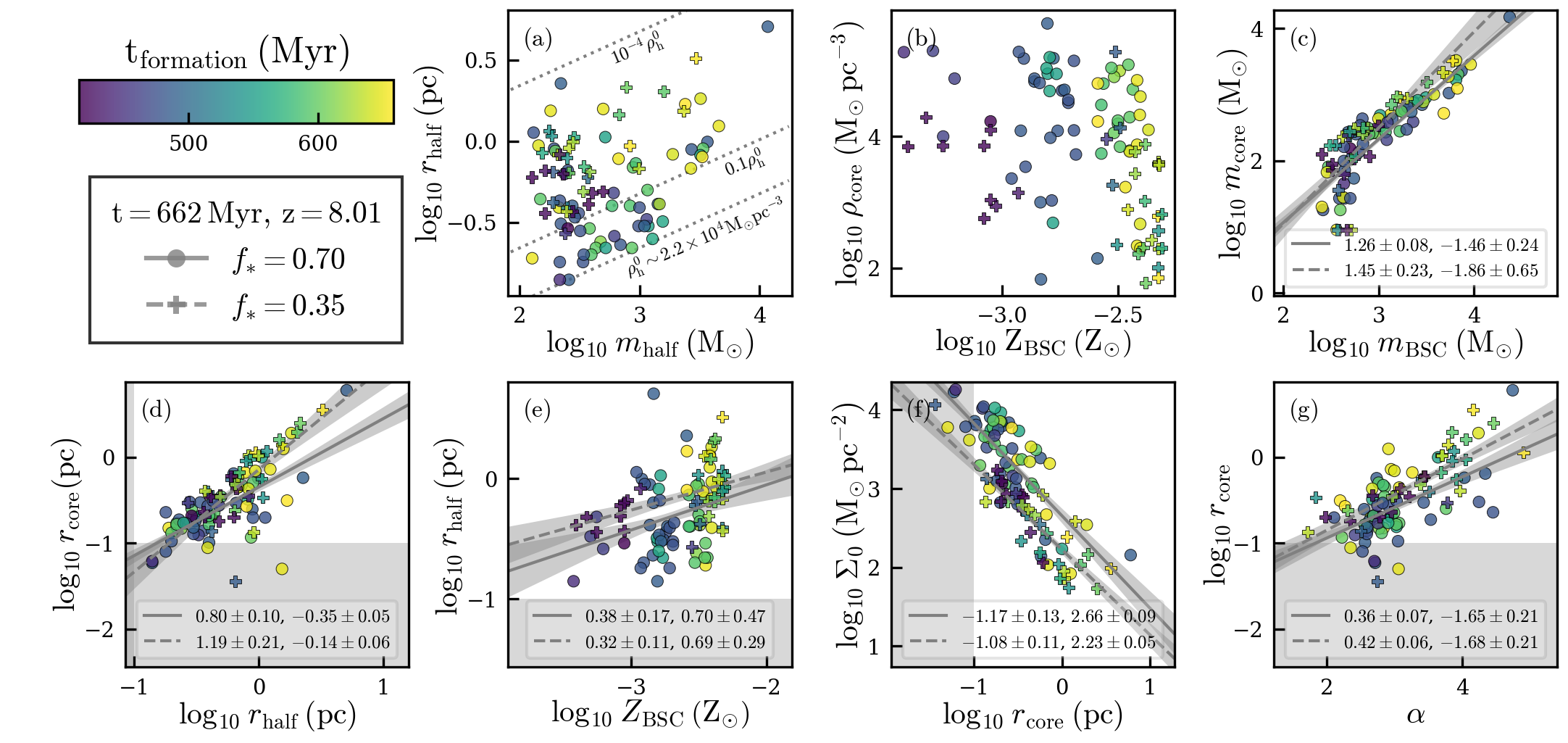}
    \caption{Scatter plots relating the BSC properties. The circles refer to the run with $f_{*} = 0.70$ and the crosses $f_{*} = 0.35$. The symbols are colored corresponding to the BSC formation times. Corresponding linear fits in log-log space are depicted in subplots (c) - (g), see legend. Along with fits are the best-fit parameters: $\mathrm{\log_{10} (slope, intercept)}$ and corresponding $1\mathrm{\sigma}$ errors. The shaded region around each line represents the $1\sigma$ confidence band for each fit; \ie, the region where the best fit line lies with a confidence of 68 per cent. The different subplots relate: (a) the half-mass radius as a function of half-mass, (b) core density as a function of metallicity, (c) core mass as a function of the cluster mass, (d) core radius as a function of half-mass radius, (e) the half-mass radius as a function of metallicity, (f) the central surface density as a function of core radius, and (g) the core radius as a function of $\alpha$. Note, the resolution limit of $\Delta x = 0.1$~pc is indicated by a grey region, where appropriate.}
    \label{fig:BSC Scatter Plots} 
\end{figure*}

\subsubsection{Properties of the bound star cluster population}
\label{ssec:bsc composition}
Differences in statistical properties of the BSC population obtained by assuming different values for the SFE in MCs are clearly conveyed in Fig.~\ref{fig:Parameter Histogram}, showing the probability distribution functions for $r_{\mathrm{core}}$, $\Sigma_0$, $\alpha$, and $m_{BSC}/(f_*M_{\mathrm{MC}})$ (the ratio of the current mass of the bound star cluster to their mass at formation) at the last output ($t \sim 632$~Myr) for the two simulations\footnote{For the case in which a single BSC has multiple progenitors, we attribute the most massive progenitor mass $f_*M_{MC}$ as its initial stellar mass.}. We see that the BSCs in the high-SFE run have consistently smaller cores and higher central surface densities by almost an order of magnitude on average. The distribution of $\alpha$ shows that the surface density profiles have a power-law slope quite close to a King's profile ($\alpha=2$), but most often with a steeper power-law index $\alpha \sim 3 \pm 1$. This could be due to deviation from a spherically symmetric cluster, perhaps due to tidal interactions or mergers/substructure within not fully relaxed star clusters. This is not surprising given the young age of the clusters and the crowded environment in which they reside. The bottom right panel in Fig.~\ref{fig:Parameter Histogram} presents further evidence of the extent by which early-stage mergers (mentioned in \S~\ref{ssec:Fraction of stars in BSCs and their mass functions}) affect the BSC population across efficiencies. Here, $m_{{SC}}=f_* M_{{MC}}$ is the mass of a cluster immediately after birth; \ie, the total mass in stars that a single MC produced. The same star cluster is tracked using its time of formation. Its mass, $m_{{BSC}}$, is then measured at $t=662$~Myr. The ratio $m_{{BSC}} / m_{{SC}}$ spans the range $1/10$ to $2$, illustrating that: (i) BSCs can lose up to 9/10 of their initial mass before being completely destroyed and, (ii) the existence of star clusters with a ratio greater than unity, especially prominent in the high-SFE run, suggests that a sizable number of BSCs is the result of mergers between smaller sub-units (see Fig.~\ref{fig:BSC Merger}). This effect is less evident in the low-efficiency run (where the probabilities are lower near a ratio of unity), likely because the more efficient natal mass loss dominates over the merger-induced mass growth. This corroborates the result shown in Fig.~\ref{fig:CMF}, where the high-mass end of the CMF in the low-SFE run is systematically lower than the ICMF high-mass end.

We now turn out attention to and explore the correlations between structure parameters of the BSC population in the two simulations, as illustrated in Fig.~\ref{fig:BSC Scatter Plots}. We see a weak correlation (panel \ref{fig:BSC Scatter Plots}a) in the mass-radius relation (MRR) at the half-mass radius $r_{{half}}$, consistent with a star cluster population that has undergone a combination of tidal disruption and two-body relaxation \citep{Gieles:2016}. Keeping our attention to panel a, we draw lines of constant mean density within the half-mass radius: $\rho_{h} \equiv 3 m_{half} / 4 \pi r^3_{half} $. We find that the densest BSC (at t = 662~Myr, $z \sim 8$) has $\rho^0_{h}\sim 2.2 \times 10^4~\msun~{\rm pc}^{-3}$ and was produced in the run with high sub-grid SFE. The lowest density BSC has a density of around $10^{-4}\rho^0_{h}$. The occupancy of the plot can offer insight into the types of dynamical evolution under which each cluster is undergoing. The bottom left of the plot is filled by older high-density objects, most of which are formed in a high-SFE run. These BSCs tend to be less affected by tidal disruption and therefore can stay bound for a longer time as they eventually succumb to evaporation via two-body relaxation. Near the $10^{-4} \rho^0_{h}$ line is a population of loosely bound, mostly young objects in the process of destruction via a combination of tidal and initial gas expulsion. 

Contrary to the tight relationship observed between MC mean density and their gas metallicity (Fig.~\ref{fig:SFC Mean Density vs Metallicity}), there is not a strong correlation between the core density and the metallicities of BSCs (panel \ref{fig:BSC Scatter Plots}b). However, we see that the densest cores are formed in the run with high-SFE in MCs. At later times, we see a larger spread in core densities. The dynamical fate of these objects has yet to be determined. However, if history foretold, we expect these young, more loosely bound objects ($\rho_{{core}} \lesssim 10^2~\msun~\mathrm{pc}^{-3}$) to ultimately meet the same fate as their older counterparts and get tidally destroyed. We see that the relationship between the core mass $m_{{core}}$ and the total mass $m_{{BSC}}$ of BSCs is similar across simulations, with $m_{{core}} = \beta (m_{{BSC}}/10^3 ~\msun)^{\gamma}$, where $\beta \sim 181$~M$_\odot$ for the high-SFE run and $\beta \sim 224$~M$_\odot$ for the low-SFE run; $\gamma$ is of the order of unity for both (panel \ref{fig:BSC Scatter Plots}c). This, coupled with the fact that the BSCs in the low-SFE run tend to have a larger core size with respect to their half-mass radius, shows that the central densities in low-SFE BSCs tend to be lower, corroborating the results in panel b, and serves to bolster the idea that low-SFE produces loosely bound BSCs. Panel (\ref{fig:BSC Scatter Plots}d) shows that the high-SFE and low-SFE runs have $r_{{core}} / r_{{half}} \sim 0.44$ and  $r_{{core}} / r_{{half}} \sim 0.72$, respectively. This compactness is further underscored by panel \ref{fig:BSC Scatter Plots}e, where we show that BSCs with higher metallicities -- most of which are formed in the low-SFE run -- have $r_{{half}}$ that are on average larger, even though the spread is about one dex. In terms of $\Sigma_0$, the trends across simulations appear to be similar as a function of $r_{{core}}$. Although, $\Sigma_0$ in the high-SFE run is roughly a factor of 3 higher (panel \ref{fig:BSC Scatter Plots}f), consistent with our aforementioned claims. This can have interesting implications for the overall morphology and observability of these galaxies which we will discuss in the next section.

We can also look at $r_{{core}}$ as a function of $\alpha$ in panel \ref{fig:BSC Scatter Plots}g. We see that the galaxy in the low-SFE run hosts BSCs with larger values of $\alpha$ which, when taken with the fact that they also have larger cores, may indicate that they are in the process of tidal disruption. This is certainly the case for the upper part of the diagram, mostly occupied by star clusters in the low-SFE run. These objects have a rather steep truncation in the outer parts of their surface-density profiles. Though, note that this could also mean that BSCs are in a crowded field, wherein the clump-finding algorithm partially fails, producing a purely numerical result.

\subsection{Galaxy Morphology in the Rest-frame UV}
\label{ssec:morpho}
In this section, we will turn our attention back to the dwarf galaxy as a whole, showing how each distinct BSC population produced by varying $f_*$ fits within the larger galactic host and what this means for existing and future observations of galaxies at high-z. 

\begin{figure*}
	\includegraphics[width=\textwidth]{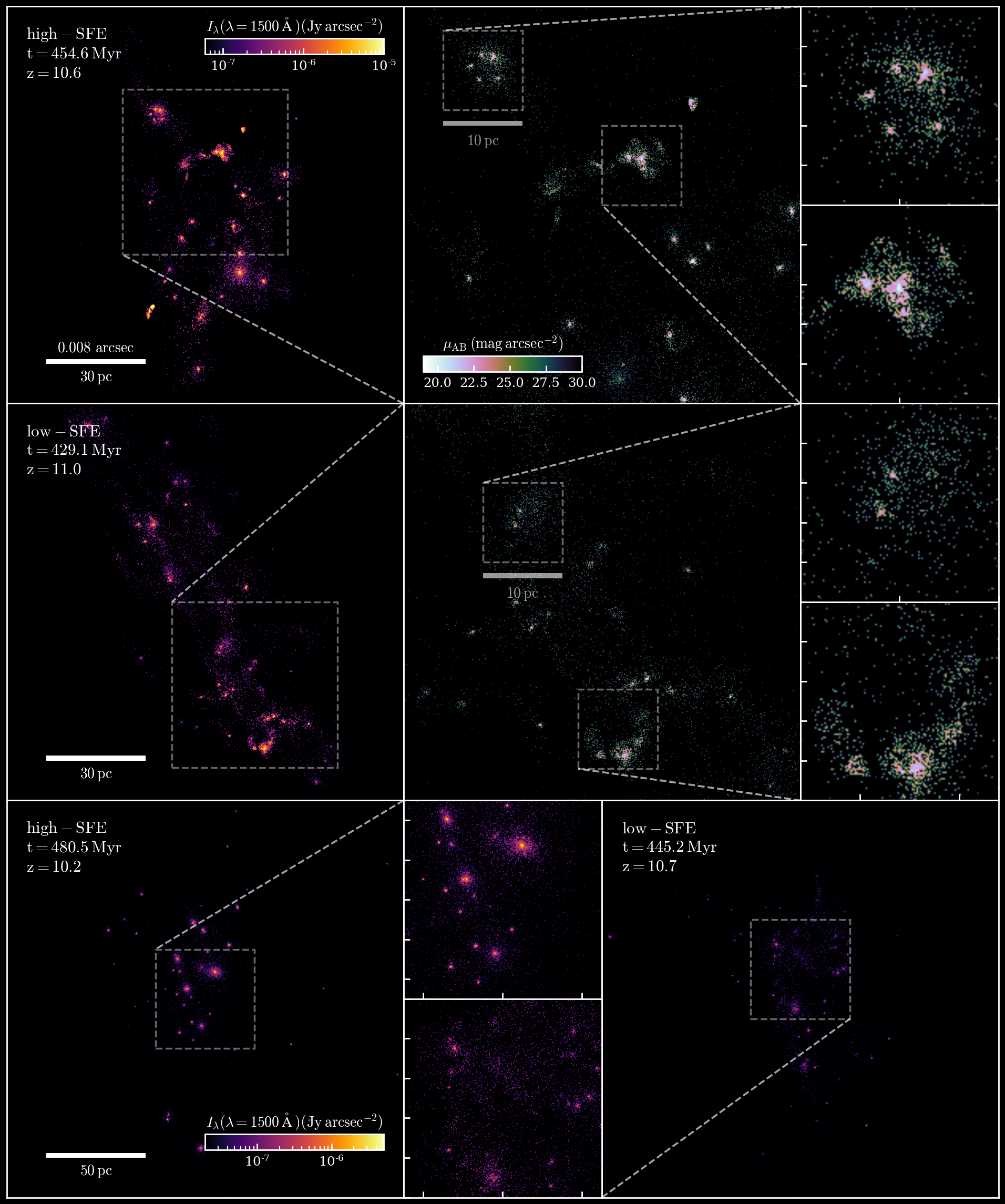}
    \caption{Redshift-corrected line-of-sight specific surface brightness, $I_{\lambda}$ (at $\mathrm{\lambda}$ = 1500 \text{\AA} rest-frame) measured in Jy~$\mrm{arcsec}^{-2}$, for the galaxies in our two simulations. The zoomed-in regions show the same quantity but in mag $\mrm{arcsec}^{-2}$  ($\mu_{AB}$). The \textit{first row} depicts the high-SFE galaxy during its first and largest star-forming episode. The \textit{second row} shows the low-SFE galaxy at a similar stage of its evolution. The \textit{third row} shows the aftermath of this event for both galaxies (see labels). Note that the third row shows the full galaxy view and differs in both spatial and surface brightness scales from the previous rows, with the latter being inherited by the zoomed-in panels for both the SFE runs. The redshift-dimmed calculations of the surface brightness are shown in Appendix~\ref{app:B}.}
    \label{fig:Light Projection}
\end{figure*}

Spatially resolved star-forming knots have been already observed by the HST in $z\sim 6$ galaxies magnified by gravitational lenses \citep[\eg,][]{Vanzella:2022arXiv221109839V, Welch:2022}. Our simulated galaxies show a morphology remarkably similar to these observations. We find that most of the galaxy UV light is emitted in compact nuggets of SF, as conveyed quite clearly in Fig.~\ref{fig:Light Projection}, showing a snapshot of each galaxy (high-SFE top row, low-SFE middle row) during its strongest starburst. 
In the figure, we show the specific surface brightness, $I_{\nu}$ at $\lambda = 1500$~\AA, in Jansky arcsec$^{-2}$ and $\mu_{AB}$ in mag arcsec$^{-2}$, corrected for cosmological redshift dimming using the relationships derived in Appendix~\ref{app:B}.

We observe that the majority of the stars are formed in very compact star clusters, though note that there is a small yet noticeable field component made of unbound stars, caused by the initial cluster expansion right after gas removal and subsequent ejection of stars with high-energy orbits. This is especially visible in the low-SFE galaxy. The bottom row, however, shows how the value of the sub-grid SFE ultimately decides the longer-term survival of the BSC population and what this means for the galactic morphology in the rest frame UV.

\begin{figure}
	\includegraphics[width=\columnwidth]{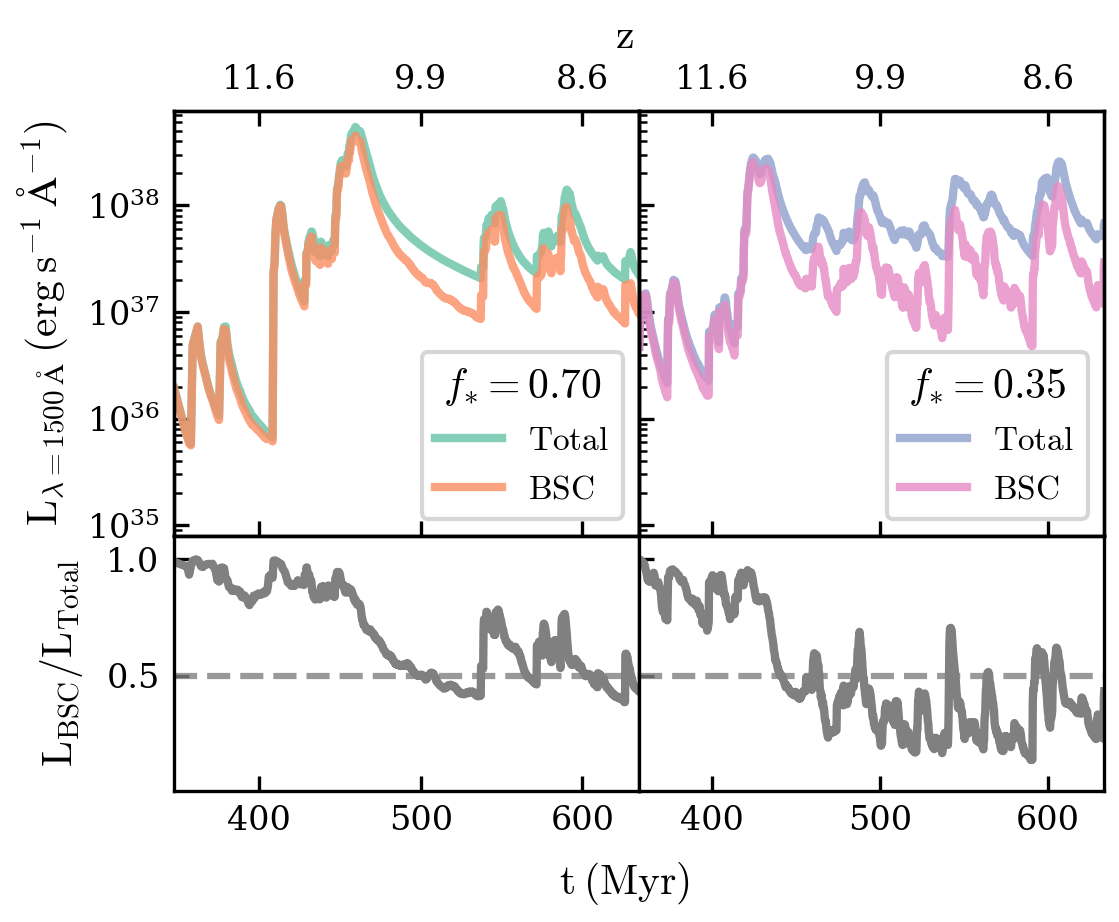}
    \caption{The total specific luminosity in the rest frame UV (at $\mathrm{\lambda}$ = 1500 \text{\AA}) of the galaxy compared to the specific luminosity emitted only from BSCs as a function of time for both simulations (see legend). The \textit{bottom row} shows the fraction of the total UV light emitted from bound star clusters (BSCs) as a function of time for each simulation. The dashed line shows $\mathrm{L_{BSC} / L_{Total} = 0.50}$, which indicates equal UV light contribution from BSCs and field stars.}
    \label{fig:Bound Unbound Luminosity Overtime} 
\end{figure}
\hide{We define the monochromatic absolute AB magnitude $M_{{AB}} = -2.5 \log_{10} f_{\nu} + 8.90$, where $f_{\nu}$ is the spectral flux density per unit frequency in Jansky, which we calculated via the relation:
\begin{equation}
    \frac{f_{\nu}}{\rm Jy} = 7.5 \times 10^{10} \left( \frac{\lambda}{\text{1500 \AA}} \right)^2 
    \frac{f_\lambda}{{erg \: s^{-1} \: \text{\AA}^{-1} \: cm^{-2}}}
    \label{Eq Flux}
\end{equation}
for a given spectral flux density per wavelength $f_{\lambda}$ obtained from \cite{1999ApJS..123....3L}. To find the surface brightness ${SB}$ in the rest frame, we use ${SB} = M_{{AB}} + 2.5 \log_{10} (\Theta)$, where $\Theta$ is the adjusted angular area subtended by an image pixel $d\phi$ in pc. In the rest frame case, $\Theta = (6.48 \times 10^5" \: d\phi  / 10 {pc} )^2$. This equation gives us SB in $M_{AB} \: (")^{-2}$, which we show in the zoomed-in star-forming regions of Fig.~\ref{fig:Light Projection}.} 
BSCs constitute the primary mode of star formation in these high-redshift galaxies, more noticeably so for the high-SFE galaxy. Given the small mass ($\sim 2\times 10^8$~M$_\odot$ at \zend) of this dwarf galaxy, it is unlikely that individual star clusters within it can be detected at $z\sim 6-8$ directly without the aid of gravitational lensing magnification. Magnification from galaxy clusters has already proven to be an important tool, with early JWST observations of the most highly-magnified $z \sim 6$ galaxy, the $\textit{ Sunrise Arc}$ \citep{Salmon:2020ApJ...889..189S}, already offering a treasure-trove of young star clusters \citep{Vanzella:2022arXiv221109839V}. In future work, we plan to study quantitatively how our simulated galaxies compare to the observation of strongly lensed high-z galaxies by applying lensing magnification/sheer maps to assess the likelihood of star cluster magnifications near caustics.

Fig.~\ref{fig:Bound Unbound Luminosity Overtime} shows the specific luminosity, $L_{\lambda}$, of the whole galaxy in the rest-frame UV ($\lambda = 1500 \text{\AA}$), and the same quantity but only from stars belonging to BSCs. The first column refers to the high-SFE run and the second column to the low-SFE run. We observe that the UV luminosity obtained assuming a Salpeter IMF and instantaneous SF-law, fades as the stars age on timescales of tens of Myr after each burst of star formation. In the bottom panels, we show the ratio between the UV light from stars that are in BSCs and the total luminosity of the galaxy. We observe that during bursts, the luminosity of the galaxy is almost exclusively supplied by BSCs. Furthermore, in the high-SFE run, the rest-frame luminosity output is either dominated or contributed for about half, by light from compact bound sources nearly all the time, regardless of the dwindling mass fraction in BSCs. This makes sense given that, although the luminosity decreases by a factor $\sim 100$ through a timescale of around 100~Myr, the repeated large bursts of star formation in BSCs and the higher retention of young stars within them make it so that the star cluster contribution to the total galaxy luminosity remains ${L_{BSC} / L_{Total}} \gtrsim 0.5$ during most of the galaxy's evolution. This is different for the low-SFE galaxy, wherein this value is only met and exceeded during the first starburst episode for $\sim 100$~Myr since the beginning of Pop~II star formation, while later the BSC contribution remains relatively steady, oscillating around ${L_{BSC} / L_{Total}} \gtrsim 0.25$. 

The absolute AB magnitude at 1500~\AA\ of the galaxy during its peak luminosity at $z\sim 11$ is roughly $M_{AB}=-15.65$:
\begin{equation}
M_{AB}= -15.65 -2.5 \log_{10} \left(\frac{L_{\lambda}}{10^{39}~{\rm erg~s}^{-1}~\text{\AA}^{-1}}\right),
\end{equation}
where we have used the formulas derived in Appendix \ref{app:B} to translate $L_{\lambda}$ to the absolute AB magnitude $M_{AB}$. Therefore the apparent magnitude, given by $m_{{AB}} = M_{{AB}} +5 \: \log_{10} (d_{{L}} / 100 \: \rm{Gpc}) + 50$, where $d_{\rm L}$ is the luminosity distance ($\sim120$~Gpc at $z=11$), would be $m_{{AB}} \approx 34.35$ at $z=11$ when our major starbursts occurs. If we were to observe this galaxy as it was at $z = 11$, it will be spread out over the redshift-adjusted angular area $\Theta = (R / d_{{A}})^{2}$, where R is the physical extent of the galaxy ($R \sim 400 $ pc) at a given angular size distance $d_{{A}} \equiv d_{{L}} (1+z)^{-2}$. Then, it is simple to show that $\Theta \sim (0.1~\text{arcsec})^2$. 

\begin{figure*}
\includegraphics[width=\textwidth]{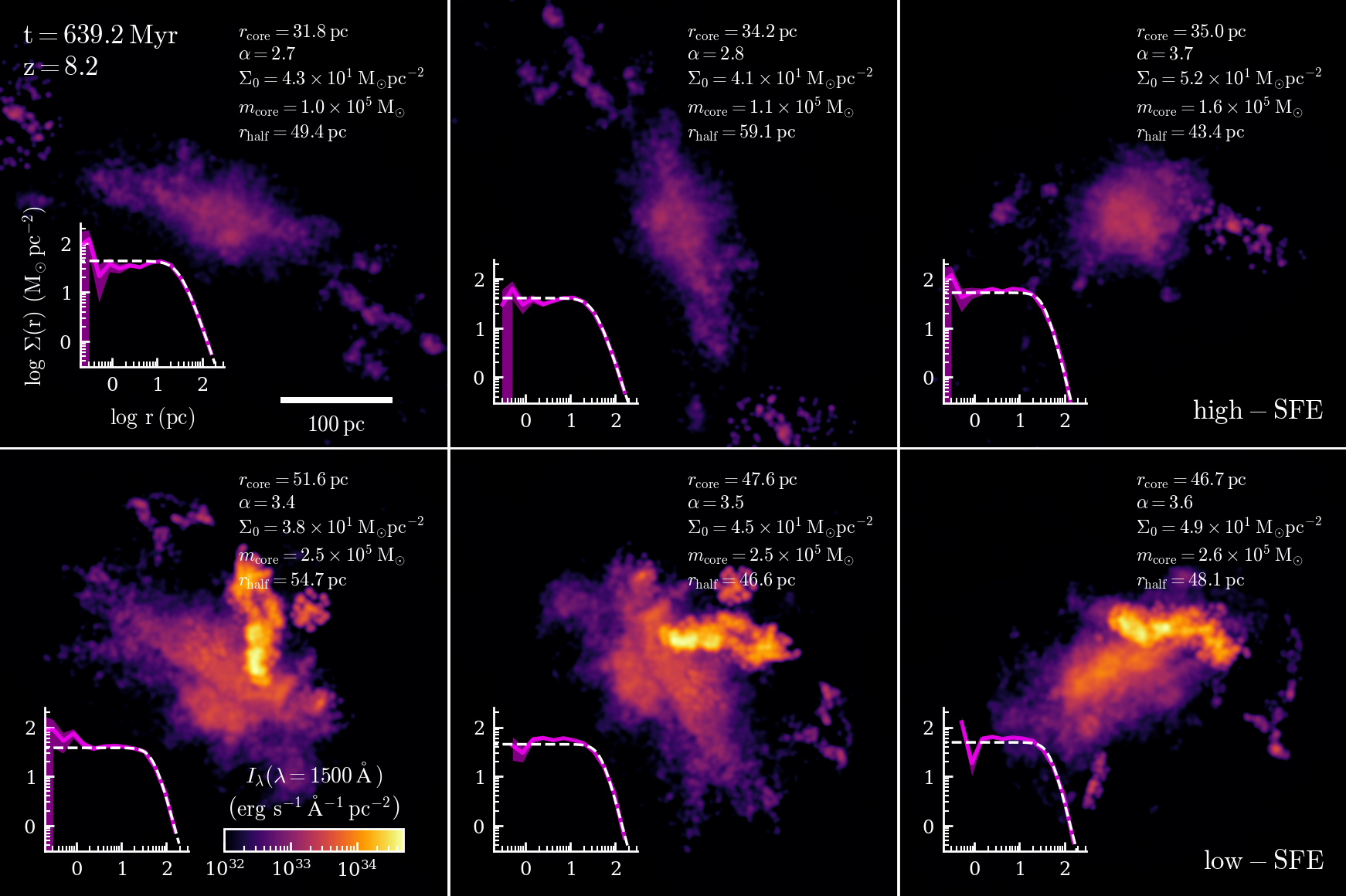}
    \caption{\textit{Top row}: Specific surface brightness of the unbound stars in the galaxy with high sub-grid SFE. Each panel, (\textit{left to right}) shows a different line-of-sight projection. Each panel inset shows the corresponding surface mass density profile (magenta line, along with 1$\sigma$ errors due to Poisson noise indicated by a shaded region of the same color) as a function of distance from the center of each panel along with a profile fit (dashed white line) and extracted parameters (\textit{top right} of each panel). \textit{Bottom row:} Same as the top row, but for  $f_{*} = 0.35$ (the axes labels are identical to the inset in the top left panel). Note that the surface brightness from unbound stars is about two orders of magnitude fainter than that from BSCs (see color bar scale for $I_\lambda$ in Fig.~\ref{fig:BSC Merger}). The galaxy is depicted near the end of our simulations when the field component of each galaxy has sufficiently developed. The time and redshift are shown in the \textit{top left} panel.}
    \label{fig:Field Stars Density Profile} 
\end{figure*}
\subsubsection{Morphology of the Diffuse Light in the Galaxy}
\label{ssec:diffuse and star cluster light}
We fitted surface density profiles to help quantify the morphologies of the unbound stars in each galaxy. To do this, we followed the procedure in Appendix~\ref{sec:Profiling BSCs}, but solely on the unbound field stars. Fig.~\ref{fig:Field Stars Density Profile} depicts the result of this for different viewing angles at the latest stage of each galaxy's evolution. We observe that the field star component in the low-SFE galaxy contains significantly more young stars (consistent with Fig.~\ref{fig:Bound Unbound Scatter}) than in the high-SFE run, as suggested by its higher specific surface brightness, which can differ by around an order of magnitude looking at the color bars. This is consistent with our view that young BSCs in the low-SFE galaxy are more likely to lose stars due to being more vulnerable to tidal perturbations at birth, an idea we will discuss in the next section (\S~\ref{sec:dis}). This vulnerability ultimately contributes to the overall morphology of the galaxy, producing a larger  $r_{{core}}$ for the field stars on average. The high-SFE galaxy also appears to have a denser central region as shown by the value of $\Sigma_0$. The age and morphology of the field stars give additional insight into the dynamical processes at play, with the rather irregular shape of the low-SFE field component suggesting a constant supply of young unbounded stars marring the tidal field. This is in contrast to the high-SFE galaxy wherein the mostly old field stars settle into a more elongated spheroid.

\section{Discussion}
\label{sec:dis}
The need for simulations of star cluster formation that go beyond idealized isolated clouds, contextualizing star clusters within the larger galactic environment is especially underscored when looking at the dynamics within and amongst the first star clusters. After formation and subsequent gas expulsion from internal feedback, a star cluster can either remain bound or become an open cluster. The subset of bound star clusters evolves through various processes, including tidal disruptions and two-body relaxation. The trends discussed in the previous section can be explained by a combination of these two processes. The commentary in the following section serves to explain the underlying physical processes responsible for the trends we observe in our simulations.

\subsection{Dynamical Evolution of BSCs}
\label{ssec:Dynamical Evolution of BSCs}
Here we will discuss the roles of natal mortality, early tidal disruptions, and two-body relaxation in the evolution of BSCs through cosmic time. 
\subsubsection{Natal mortality}
Early dynamical evolution of star clusters ($\lesssim 100$ Myr since formation) is mainly dictated by the natal cradle. Models have shown that mass loss through gas outflows leads to a complete dissolution of star clusters at birth in a process known as infant mortality \citep[\eg,][]{1980ApJ...235..986H, Lada:2003}. Early work by \cite{1980ApJ...235..986H} suggested that instantaneous gas ejection ($t_{{loss}} \ll t_{{dyn}}$ ) upon birth dooms any star cluster with $f_* < 0.5$ to an unbound configuration. Though, simulations by \cite{Kroupa:2001} found BSCs for $f_* = 0.30$, and AU-scale high-resolution simulations of giant molecular clouds (GMCs) at solar metallicities by \cite{HeRG:2019} show the production of BSCs for $f_*$ as low as 0.25. For $f_* > 0.5$, the star cluster expands from the initial radius of the star cluster -- which in our simulations is defined as $R_{{MC}}$ -- to the final radius after complete gas removal $r_{{BSC}}$, where the relationship is expressed as:
\begin{equation}
   \frac{r_{{BSC}}}{R_{{MC}}} = \frac{f_* }{ 2f_* - 1}  
   \qquad {\rm for} \quad  0.5 < f_* < 1 .
    \label{Eq Natal Death}
\end{equation}
Note that if this process occurs gradually over many dynamical times (\ie, $t_{{loss}} \gg t_{{dyn}}$ ), the cluster can remain bound, irrespective of the cloud's $f_*$. In this limit, the radius expands by a factor $r_{{BSC}} / R_{{MC}} = 1 / f_*$, with low-SFEs causing greater expansion. Assuming these two limits bracket the range of star cluster expansion, we expect  $r_{{BSC}} / R_{{MC}} \sim 1.4-1.5$ in the high-SFE run, and $r_{{BSC}} / R_{{MC}} \gtrsim 2.9$ in the low-SFE run.
 
This feedback-driven process is suspected to be the sole form of disruption only in environments with low ambient densities \citep[\eg,][]{Lamers:2005}, while in high-density environments, characteristic of our high-redshift galaxies, its effect is compounded with that of tidal forces. Rather than simple natal mortality, \cite{Kruijssen:2012} suggests a "cruel cradle", whereby the external tidal disruptions add to the effects of cloud expansion due to the initial gas expulsion, which only ejects the stars already in high energy orbits. It has long been known that tidal fields can disrupt star clusters \citep[\eg,][]{Spitzer:1958}, even very young ones \citep[\eg,][]{Elmegreen:2010}, suggesting that tidal disruptions play a major role in the early evolution of star clusters. A tidally crowded environment with dense GMC is present in both low and high-SFE galaxies immediately after each starburst, but more so in the low-SFE galaxy, wherein there is a larger amount of gas left over from the star formation period. For example, the mean ambient gas density right after the largest star-forming period (second row in Fig.~\ref{fig:Gas Projection}) is marginally higher in the low-SFE case ($\overline{\rho}_{amb} = 0.021 \: {\msun \: \mrm{pc}^{-3}}$, at $z = 9.8$), compared to the value in the high-SFE run ($\overline{\rho}_{amb} = 0.014 \: {\msun \: \mrm{pc}^{-3}}$, at $z = 9.3$). Though infant death is likely the main mechanism dictating the early life of BSCs in the low-SFE run, the larger fractional expansion after gas dispersal makes the BSCs especially vulnerable to tidal perturbations, stripping stars from the already looser-bound star clusters. 

In the environment of dwarf galaxies in the early universe, the process of star cluster tidal stripping has yet to be fully understood through observations and simulations. However, star cluster populations in the Milky Way neighborhood (in the SMC, M51, and M33), show that the empirically derived disruption time, $t_{dis}$, scales roughly with $m_{BSC}^{0.6}$ for clusters between $10^3$ to $10^6$~M$_{\odot}$. Using these observations and N-body simulations, \cite{Lamers:2005} derived a best-fit expression dependent on the mean ambient density of the field $\rho_{amb}$, and the mass of the star cluster $m_{BSC}$:
\begin{equation}
    \frac{t_{dis}}{\mathrm{Myr}} = 20-60\left(\frac{m_{BSC}}{10^3 \: \mathrm{M_{\odot}}}\right)^{0.62} \: \left(\frac{\rho_{amb}}{ \mathrm{10~M_{\odot} \: pc^{-3}}}\right)^{-0.5}.
    \label{Eq DisruptionLamers}
\end{equation}

Given that our galaxy with low sub-grid SFE births star clusters in an environment with $\rho_{amb}$ roughly 35 per cent higher than in the high-SFE run, and BSCs have a mean mass $m_{BSC} \sim 10^{2.9}  \: \mathrm{M_{\odot}}$ in comparison to $\sim 10^{3.4}~\msun$ in the high-SFE run, applying Eq.~(\ref{Eq DisruptionLamers}) we should expect the bound star fraction in the low-SFE galaxy to decrease at a rate of around 2 - 3 times faster than that in the high-SFE galaxy, during evolutionary stages when tidal disruption is dominant. This is consistent with our results in Fig.~\ref{fig:Mass Bound Unbound}, where the low-SFE galaxy reaches ${M_{BSC} / M_{Total}} = 0.5$ after only $\sim 75$~Myr in comparison to $\sim 150$~Myr for the high-SFE galaxy. A caveat of this result is the fact that Eq.~(\ref{Eq DisruptionLamers}) is derived from observations at $z = 0$, and does not take into account the different values of the sub-grid $f_*$.

In summary, we are confident that in low-SFE run, the initial mass loss of BSCs is caused by a two-pronged process: (i) natal death, which results from the lower SFE in the clouds losing nearly twice their initial gas mass when compared to their high-SFE counterparts, hence leading to a greater fractional expansion of the star cluster, and (ii) tidal perturbations that -- in conjunction -- are what ultimately shape the early BSC population in both galaxies, with the low-SFE galaxy being especially vulnerable to the additional tidal effects. After the initial mass loss, what is left over are the denser central regions of the BSCs. These too are subject to mass loss; however, the main process is entirely different.

\subsubsection{Two-body relaxation}\label{ssec:2body}
Internal processes are thought to mainly dictate the later evolution ($\gtrsim 100$ Myr since formation) of BSCs. As stars escape the potential randomly, two-body interaction help refill the tails of the clusters' velocity dispersion distribution, leading to the gradual evaporation of star clusters. Two-body relaxation has long been thought to be a dominant process dictating the dynamical evolution of low-mass ($\lesssim 10^5 \msun$) GC \citep[\eg,][]{Henon:1965, Gnedin:1997, Fall:2001}.
A typical timescale for this process, introduced by \cite{Spitzer:1987}, is the two-body relaxation time at the half-mass radius $t_{{rh}} \sim 0.138 N (\ln \gamma N )^{-1} t_{{cross}}$, for a cluster of $N$ stars, with $\gamma = 0.1 - 0.4$ for equal mass stars, and crossing time $t_{{cross}} \equiv ( G \rho_{{h}})^{-1/2} $. This can then be conveniently expressed in terms of the BSC mass and half-mass radius as:
\begin{equation}
    \frac{t_{{rh}}}{{\rm Myr}} = 
     \frac{205.9}{\ln (0.4 N)} \left( \frac{r_{{half}}}{{1~\mathrm{pc}}} \right)^{3/2} \left( \frac{m_{{BSC}}}{10^4~\msun} \right)^{1/2} \left(\frac{1~\msun}{\overline{m_*}} \right),
    \label{Eq Relaxation}
\end{equation}
for a system of $N$ stars with mean mass $\overline{m_*}$.  
The evaporation time for a star cluster is roughly $t_{{evap}} \sim 100 t_{{rh}}$, with core collapse starting at $t_{{cc}} \sim 15 t_{{rh}}$ \citep{Combes:1999}. This is a reasonable approximation to the first order, though it is important to note that N-body simulations along with observations of GCs show that this time-scale is affected by the broader tidal field and background potential \citep[\eg,][]{Lamers:2010, Reinoso:2020}, scaling as $t_{{evap}} \propto \Omega^{-1}$ where $\Omega$ is the angular velocity of the cluster's orbit around the host galaxy, uncharacteristic of the galactic environments in our simulations since there is no central disk or bulge.

The fact that two-body relaxation happens on a time scale much longer than the dynamical timescale implies a nearly steady-state bound fraction. We see this for the low-SFE galaxy in Fig.~\ref{fig:Mass Bound Unbound}, with oscillations centered at ${M_{BSC} / M_{{Total}}} \approx 0.1$. 

In our simulations, we expect that mass-loss via two-body interactions for low-mass BSCs ($\lesssim 10^3~\msun$) is due to numerical effects (since $m_*=10~\msun$ in our simulations); however, it should be negligible at the high-mass end of the BSC distribution, as motivated below.
At $t \sim 662$ Myr ($z \sim 8.0$) in the low-SFE galaxy, the most massive BSC has a mass of $m_{{BSC}} = 5.9 \times 10^{3}~\msun$ and half-mass radius of $r_{{half}} = 3.3$ pc. In the high-SFE run, the most massive BSC has mass $M_{{BSC}} = 2.4 \times 10^{4}~\msun$ and radius $r_{{half}} = 5.1$~pc. Eq.~(\ref{Eq Relaxation}) gives us insight into the survival timescales for these clusters. Using the actual mass of the stars in our simulations ($\overline{m_*}=10~\msun$), in the low-SFE galaxy, the most massive cluster has a relaxation time of $t_{{rh}} = 17$~Myr, corresponding to $t_{{evap}} \sim 1.7$~Gyr. In the high-SFE run, the most massive BSC has $t_{{rh}} = 53$~Myr and $t_{{evap}} \sim 5.3$~Gyr, hence we do not expect that star clusters at the high-mass end of the mass function are affected by "numerical" two-body relaxation, while star clusters smaller by a factor of $\sim 30-100$ can be artificially destroyed by this effect during the duration of our simulation ($\sim 300$~Myr).  

\subsection{Functional Form and Evolution of Cluster Mass Functions}
Previous simulations of BSC formation in galaxies have resolved star clusters with masses as small as $\sim 10^3 \:\msun$, giving insights into the functional form of the CMFs. Early hierarchical arguments by \cite{Elmegreen:2006ApJ...648..572E} suggest that the power-law functional form of the CMFs is a consequence of stellar IMFs within individual star clusters affecting the broader galactic IMF \citep[see also][]{Krumholz:2019}, which has a slope $\Gamma \sim -2.3$ at the high-mass end. Indeed, observations of the local universe \citep[see review by][]{Zwart:2010ARA&A..48..431P}, as well as simulations of early star clusters \citep[\eg,][]{Lahen:2020, Ma:2020, Hislop:2022} find a power-law slope $\Gamma \sim -2$ for the CMF of young star clusters, with observational estimates going as flat as $\Gamma \sim -1.5$ \citep{Adamo:2020MNRAS.499.3267A}. Furthermore, \cite{Hislop:2022} have recovered CMFs in simulations of dwarf galaxies with $\Gamma$ ranging from $-3$ to $-2$. In agreement with the results of this work, they also see that increasing the SFE per free-fall time results in shallower CMFs. This agrees with the idea that higher SFE produces a sub-population of low-mass, compact BSCs that survive natal death, filling the high-mass end as the low-mass end is depleted, flattening the power law slope. Simulations by \cite{Lahen:2020} also observe a similar CMF slope of $\Gamma = -2$ for $f_*$ values ranging from 20 to 80 per cent, with BSC masses ranging from $8 \times 10^5 \: \msun$ down to a few hundred solar masses. However, a difference between our results and previous studies, and also with respect to the interpretation of the origin of the CMF, is that we find that the ICMF has a log-normal shape that reflects the mass function of the natal star-forming gas clouds. This is certainly true for our assumed constant $f_*$, but also using $f_*$ based on Eq.~(\ref{Eq SFE}). Despite our simple assumption of constant sub-grid $f_*$, this result is in agreement with other previous studies \citep[\eg,][]{Parmentier:2008ApJ...678..347P, Maji:2017ApJ...844..108M}, also finding bell-shaped or log-normal ICMFs. Although the true form of the ICMF is still undetermined \citep[see review by][]{Krumholz:2019}, our work shows how the combination of natal death and tidal destruction processes can transform the original log-normal ICMF (found in both simulations with different $f_*$) into a power-law CMF with slopes $\Gamma$ that depend on time and $f_*$, on very short timescales. Though, a considerable amount of work is still needed to fully understand how the CMF evolves on much longer timescales, comparable to the Hubble time.

\subsection{BSCs as proto-GC Candidates}
A combination of tidal disruptions and two-body relaxation -- where perturbations disrupt the low-mass end and two-body relaxation dictating the evolution of the higher-mass end -- are likely when star cluster radii have a weak correlation with their masses \citep{Gieles:2016}. We observe this weak correlation in our simulations (see \S~\ref{ssec:bsc composition}), hence we expect that the long-term evolution of BSCs in our galaxy will be governed by these two main mechanisms. This claim is fairly established, with surveys of the local universe \citep[\eg,][]{Kamann:2018} and reviews \citep[\eg,][]{Adamo:2020} suggesting this two-pronged view.

Similar to this work, recent efforts have also looked at this process. \cite{Sameie:2022} presents cosmological simulation of dwarf galaxies at $z > 4$ that resolve and track proto-GC candidates. They form BSCs with masses ranging from $5 \times 10^4~\msun$ to $5 \times 10^5~\msun$ between $z = 11$ and $5$, with BSCs surviving for $0.2 - 2.5$~Gyr. However, the parsec-scale resolution in the simulations prevents them from forming compact clusters, with most of their clusters having $r_{half}$ between 6 and 30~pc. Previous simulations by \cite{Kim:2018} looked at a different scenario; namely, high-z ($z \gtrsim 5$) mergers of proto-galaxies as ideal environments for progenitors of metal-poor GCs at $z = 0$. They formed BSCs with dense cores and lower-density envelopes that are stripped by tidal effects of other clusters. The cores then dissolved on a timescale $\lesssim 0.5$~Gyr, likely because of numerically-driven two-body relaxation effects produced by the relatively large ($800~\msun$) masses of the stars.

Given the demographics of BSCs in our simulations and our previous discussion about their dynamical evolution, we can place rough estimates on the overall long-term fate of the star clusters and their likelihood to be proto-GC candidates. In our simulations, Pop~II particles are mere tracers of the gravitational potential with masses of $\overline{m_*} = 10$. However, assuming a realistic IMFs \citep[\eg][]{Salpeter:1955, Kroupa:2001}, $\overline{m_*} \sim 1$, which results in timescales for two-body relaxation and evaporation roughly seven times longer than our estimates at the end of \S~\ref{ssec:2body}. Hence, neglecting the effect of tidal forces, the most massive clusters in the high-SFE run should easily survive to $z=0$ ($t_{evap}\sim 40$~Gyr), while the most massive cluster in the low-SFE run will certainly undergo core collapse before $z=0$ and may not survive complete evaporation to present days ($t_{evap}\sim 12$~Gyr). 

\section{Summary and Conclusions}
\label{sec:sum}
We presented the results of high-resolution radiation hydrodynamic (RHD) cosmological simulations of a low-mass dwarf galaxy at redshift $z>8$ using the AMR code \ramsesrt. The galaxy is an analog of the dwarf galaxy WLM in the Local Group, with a DM mass of $\sim 2 \times 10^8~\msun$ at $z=8$. In order to reliably capture the formation and evolution of bound/unbound star clusters in the dwarf galaxy, we form and track the dynamics of individual star particles with masses of 10~$\msun$. We achieve a spatial resolution for the gas cells (and the softening length) of $0.1$~pc in physical units at $z\sim 10$ and form stars by identifying, on-the-fly, dense gas clouds with a central density exceeding a critical value ($n_{crit} \sim 5 \times 10^4$~cm$^{-3}$), defining their radii and masses where the density -- in their spherically averaged density profiles -- drops below $1/N_{cut}$ of the central density (with $N_{cut}=10$ as a fiducial value). We then form stars in the cloud stochastically, but density-weighted, with a prescribed sub-grid star formation efficiency (SFE) $f_*$. In this first paper, we focus on understating how the assumed sub-grid SFE affects the properties of the star cluster populations; hence, we run two identical simulations in which we only vary $f_*$ in star-forming gas clouds. We adopt two values of $f_*$ (0.70 and 0.35) that roughly bracket the theoretically expected range of efficiencies given the properties (metallicity, density, and mass) of the gas clouds. An educated guess on the range of SFE in the molecular clouds formed in our galaxy simulations is derived using published radiation-MHD simulations of star formation in molecular clouds at AU-scale resolution producing a realistic stellar IMF \citep{HeRG:2019}.

In this work we have been able to resolve and track the formation and evolution of bound star clusters on galactic scales with masses as small as a few $100~\msun$ and with core radii of a few $\sim 0.1$~pc, producing a remarkably detailed and physically motivated portrait of what one of the first galaxies formed before the epoch of reionization may look like. This is very exciting, especially in light of recent and upcoming detailed observations of similar galaxies at $z\sim 6$ and beyond, aided by the magnifying power of gravitational lenses \citep{Welch:2022ApJ...940L...1W, Vanzella:2022arXiv221109839V}. Thus far, astronomers have been able to image young star clusters in the process of forming with parsec resolution and magnification of over a factor of 100, demonstrating that small mass galaxies at $z\sim 6$, such as the one presented in this work, can be directly observed by HST and JWST.

Below is a list of the main results and conclusions inferred from the analysis of our simulations of a typical-mass dwarf galaxy found at redshifts $z>8$. Note that this is not a rare galaxy at this redshift, but rather one of the most common sufficiently luminous to be detected by JWST if magnified by a gravitational lens by a factor of $\sim 30-100$.

\begin{enumerate}
\item Star formation in the first dwarf galaxies is characterized by repeated short bursts of star formation produced by compact star-forming regions forming stars with high efficiency, thus producing large populations of bound star clusters, analogs to globular cluster progenitors but typically of smaller mass, with masses between a few 100 to $2\times 10^4$~M$_\odot$ and sizes of $0.1-3$~pc. We expect that a galaxy more massive than what we have simulated here will have a proportionally larger number of bound star clusters, hence forming a few GC progenitors with masses $>10^5~\msun$.

\item Star formation is self-regulated on galactic scales, hence the total mass in stars in the galaxy at \zend is completely independent of the assumed sub-grid SFE: assuming a higher SFE, the bursts of star formation are more intense but the quench time between bursts is longer, and vice versa for lower SFE. However, the metallicity distribution of the stars and the fraction of stars in bound star clusters depends sensitively on $f_*$. Not surprisingly, increasing $f_*$ leads to fewer metals locked in stars and a higher fraction of surviving bound star clusters.

\item In the rest-frame UV, the galaxy has an irregular morphology dominated by light from bound star clusters. There is no evidence of a central bulge or a galactic disk. During a major starburst episode, nearly 100 per cent of the light is emitted from stars in bound star clusters in compact star-forming knots that can be as small as 0.1--2~pc. Assuming $f_*=0.7$, by the end of the simulation at \zend while the fraction of mass locked in BSCs is 25 per cent, more than 50 per cent of the UV light is emitted by stars in BSC. In the $f_*=0.35$ run, the proportions are 10 per cent by mass and >25 per cent of UV light.

\item We find a population of very low metallicity BSCs with $Z\sim 10^{-3}$~Z$_\odot$ and masses $\sim 10^4~\msun$. The most massive of these have a good probability of surviving to $z=0$ and therefore be a GC progenitor; however, no GCs with metallicities $\lesssim 10^{-2.5}$ have been observed so far. Perhaps this has to do with the erroneous assumption that the stars in them have a Salpeter IMF even at these low metallicities, and/or too optimistic assumptions on their longer timescale survival.

\item Star-forming gas clouds have densities $\sim 10-100$ times higher than the typical GMC densities in the Milky Way. This is mainly because of the lower metallicity of the gas and its higher temperature in the strong Lyman-Werner radiation background produced by the stars in the galaxy. Based on their density and masses we expect that the star formation efficiency in these clouds ranges between 10 per cent and 70 per cent, hence producing mostly bound star clusters after feedback destroys their natal clouds.

\item The initial (at formation) mass function of the star-forming clouds (and the initial cluster mass function, given that we assumed constant $f_*$) is a log-normal distribution with peak mass around $5 \times 10^3~\msun$, weakly dependent on $f_*$. The bound star cluster mass function is instead best described by a power-law d$N/\mrm{d}\log m_{BSC} \propto  m_{BSC}^\Gamma$ with a slope that depends mainly on the sub-grid SFE but also on the time since a burst of star formation: the slope becomes flatter with increasing sub-grid SFE and during a burst of star formation, which ranges between $\Gamma=-0.5$ during a burst in the $f_*=0.7$ run to $\Gamma=-2.5$ during a quiescent phase in the low-SFE case ($f_*=0.35$).

\item We find that the most massive bound star clusters are often produced by the mergers of nearby sub-units that survive initial mass loss, merging through timescales $\lesssim 10$ Myr.

\item The majority of BSCs existing at \zend in both galaxies, though compact enough to be resilient to tidal destruction in their galactic environment, have masses that are too small to survive core collapse and complete evaporation to $z=0$. For a galaxy of the mass simulated in this work (with a stellar mass of $\sim 10^6~\msun$ at $z=8$), only a handful of BSCs with masses $\gtrsim 10^4~\msun$ can be considered as true GC progenitors. Hence, candidates for today's GC progenitors would only be a few per cent of the total mass in stars in the galaxy, even though BSCs account for about 25 -- 50 per cent of the stellar mass in the galaxy at high redshift (with estimates strongly depending on the assumed sub-grid $f_*$). For this reason, throughout this paper, we referred to these smaller mass analogs of GCs as bound star clusters (BSCs). However, even though these mini-GCs -- which constitute the low-mass end of the GC progenitor population -- may be objects that exist only at high redshifts, they can be of fundamental importance as sources of reionization of the intergalactic medium because we expect that most of the ionizing radiation they emit escapes their natal cloud \citep{HeRG:2020}.
\end{enumerate}

In this paper, we mostly focused on the stellar component of the galaxy. However, in future work, we plan to study in more detail the gas physics and the fraction of ionizing radiation leaking from the galaxy into the IGM in the two runs with different sub-grid $f_*$. We also plan to run more simulations assuming a sub-grid SFE based on the individual masses, densities, and metallicities of the natal clouds using a recipe informed by high-resolution AU-scale radiation MHD simulations of individual GMCs. In addition, we plan to make more direct comparisons to JWST observations of lensed galaxies at high-z by creating random realizations of mock images and IFU spectra of our simulated galaxies lensed by real clusters, adopting magnification and shear maps obtained from lens modeling.

\section*{Acknowledgements}
MR acknowledges the support of grant JWSTGO01908017A. This research is partially supported by Grants-in-Aid for Scientific Research (KS: 21K20373) from the Japan Society for the Promotion of Science and the Hakubi Project Funding of Kyoto University (KS). The authors acknowledge the University of Maryland supercomputing resources (http://hpcc.umd.edu) made available for conducting the research reported in this paper. 
\section*{Data Availability}
The data underlying this article is stored and accessible on the University of Maryland supercomputing resources (http://hpcc.umd.edu). The data generated in this research will be shared on reasonable request to the corresponding author.
 
\bibliographystyle{mnras}
\bibliography{main} 

\begin{thebibliography}{}
\makeatletter
\relax
\def\mn@urlcharsother{\let\do\@makeother \do\$\do\&\do\#\do\^\do\_\do\%\do\~}
\def\mn@doi{\begingroup\mn@urlcharsother \@ifnextchar [ {\mn@doi@}
  {\mn@doi@[]}}
\def\mn@doi@[#1]#2{\def\@tempa{#1}\ifx\@tempa\@empty \href
  {http://dx.doi.org/#2} {doi:#2}\else \href {http://dx.doi.org/#2} {#1}\fi
  \endgroup}
\def\mn@eprint#1#2{\mn@eprint@#1:#2::\@nil}
\def\mn@eprint@arXiv#1{\href {http://arxiv.org/abs/#1} {{\tt arXiv:#1}}}
\def\mn@eprint@dblp#1{\href {http://dblp.uni-trier.de/rec/bibtex/#1.xml}
  {dblp:#1}}
\def\mn@eprint@#1:#2:#3:#4\@nil{\def\@tempa {#1}\def\@tempb {#2}\def\@tempc
  {#3}\ifx \@tempc \@empty \let \@tempc \@tempb \let \@tempb \@tempa \fi \ifx
  \@tempb \@empty \def\@tempb {arXiv}\fi \@ifundefined
  {mn@eprint@\@tempb}{\@tempb:\@tempc}{\expandafter \expandafter \csname
  mn@eprint@\@tempb\endcsname \expandafter{\@tempc}}}

\bibitem[\protect\citeauthoryear{{Abe}, {Yajima}, {Khochfar}, {Dalla Vecchia}
  \& {Omukai}}{{Abe} et~al.}{2021}]{Abe:2021}
{Abe} M.,  {Yajima} H.,  {Khochfar} S.,  {Dalla Vecchia} C.,   {Omukai} K.,
  2021, \mn@doi [\mnras] {10.1093/mnras/stab2637}, \href
  {https://ui.adsabs.harvard.edu/abs/2021MNRAS.508.3226A} {508, 3226}

\bibitem[\protect\citeauthoryear{{Adamo} et~al.,}{{Adamo}
  et~al.}{2020a}]{Adamo:2020}
{Adamo} A.,  et~al., 2020a, \mn@doi [\ssr] {10.1007/s11214-020-00690-x}, \href
  {https://ui.adsabs.harvard.edu/abs/2020SSRv..216...69A} {216, 69}

\bibitem[\protect\citeauthoryear{{Adamo} et~al.,}{{Adamo}
  et~al.}{2020b}]{Adamo:2020MNRAS.499.3267A}
{Adamo} A.,  et~al., 2020b, \mn@doi [\mnras] {10.1093/mnras/staa2380}, \href
  {https://ui.adsabs.harvard.edu/abs/2020MNRAS.499.3267A} {499, 3267}

\bibitem[\protect\citeauthoryear{{Arata}, {Yajima}  \& {Nagamine}}{{Arata}
  et~al.}{2018}]{Arata:2018MNRAS.475.4252A}
{Arata} S.,  {Yajima} H.,   {Nagamine} K.,  2018, \mn@doi [\mnras]
  {10.1093/mnras/sty122}, \href
  {https://ui.adsabs.harvard.edu/abs/2018MNRAS.475.4252A} {475, 4252}

\bibitem[\protect\citeauthoryear{{Beasley}, {Leaman}, {Gallart}, {Larsen},
  {Battaglia}, {Monelli}  \& {Pedreros}}{{Beasley}
  et~al.}{2019}]{Beasley:2019MNRAS.487.1986B}
{Beasley} M.~A.,  {Leaman} R.,  {Gallart} C.,  {Larsen} S.~S.,  {Battaglia} G.,
   {Monelli} M.,   {Pedreros} M.~H.,  2019, \mn@doi [\mnras]
  {10.1093/mnras/stz1349}, \href
  {https://ui.adsabs.harvard.edu/abs/2019MNRAS.487.1986B} {487, 1986}

\bibitem[\protect\citeauthoryear{{Behroozi}, {Wechsler}  \&
  {Conroy}}{{Behroozi} et~al.}{2013}]{Behroozi:2013ApJ...762L..31B}
{Behroozi} P.~S.,  {Wechsler} R.~H.,   {Conroy} C.,  2013, \mn@doi [\apjl]
  {10.1088/2041-8205/762/2/L31}, \href
  {https://ui.adsabs.harvard.edu/abs/2013ApJ...762L..31B} {762, L31}

\bibitem[\protect\citeauthoryear{{Besla}, {Kallivayalil}, {Hernquist}, {van der
  Marel}, {Cox}  \& {Kere{\v{s}}}}{{Besla}
  et~al.}{2012}]{Besla:2012MNRAS.421.2109B}
{Besla} G.,  {Kallivayalil} N.,  {Hernquist} L.,  {van der Marel} R.~P.,  {Cox}
  T.~J.,   {Kere{\v{s}}} D.,  2012, \mn@doi [\mnras]
  {10.1111/j.1365-2966.2012.20466.x}, \href
  {https://ui.adsabs.harvard.edu/abs/2012MNRAS.421.2109B} {421, 2109}

\bibitem[\protect\citeauthoryear{{Bonnor}}{{Bonnor}}{1956}]{Bonnor:1956MNRAS.116..351B}
{Bonnor} W.~B.,  1956, \mn@doi [\mnras] {10.1093/mnras/116.3.351}, \href
  {https://ui.adsabs.harvard.edu/abs/1956MNRAS.116..351B} {116, 351}

\bibitem[\protect\citeauthoryear{{Bournaud}, {Duc}  \& {Emsellem}}{{Bournaud}
  et~al.}{2008}]{Bournaud:2008MNRAS.389L...8B}
{Bournaud} F.,  {Duc} P.~A.,   {Emsellem} E.,  2008, \mn@doi [\mnras]
  {10.1111/j.1745-3933.2008.00511.x}, \href
  {https://ui.adsabs.harvard.edu/abs/2008MNRAS.389L...8B} {389, L8}

\bibitem[\protect\citeauthoryear{{Bovill} \& {Ricotti}}{{Bovill} \&
  {Ricotti}}{2009}]{BovillR:2009}
{Bovill} M.~S.,  {Ricotti} M.,  2009, \mn@doi [\apj]
  {10.1088/0004-637X/693/2/1859}, \href
  {https://ui.adsabs.harvard.edu/abs/2009ApJ...693.1859B} {693, 1859}

\bibitem[\protect\citeauthoryear{{Boylan-Kolchin}}{{Boylan-Kolchin}}{2018}]{Boylan-Kolchin:2018}
{Boylan-Kolchin} M.,  2018, \mn@doi [\mnras] {10.1093/mnras/sty1490}, \href
  {https://ui.adsabs.harvard.edu/abs/2018MNRAS.479..332B} {479, 332}

\bibitem[\protect\citeauthoryear{{Bromm}, {Yoshida}, {Hernquist}  \&
  {McKee}}{{Bromm} et~al.}{2009}]{Bromm:2009Natur.459...49B}
{Bromm} V.,  {Yoshida} N.,  {Hernquist} L.,   {McKee} C.~F.,  2009, \mn@doi
  [\nat] {10.1038/nature07990}, \href
  {https://ui.adsabs.harvard.edu/abs/2009Natur.459...49B} {459, 49}

\bibitem[\protect\citeauthoryear{{Brown} et~al.,}{{Brown}
  et~al.}{2012}]{Brown:2012}
{Brown} T.~M.,  et~al., 2012, \mn@doi [\apjl] {10.1088/2041-8205/753/1/L21},
  \href {https://ui.adsabs.harvard.edu/abs/2012ApJ...753L..21B} {753, L21}

\bibitem[\protect\citeauthoryear{{Brown} et~al.,}{{Brown}
  et~al.}{2014}]{Brown:2014}
{Brown} T.~M.,  et~al., 2014, \mn@doi [\apj] {10.1088/0004-637X/796/2/91},
  \href {https://ui.adsabs.harvard.edu/abs/2014ApJ...796...91B} {796, 91}

\bibitem[\protect\citeauthoryear{{Chan}, {Kere{\v{s}}}, {O{\~n}orbe},
  {Hopkins}, {Muratov}, {Faucher-Gigu{\`e}re}  \& {Quataert}}{{Chan}
  et~al.}{2015}]{Chan:2015}
{Chan} T.~K.,  {Kere{\v{s}}} D.,  {O{\~n}orbe} J.,  {Hopkins} P.~F.,  {Muratov}
  A.~L.,  {Faucher-Gigu{\`e}re} C.~A.,   {Quataert} E.,  2015, \mn@doi [\mnras]
  {10.1093/mnras/stv2165}, \href
  {https://ui.adsabs.harvard.edu/abs/2015MNRAS.454.2981C} {454, 2981}

\bibitem[\protect\citeauthoryear{{Chiaki}, {Susa}  \& {Hirano}}{{Chiaki}
  et~al.}{2018}]{Chiaki:2018}
{Chiaki} G.,  {Susa} H.,   {Hirano} S.,  2018, \mn@doi [\mnras]
  {10.1093/mnras/sty040}, \href
  {https://ui.adsabs.harvard.edu/abs/2018MNRAS.475.4378C} {475, 4378}

\bibitem[\protect\citeauthoryear{{Combes}, {Leon}  \& {Meylan}}{{Combes}
  et~al.}{1999}]{Combes:1999}
{Combes} F.,  {Leon} S.,   {Meylan} G.,  1999, \aap, \href
  {https://ui.adsabs.harvard.edu/abs/1999A&A...352..149C} {352, 149}

\bibitem[\protect\citeauthoryear{{Draine} \& {Bertoldi}}{{Draine} \&
  {Bertoldi}}{1996}]{Draine:1996ApJ...468..269D}
{Draine} B.~T.,  {Bertoldi} F.,  1996, \mn@doi [\apj] {10.1086/177689}, \href
  {https://ui.adsabs.harvard.edu/abs/1996ApJ...468..269D} {468, 269}

\bibitem[\protect\citeauthoryear{{Duc}, {Bournaud}  \& {Masset}}{{Duc}
  et~al.}{2004}]{Duc:2004A&A...427..803D}
{Duc} P.~A.,  {Bournaud} F.,   {Masset} F.,  2004, \mn@doi [\aap]
  {10.1051/0004-6361:20041410}, \href
  {https://ui.adsabs.harvard.edu/abs/2004A&A...427..803D} {427, 803}

\bibitem[\protect\citeauthoryear{{Ebert}}{{Ebert}}{1955}]{Ebert:1955ZA.....37..217E}
{Ebert} R.,  1955, \zap, \href
  {https://ui.adsabs.harvard.edu/abs/1955ZA.....37..217E} {37, 217}

\bibitem[\protect\citeauthoryear{{Efstathiou}, {Davis}, {White}  \&
  {Frenk}}{{Efstathiou} et~al.}{1985}]{Efstathiou:1985}
{Efstathiou} G.,  {Davis} M.,  {White} S.~D.~M.,   {Frenk} C.~S.,  1985,
  \mn@doi [\apjs] {10.1086/191003}, \href
  {https://ui.adsabs.harvard.edu/abs/1985ApJS...57..241E} {57, 241}

\bibitem[\protect\citeauthoryear{{Elmegreen}}{{Elmegreen}}{2006}]{Elmegreen:2006ApJ...648..572E}
{Elmegreen} B.~G.,  2006, \mn@doi [\apj] {10.1086/505785}, \href
  {https://ui.adsabs.harvard.edu/abs/2006ApJ...648..572E} {648, 572}

\bibitem[\protect\citeauthoryear{{Elmegreen} \& {Hunter}}{{Elmegreen} \&
  {Hunter}}{2010}]{Elmegreen:2010}
{Elmegreen} B.~G.,  {Hunter} D.~A.,  2010, \mn@doi [\apj]
  {10.1088/0004-637X/712/1/604}, \href
  {https://ui.adsabs.harvard.edu/abs/2010ApJ...712..604E} {712, 604}

\bibitem[\protect\citeauthoryear{{Fall} \& {Zhang}}{{Fall} \&
  {Zhang}}{2001}]{Fall:2001}
{Fall} S.~M.,  {Zhang} Q.,  2001, \mn@doi [\apj] {10.1086/323358}, \href
  {https://ui.adsabs.harvard.edu/abs/2001ApJ...561..751F} {561, 751}

\bibitem[\protect\citeauthoryear{{Fern{\'a}ndez-Trincado}
  et~al.,}{{Fern{\'a}ndez-Trincado}
  et~al.}{2021}]{Fernandex:2021ApJ...908L..42F}
{Fern{\'a}ndez-Trincado} J.~G.,  et~al., 2021, \mn@doi [\apjl]
  {10.3847/2041-8213/abdf47}, \href
  {https://ui.adsabs.harvard.edu/abs/2021ApJ...908L..42F} {908, L42}

\bibitem[\protect\citeauthoryear{{Forbes} et~al.,}{{Forbes}
  et~al.}{2018}]{Forbes:2018RSPSA.47470616F}
{Forbes} D.~A.,  et~al., 2018, \mn@doi [Proceedings of the Royal Society of
  London Series A] {10.1098/rspa.2017.0616}, \href
  {https://ui.adsabs.harvard.edu/abs/2018RSPSA.47470616F} {474, 20170616}

\bibitem[\protect\citeauthoryear{{Fu} et~al.,}{{Fu}
  et~al.}{2022}]{Fu:2022ApJ...925....6F}
{Fu} S.~W.,  et~al., 2022, \mn@doi [\apj] {10.3847/1538-4357/ac3665}, \href
  {https://ui.adsabs.harvard.edu/abs/2022ApJ...925....6F} {925, 6}

\bibitem[\protect\citeauthoryear{{Fukushima}, {Yajima}, {Sugimura}, {Hosokawa},
  {Omukai}  \& {Matsumoto}}{{Fukushima} et~al.}{2020a}]{Fukushima2019}
{Fukushima} H.,  {Yajima} H.,  {Sugimura} K.,  {Hosokawa} T.,  {Omukai} K.,
  {Matsumoto} T.,  2020a, \mn@doi [\mnras] {10.1093/mnras/staa2062}, \href
  {https://ui.adsabs.harvard.edu/abs/2020MNRAS.497.3830F} {497, 3830}

\bibitem[\protect\citeauthoryear{{Fukushima}, {Yajima}, {Sugimura}, {Hosokawa},
  {Omukai}  \& {Matsumoto}}{{Fukushima} et~al.}{2020b}]{FukushimaHajime:2020}
{Fukushima} H.,  {Yajima} H.,  {Sugimura} K.,  {Hosokawa} T.,  {Omukai} K.,
  {Matsumoto} T.,  2020b, \mn@doi [\mnras] {10.1093/mnras/staa2062}, \href
  {https://ui.adsabs.harvard.edu/abs/2020MNRAS.497.3830F} {497, 3830}

\bibitem[\protect\citeauthoryear{{Gieles} \& {Renaud}}{{Gieles} \&
  {Renaud}}{2016}]{Gieles:2016}
{Gieles} M.,  {Renaud} F.,  2016, \mn@doi [\mnras] {10.1093/mnrasl/slw163},
  \href {https://ui.adsabs.harvard.edu/abs/2016MNRAS.463L.103G} {463, L103}

\bibitem[\protect\citeauthoryear{{Gnedin} \& {Ostriker}}{{Gnedin} \&
  {Ostriker}}{1997}]{Gnedin:1997}
{Gnedin} O.~Y.,  {Ostriker} J.~P.,  1997, \mn@doi [\apj] {10.1086/303441},
  \href {https://ui.adsabs.harvard.edu/abs/1997ApJ...474..223G} {474, 223}

\bibitem[\protect\citeauthoryear{{Governato} et~al.,}{{Governato}
  et~al.}{2010}]{Governato:2010}
{Governato} F.,  et~al., 2010, \mn@doi [\nat] {10.1038/nature08640}, \href
  {https://ui.adsabs.harvard.edu/abs/2010Natur.463..203G} {463, 203}

\bibitem[\protect\citeauthoryear{{Griffen}, {Drinkwater}, {Thomas}, {Helly}  \&
  {Pimbblet}}{{Griffen} et~al.}{2010}]{Griffen:2010MNRAS.405..375G}
{Griffen} B.~F.,  {Drinkwater} M.~J.,  {Thomas} P.~A.,  {Helly} J.~C.,
  {Pimbblet} K.~A.,  2010, \mn@doi [\mnras] {10.1111/j.1365-2966.2010.16458.x},
  \href {https://ui.adsabs.harvard.edu/abs/2010MNRAS.405..375G} {405, 375}

\bibitem[\protect\citeauthoryear{{Grudi{\'c}}, {Hafen}, {Rodriguez},
  {Guszejnov}, {Lamberts}, {Wetzel}, {Boylan-Kolchin}  \&
  {Faucher-Gigu{\`e}re}}{{Grudi{\'c}}
  et~al.}{2022}]{Grudic:2022arXiv220305732G}
{Grudi{\'c}} M.~Y.,  {Hafen} Z.,  {Rodriguez} C.~L.,  {Guszejnov} D.,
  {Lamberts} A.,  {Wetzel} A.,  {Boylan-Kolchin} M.,   {Faucher-Gigu{\`e}re}
  C.-A.,  2022, arXiv e-prints, \href
  {https://ui.adsabs.harvard.edu/abs/2022arXiv220305732G} {p. arXiv:2203.05732}

\bibitem[\protect\citeauthoryear{{He} \& {Ricotti}}{{He} \&
  {Ricotti}}{2022}]{HeR2022}
{He} C.-C.,  {Ricotti} M.,  2022, arXiv e-prints, \href
  {https://ui.adsabs.harvard.edu/abs/2022arXiv221011629H} {p. arXiv:2210.11629}

\bibitem[\protect\citeauthoryear{{He}, {Ricotti}  \& {Geen}}{{He}
  et~al.}{2019}]{HeRG:2019}
{He} C.-C.,  {Ricotti} M.,   {Geen} S.,  2019, \mn@doi [\mnras]
  {10.1093/mnras/stz2239}, \href
  {https://ui.adsabs.harvard.edu/abs/2019MNRAS.489.1880H} {489, 1880}

\bibitem[\protect\citeauthoryear{{He}, {Ricotti}  \& {Geen}}{{He}
  et~al.}{2020}]{HeRG:2020}
{He} C.-C.,  {Ricotti} M.,   {Geen} S.,  2020, \mn@doi [\mnras]
  {10.1093/mnras/staa165}, \href
  {https://ui.adsabs.harvard.edu/abs/2020MNRAS.492.4858H} {492, 4858}

\bibitem[\protect\citeauthoryear{{H{\'e}non}}{{H{\'e}non}}{1965}]{Henon:1965}
{H{\'e}non} M.,  1965, Annales d'Astrophysique, \href
  {https://ui.adsabs.harvard.edu/abs/1965AnAp...28...62H} {28, 62}

\bibitem[\protect\citeauthoryear{{Hills}}{{Hills}}{1980}]{1980ApJ...235..986H}
{Hills} J.~G.,  1980, \mn@doi [\apj] {10.1086/157703}, \href
  {https://ui.adsabs.harvard.edu/abs/1980ApJ...235..986H} {235, 986}

\bibitem[\protect\citeauthoryear{{Hislop}, {Naab}, {Steinwandel}, {Lah{\'e}n},
  {Irodotou}, {Johansson}  \& {Walch}}{{Hislop} et~al.}{2022}]{Hislop:2022}
{Hislop} J.~M.,  {Naab} T.,  {Steinwandel} U.~P.,  {Lah{\'e}n} N.,  {Irodotou}
  D.,  {Johansson} P.~H.,   {Walch} S.,  2022, \mn@doi [\mnras]
  {10.1093/mnras/stab3347}, \href
  {https://ui.adsabs.harvard.edu/abs/2022MNRAS.509.5938H} {509, 5938}

\bibitem[\protect\citeauthoryear{{Hix}, {He}  \& {Ricotti}}{{Hix}
  et~al.}{2022}]{Hix:preprint}
{Hix} R.,  {He} C.-C.,   {Ricotti} M.,  2022, arXiv e-prints, \href
  {https://ui.adsabs.harvard.edu/abs/2022arXiv221204411H} {p. arXiv:2212.04411}

\bibitem[\protect\citeauthoryear{{Hopkins}, {Quataert}  \& {Murray}}{{Hopkins}
  et~al.}{2011}]{Hopkins:2011MNRAS.417..950H}
{Hopkins} P.~F.,  {Quataert} E.,   {Murray} N.,  2011, \mn@doi [\mnras]
  {10.1111/j.1365-2966.2011.19306.x}, \href
  {https://ui.adsabs.harvard.edu/abs/2011MNRAS.417..950H} {417, 950}

\bibitem[\protect\citeauthoryear{{Kamann} et~al.,}{{Kamann}
  et~al.}{2018}]{Kamann:2018}
{Kamann} S.,  et~al., 2018, \mn@doi [\mnras] {10.1093/mnras/stx2719}, \href
  {https://ui.adsabs.harvard.edu/abs/2018MNRAS.473.5591K} {473, 5591}

\bibitem[\protect\citeauthoryear{{Katz} \& {Ricotti}}{{Katz} \&
  {Ricotti}}{2013}]{KatzR:2013}
{Katz} H.,  {Ricotti} M.,  2013, \mn@doi [\mnras] {10.1093/mnras/stt676}, \href
  {https://ui.adsabs.harvard.edu/abs/2013MNRAS.432.3250K} {432, 3250}

\bibitem[\protect\citeauthoryear{{Katz} \& {Ricotti}}{{Katz} \&
  {Ricotti}}{2014}]{KatzR:2014}
{Katz} H.,  {Ricotti} M.,  2014, \mn@doi [\mnras] {10.1093/mnras/stu1489},
  \href {https://ui.adsabs.harvard.edu/abs/2014MNRAS.444.2377K} {444, 2377}

\bibitem[\protect\citeauthoryear{Katz, Kimm, Sijacki  \& Haehnelt}{Katz
  et~al.}{2017}]{katz2017}
Katz H.,  Kimm T.,  Sijacki D.,   Haehnelt M.~G.,  2017, \mn@doi [\mnras]
  {10.1093/mnras/stx608}, 468, 4831

\bibitem[\protect\citeauthoryear{{Kim} et~al.,}{{Kim} et~al.}{2018}]{Kim:2018}
{Kim} J.-h.,  et~al., 2018, \mn@doi [\mnras] {10.1093/mnras/stx2994}, \href
  {https://ui.adsabs.harvard.edu/abs/2018MNRAS.474.4232K} {474, 4232}

\bibitem[\protect\citeauthoryear{{Kim}, {Ostriker}  \& {Filippova}}{{Kim}
  et~al.}{2021}]{Kim:2021}
{Kim} J.-G.,  {Ostriker} E.~C.,   {Filippova} N.,  2021, \mn@doi [\apj]
  {10.3847/1538-4357/abe934}, \href
  {https://ui.adsabs.harvard.edu/abs/2021ApJ...911..128K} {911, 128}

\bibitem[\protect\citeauthoryear{Kimm, Katz, Haehnelt, Rosdahl, Devriendt  \&
  Slyz}{Kimm et~al.}{2017}]{kimm2017}
Kimm T.,  Katz H.,  Haehnelt M.,  Rosdahl J.,  Devriendt J.,   Slyz A.,  2017,
  \mn@doi [\mnras] {10.1093/mnras/stx052}, 466, 4826

\bibitem[\protect\citeauthoryear{{King}}{{King}}{1962}]{1962AJ.....67..471K}
{King} I.,  1962, \mn@doi [\aj] {10.1086/108756}, \href
  {https://ui.adsabs.harvard.edu/abs/1962AJ.....67..471K} {67, 471}

\bibitem[\protect\citeauthoryear{{Kroupa}, {Aarseth}  \& {Hurley}}{{Kroupa}
  et~al.}{2001}]{Kroupa:2001}
{Kroupa} P.,  {Aarseth} S.,   {Hurley} J.,  2001, \mn@doi [\mnras]
  {10.1046/j.1365-8711.2001.04050.x}, \href
  {https://ui.adsabs.harvard.edu/abs/2001MNRAS.321..699K} {321, 699}

\bibitem[\protect\citeauthoryear{{Kruijssen}, {Maschberger}, {Moeckel},
  {Clarke}, {Bastian}  \& {Bonnell}}{{Kruijssen} et~al.}{2012}]{Kruijssen:2012}
{Kruijssen} J.~M.~D.,  {Maschberger} T.,  {Moeckel} N.,  {Clarke} C.~J.,
  {Bastian} N.,   {Bonnell} I.~A.,  2012, \mn@doi [\mnras]
  {10.1111/j.1365-2966.2011.19748.x}, \href
  {https://ui.adsabs.harvard.edu/abs/2012MNRAS.419..841K} {419, 841}

\bibitem[\protect\citeauthoryear{{Kruijssen}, {Pfeffer}, {Crain}  \&
  {Bastian}}{{Kruijssen} et~al.}{2019}]{2019MNRAS.486.3134K}
{Kruijssen} J.~M.~D.,  {Pfeffer} J.~L.,  {Crain} R.~A.,   {Bastian} N.,  2019,
  \mn@doi [\mnras] {10.1093/mnras/stz968}, \href
  {https://ui.adsabs.harvard.edu/abs/2019MNRAS.486.3134K} {486, 3134}

\bibitem[\protect\citeauthoryear{{Krumholz}, {McKee}  \&
  {Bland-Hawthorn}}{{Krumholz} et~al.}{2019}]{Krumholz:2019}
{Krumholz} M.~R.,  {McKee} C.~F.,   {Bland-Hawthorn} J.,  2019, \mn@doi [\araa]
  {10.1146/annurev-astro-091918-104430}, \href
  {https://ui.adsabs.harvard.edu/abs/2019ARA&A..57..227K} {57, 227}

\bibitem[\protect\citeauthoryear{{Lada} \& {Lada}}{{Lada} \&
  {Lada}}{2003}]{Lada:2003}
{Lada} C.~J.,  {Lada} E.~A.,  2003, \mn@doi [\araa]
  {10.1146/annurev.astro.41.011802.094844}, \href
  {https://ui.adsabs.harvard.edu/abs/2003ARA&A..41...57L} {41, 57}

\bibitem[\protect\citeauthoryear{{Lah{\'e}n}, {Naab}, {Johansson}, {Elmegreen},
  {Hu}, {Walch}, {Steinwandel}  \& {Moster}}{{Lah{\'e}n}
  et~al.}{2020}]{Lahen:2020}
{Lah{\'e}n} N.,  {Naab} T.,  {Johansson} P.~H.,  {Elmegreen} B.,  {Hu} C.-Y.,
  {Walch} S.,  {Steinwandel} U.~P.,   {Moster} B.~P.,  2020, \mn@doi [\apj]
  {10.3847/1538-4357/ab7190}, \href
  {https://ui.adsabs.harvard.edu/abs/2020ApJ...891....2L} {891, 2}

\bibitem[\protect\citeauthoryear{{Lamers}, {Gieles}  \& {Portegies
  Zwart}}{{Lamers} et~al.}{2005}]{Lamers:2005}
{Lamers} H.~J.~G.~L.~M.,  {Gieles} M.,   {Portegies Zwart} S.~F.,  2005,
  \mn@doi [\aap] {10.1051/0004-6361:20041476}, \href
  {https://ui.adsabs.harvard.edu/abs/2005A&A...429..173L} {429, 173}

\bibitem[\protect\citeauthoryear{{Lamers}, {Baumgardt}  \& {Gieles}}{{Lamers}
  et~al.}{2010}]{Lamers:2010}
{Lamers} H. J.~G.~L.~M.,  {Baumgardt} H.,   {Gieles} M.,  2010, \mn@doi
  [\mnras] {10.1111/j.1365-2966.2010.17309.x}, \href
  {https://ui.adsabs.harvard.edu/abs/2010MNRAS.409..305L} {409, 305}

\bibitem[\protect\citeauthoryear{{Leitherer} et~al.,}{{Leitherer}
  et~al.}{1999}]{1999ApJS..123....3L}
{Leitherer} C.,  et~al., 1999, \mn@doi [\apjs] {10.1086/313233}, \href
  {https://ui.adsabs.harvard.edu/abs/1999ApJS..123....3L} {123, 3}

\bibitem[\protect\citeauthoryear{{Li} \& {Gnedin}}{{Li} \&
  {Gnedin}}{2019}]{Li:2019MNRAS.486.4030L}
{Li} H.,  {Gnedin} O.~Y.,  2019, \mn@doi [\mnras] {10.1093/mnras/stz1114},
  \href {https://ui.adsabs.harvard.edu/abs/2019MNRAS.486.4030L} {486, 4030}

\bibitem[\protect\citeauthoryear{{Li}, {Vogelsberger}, {Bryan}, {Marinacci},
  {Sales}  \& {Torrey}}{{Li} et~al.}{2022}]{Li:2022MNRAS.514..265L}
{Li} H.,  {Vogelsberger} M.,  {Bryan} G.~L.,  {Marinacci} F.,  {Sales} L.~V.,
  {Torrey} P.,  2022, \mn@doi [\mnras] {10.1093/mnras/stac1136}, \href
  {https://ui.adsabs.harvard.edu/abs/2022MNRAS.514..265L} {514, 265}

\bibitem[\protect\citeauthoryear{{Ma} et~al.,}{{Ma} et~al.}{2020}]{Ma:2020}
{Ma} X.,  et~al., 2020, \mn@doi [\mnras] {10.1093/mnras/staa527}, \href
  {https://ui.adsabs.harvard.edu/abs/2020MNRAS.493.4315M} {493, 4315}

\bibitem[\protect\citeauthoryear{{Madau}, {Weisz}  \& {Conroy}}{{Madau}
  et~al.}{2014}]{Madau:2014ApJ...790L..17M}
{Madau} P.,  {Weisz} D.~R.,   {Conroy} C.,  2014, \mn@doi [\apjl]
  {10.1088/2041-8205/790/2/L17}, \href
  {https://ui.adsabs.harvard.edu/abs/2014ApJ...790L..17M} {790, L17}

\bibitem[\protect\citeauthoryear{{Maji}, {Zhu}, {Li}, {Charlton}, {Hernquist}
  \& {Knebe}}{{Maji} et~al.}{2017}]{Maji:2017ApJ...844..108M}
{Maji} M.,  {Zhu} Q.,  {Li} Y.,  {Charlton} J.,  {Hernquist} L.,   {Knebe} A.,
  2017, \mn@doi [\apj] {10.3847/1538-4357/aa7aa1}, \href
  {https://ui.adsabs.harvard.edu/abs/2017ApJ...844..108M} {844, 108}

\bibitem[\protect\citeauthoryear{{Martin} et~al.,}{{Martin}
  et~al.}{2022}]{Martin:2022Natur.601...45M}
{Martin} N.~F.,  et~al., 2022, \mn@doi [\nat] {10.1038/s41586-021-04162-2},
  \href {https://ui.adsabs.harvard.edu/abs/2022Natur.601...45M} {601, 45}

\bibitem[\protect\citeauthoryear{{Mastropietro}, {De Rijcke}  \&
  {Peletier}}{{Mastropietro} et~al.}{2021}]{Mastropietro:2021MNRAS.504.3387M}
{Mastropietro} M.,  {De Rijcke} S.,   {Peletier} R.~F.,  2021, \mn@doi [\mnras]
  {10.1093/mnras/stab1091}, \href
  {https://ui.adsabs.harvard.edu/abs/2021MNRAS.504.3387M} {504, 3387}

\bibitem[\protect\citeauthoryear{{McConnachie}}{{McConnachie}}{2012}]{McConnachie:2012}
{McConnachie} A.~W.,  2012, \mn@doi [\aj] {10.1088/0004-6256/144/1/4}, \href
  {https://ui.adsabs.harvard.edu/abs/2012AJ....144....4M} {144, 4}

\bibitem[\protect\citeauthoryear{{Muratov} \& {Gnedin}}{{Muratov} \&
  {Gnedin}}{2010}]{Muratov:2010}
{Muratov} A.~L.,  {Gnedin} O.~Y.,  2010, \mn@doi [\apj]
  {10.1088/0004-637X/718/2/1266}, \href
  {https://ui.adsabs.harvard.edu/abs/2010ApJ...718.1266M} {718, 1266}

\bibitem[\protect\citeauthoryear{{Park}, {Ricotti}  \& {Sugimura}}{{Park}
  et~al.}{2021a}]{parkR2021a}
{Park} J.,  {Ricotti} M.,   {Sugimura} K.,  2021a, \mn@doi [\mnras]
  {10.1093/mnras/stab2999}, \href
  {https://ui.adsabs.harvard.edu/abs/2021MNRAS.508.6176P} {508, 6176}

\bibitem[\protect\citeauthoryear{{Park}, {Ricotti}  \& {Sugimura}}{{Park}
  et~al.}{2021b}]{parkR2021b}
{Park} J.,  {Ricotti} M.,   {Sugimura} K.,  2021b, \mn@doi [\mnras]
  {10.1093/mnras/stab3000}, \href
  {https://ui.adsabs.harvard.edu/abs/2021MNRAS.508.6193P} {508, 6193}

\bibitem[\protect\citeauthoryear{{Park}, {Ricotti}  \& {Sugimura}}{{Park}
  et~al.}{2022}]{parkR2022}
{Park} J.,  {Ricotti} M.,   {Sugimura} K.,  2022, arXiv e-prints, \href
  {https://ui.adsabs.harvard.edu/abs/2022arXiv221204564P} {p. arXiv:2212.04564}

\bibitem[\protect\citeauthoryear{{Parmentier}, {Goodwin}, {Kroupa}  \&
  {Baumgardt}}{{Parmentier} et~al.}{2008}]{Parmentier:2008ApJ...678..347P}
{Parmentier} G.,  {Goodwin} S.~P.,  {Kroupa} P.,   {Baumgardt} H.,  2008,
  \mn@doi [\apj] {10.1086/587137}, \href
  {https://ui.adsabs.harvard.edu/abs/2008ApJ...678..347P} {678, 347}

\bibitem[\protect\citeauthoryear{{Pawlik}, {Milosavljevi{\'c}}  \&
  {Bromm}}{{Pawlik} et~al.}{2013}]{Pawlik:2013ApJ...767...59P}
{Pawlik} A.~H.,  {Milosavljevi{\'c}} M.,   {Bromm} V.,  2013, \mn@doi [\apj]
  {10.1088/0004-637X/767/1/59}, \href
  {https://ui.adsabs.harvard.edu/abs/2013ApJ...767...59P} {767, 59}

\bibitem[\protect\citeauthoryear{{Peng} et~al.,}{{Peng}
  et~al.}{2006}]{Peng:2006}
{Peng} E.~W.,  et~al., 2006, \mn@doi [\apj] {10.1086/498210}, \href
  {https://ui.adsabs.harvard.edu/abs/2006ApJ...639...95P} {639, 95}

\bibitem[\protect\citeauthoryear{{Phipps}, {Khochfar}, {Varri}  \& {Dalla
  Vecchia}}{{Phipps} et~al.}{2020}]{Phipps:2020A&A...641A.132P}
{Phipps} F.,  {Khochfar} S.,  {Varri} A.~L.,   {Dalla Vecchia} C.,  2020,
  \mn@doi [\aap] {10.1051/0004-6361/202037884}, \href
  {https://ui.adsabs.harvard.edu/abs/2020A&A...641A.132P} {641, A132}

\bibitem[\protect\citeauthoryear{Phipps, Khochfar, Varri  \&
  Dalla Vecchia}{Phipps et~al.}{2022}]{Phipps:2022}
Phipps F.,  Khochfar S.,  Varri A.~L.,   Dalla Vecchia C.,  2022, \mn@doi
  [Monthly Notices of the Royal Astronomical Society] {10.1093/mnras/stac3399},
  518, 4606

\bibitem[\protect\citeauthoryear{{Portegies Zwart}, {McMillan}  \&
  {Gieles}}{{Portegies Zwart} et~al.}{2010}]{Zwart:2010ARA&A..48..431P}
{Portegies Zwart} S.~F.,  {McMillan} S. L.~W.,   {Gieles} M.,  2010, \mn@doi
  [\araa] {10.1146/annurev-astro-081309-130834}, \href
  {https://ui.adsabs.harvard.edu/abs/2010ARA&A..48..431P} {48, 431}

\bibitem[\protect\citeauthoryear{{Reinoso}, {Schleicher}, {Fellhauer}, {Leigh}
  \& {Klessen}}{{Reinoso} et~al.}{2020}]{Reinoso:2020}
{Reinoso} B.,  {Schleicher} D.~R.~G.,  {Fellhauer} M.,  {Leigh} N.~W.~C.,
  {Klessen} R.~S.,  2020, \mn@doi [\aap] {10.1051/0004-6361/202037843}, \href
  {https://ui.adsabs.harvard.edu/abs/2020A&A...639A..92R} {639, A92}

\bibitem[\protect\citeauthoryear{{Renzini}}{{Renzini}}{2017}]{Renzini:2017MNRAS.469L..63R}
{Renzini} A.,  2017, \mn@doi [\mnras] {10.1093/mnrasl/slx057}, \href
  {https://ui.adsabs.harvard.edu/abs/2017MNRAS.469L..63R} {469, L63}

\bibitem[\protect\citeauthoryear{{Ricotti}}{{Ricotti}}{2002}]{Ricotti:2002}
{Ricotti} M.,  2002, \mn@doi [\mnras] {10.1046/j.1365-8711.2002.05990.x}, \href
  {https://ui.adsabs.harvard.edu/abs/2002MNRAS.336L..33R} {336, L33}

\bibitem[\protect\citeauthoryear{{Ricotti}}{{Ricotti}}{2016}]{Ricotti:2016}
{Ricotti} M.,  2016, \mn@doi [\mnras] {10.1093/mnras/stw1672}, \href
  {https://ui.adsabs.harvard.edu/abs/2016MNRAS.462..601R} {462, 601}

\bibitem[\protect\citeauthoryear{{Ricotti} \& {Gnedin}}{{Ricotti} \&
  {Gnedin}}{2005}]{RicottiG:2005}
{Ricotti} M.,  {Gnedin} N.~Y.,  2005, \mn@doi [\apj] {10.1086/431415}, \href
  {https://ui.adsabs.harvard.edu/abs/2005ApJ...629..259R} {629, 259}

\bibitem[\protect\citeauthoryear{{Ricotti}, {Gnedin}  \& {Shull}}{{Ricotti}
  et~al.}{2002a}]{RicottiGS:2002a}
{Ricotti} M.,  {Gnedin} N.~Y.,   {Shull} J.~M.,  2002a, \mn@doi [\apj]
  {10.1086/341255}, \href
  {https://ui.adsabs.harvard.edu/abs/2002ApJ...575...33R} {575, 33}

\bibitem[\protect\citeauthoryear{{Ricotti}, {Gnedin}  \& {Shull}}{{Ricotti}
  et~al.}{2002b}]{RicottiGS:2002b}
{Ricotti} M.,  {Gnedin} N.~Y.,   {Shull} J.~M.,  2002b, \mn@doi [\apj]
  {10.1086/341256}, \href
  {https://ui.adsabs.harvard.edu/abs/2002ApJ...575...49R} {575, 49}

\bibitem[\protect\citeauthoryear{{Ricotti}, {Parry}  \& {Gnedin}}{{Ricotti}
  et~al.}{2016}]{RicottiPG:2016}
{Ricotti} M.,  {Parry} O.~H.,   {Gnedin} N.~Y.,  2016, \mn@doi [\apj]
  {10.3847/0004-637X/831/2/204}, \href
  {https://ui.adsabs.harvard.edu/abs/2016ApJ...831..204R} {831, 204}

\bibitem[\protect\citeauthoryear{{Ricotti}, {Polisensky}  \&
  {Cleland}}{{Ricotti} et~al.}{2022}]{RicottiPC:2022}
{Ricotti} M.,  {Polisensky} E.,   {Cleland} E.,  2022, \mn@doi [\mnras]
  {10.1093/mnras/stac1485}, \href
  {https://ui.adsabs.harvard.edu/abs/2022MNRAS.515..302R} {515, 302}

\bibitem[\protect\citeauthoryear{Rosdahl, Blaizot, Aubert, Stranex  \&
  Teyssier}{Rosdahl et~al.}{2013}]{rosdahl2013}
Rosdahl J.,  Blaizot J.,  Aubert D.,  Stranex T.,   Teyssier R.,  2013, \mn@doi
  [\mnras] {10.1093/mnras/stt1722}, 436, 2188

\bibitem[\protect\citeauthoryear{{Salmon} et~al.,}{{Salmon}
  et~al.}{2020}]{Salmon:2020ApJ...889..189S}
{Salmon} B.,  et~al., 2020, \mn@doi [\apj] {10.3847/1538-4357/ab5a8b}, \href
  {https://ui.adsabs.harvard.edu/abs/2020ApJ...889..189S} {889, 189}

\bibitem[\protect\citeauthoryear{{Salpeter}}{{Salpeter}}{1955}]{Salpeter:1955}
{Salpeter} E.~E.,  1955, \mn@doi [\apj] {10.1086/145971}, \href
  {https://ui.adsabs.harvard.edu/abs/1955ApJ...121..161S} {121, 161}

\bibitem[\protect\citeauthoryear{{Salvadori} \& {Ferrara}}{{Salvadori} \&
  {Ferrara}}{2009}]{Salvadori:2009MNRAS.395L...6S}
{Salvadori} S.,  {Ferrara} A.,  2009, \mn@doi [\mnras]
  {10.1111/j.1745-3933.2009.00627.x}, \href
  {https://ui.adsabs.harvard.edu/abs/2009MNRAS.395L...6S} {395, L6}

\bibitem[\protect\citeauthoryear{{Sameie} et~al.,}{{Sameie}
  et~al.}{2022}]{Sameie:2022}
{Sameie} O.,  et~al., 2022, arXiv e-prints, \href
  {https://ui.adsabs.harvard.edu/abs/2022arXiv220400638S} {p. arXiv:2204.00638}

\bibitem[\protect\citeauthoryear{{Schaerer} \& {Charbonnel}}{{Schaerer} \&
  {Charbonnel}}{2011}]{Schaerer:2011MNRAS.413.2297S}
{Schaerer} D.,  {Charbonnel} C.,  2011, \mn@doi [\mnras]
  {10.1111/j.1365-2966.2011.18304.x}, \href
  {https://ui.adsabs.harvard.edu/abs/2011MNRAS.413.2297S} {413, 2297}

\bibitem[\protect\citeauthoryear{{Shukirgaliyev}, {Parmentier}, {Berczik}  \&
  {Just}}{{Shukirgaliyev} et~al.}{2017}]{Shukirgaliyev:2017A&A...605A.119S}
{Shukirgaliyev} B.,  {Parmentier} G.,  {Berczik} P.,   {Just} A.,  2017,
  \mn@doi [\aap] {10.1051/0004-6361/201730607}, \href
  {https://ui.adsabs.harvard.edu/abs/2017A&A...605A.119S} {605, A119}

\bibitem[\protect\citeauthoryear{{Simon}}{{Simon}}{2019}]{Simon:2019UFDReview}
{Simon} J.~D.,  2019, \mn@doi [\araa] {10.1146/annurev-astro-091918-104453},
  \href {https://ui.adsabs.harvard.edu/abs/2019ARA&A..57..375S} {57, 375}

\bibitem[\protect\citeauthoryear{{Smith} et~al.,}{{Smith}
  et~al.}{2017}]{Grackle:2017}
{Smith} B.~D.,  et~al., 2017, \mn@doi [\mnras] {10.1093/mnras/stw3291}, \href
  {https://ui.adsabs.harvard.edu/abs/2017MNRAS.466.2217S} {466, 2217}

\bibitem[\protect\citeauthoryear{{Spitzer}}{{Spitzer}}{1958}]{Spitzer:1958}
{Spitzer} Lyman J.,  1958, \mn@doi [\apj] {10.1086/146435}, \href
  {https://ui.adsabs.harvard.edu/abs/1958ApJ...127...17S} {127, 17}

\bibitem[\protect\citeauthoryear{{Spitzer}}{{Spitzer}}{1987}]{Spitzer:1987}
{Spitzer} L.,  1987, {Dynamical evolution of globular clusters}.
Princeton University Press

\bibitem[\protect\citeauthoryear{Sugimura, Matsumoto, Hosokawa, Hirano  \&
  Omukai}{Sugimura et~al.}{2020}]{sugimura:2020}
Sugimura K.,  Matsumoto T.,  Hosokawa T.,  Hirano S.,   Omukai K.,  2020,
  \mn@doi [\apjl] {10.3847/2041-8213/ab7d37}, 892, L14

\bibitem[\protect\citeauthoryear{{Sun} \& {Furlanetto}}{{Sun} \&
  {Furlanetto}}{2016}]{Sun:2016MNRAS.460..417S}
{Sun} G.,  {Furlanetto} S.~R.,  2016, \mn@doi [\mnras] {10.1093/mnras/stw980},
  \href {https://ui.adsabs.harvard.edu/abs/2016MNRAS.460..417S} {460, 417}

\bibitem[\protect\citeauthoryear{{Tassis}, {Gnedin}  \& {Kravtsov}}{{Tassis}
  et~al.}{2012}]{Tassis:2012ApJ...745...68T}
{Tassis} K.,  {Gnedin} N.~Y.,   {Kravtsov} A.~V.,  2012, \mn@doi [\apj]
  {10.1088/0004-637X/745/1/68}, \href
  {https://ui.adsabs.harvard.edu/abs/2012ApJ...745...68T} {745, 68}

\bibitem[\protect\citeauthoryear{Teyssier}{Teyssier}{2002}]{teyssier2002}
Teyssier R.,  2002, \mn@doi [\aap] {10.1051/0004-6361:20011817}, 385, 337

\bibitem[\protect\citeauthoryear{{VandenBerg}, {Brogaard}, {Leaman}  \&
  {Casagrande}}{{VandenBerg} et~al.}{2013}]{2013ApJ...775..134V}
{VandenBerg} D.~A.,  {Brogaard} K.,  {Leaman} R.,   {Casagrande} L.,  2013,
  \mn@doi [\apj] {10.1088/0004-637X/775/2/134}, \href
  {https://ui.adsabs.harvard.edu/abs/2013ApJ...775..134V} {775, 134}

\bibitem[\protect\citeauthoryear{{Vanzella} et~al.,}{{Vanzella}
  et~al.}{2022}]{Vanzella:2022arXiv221109839V}
{Vanzella} E.,  et~al., 2022, arXiv e-prints, \href
  {https://ui.adsabs.harvard.edu/abs/2022arXiv221109839V} {p. arXiv:2211.09839}

\bibitem[\protect\citeauthoryear{{V{\'a}zquez-Semadeni}, {Banerjee},
  {G{\'o}mez}, {Hennebelle}, {Duffin}  \& {Klessen}}{{V{\'a}zquez-Semadeni}
  et~al.}{2011}]{Vazquez:2011}
{V{\'a}zquez-Semadeni} E.,  {Banerjee} R.,  {G{\'o}mez} G.~C.,  {Hennebelle}
  P.,  {Duffin} D.,   {Klessen} R.~S.,  2011, \mn@doi [\mnras]
  {10.1111/j.1365-2966.2011.18569.x}, \href
  {https://ui.adsabs.harvard.edu/abs/2011MNRAS.414.2511V} {414, 2511}

\bibitem[\protect\citeauthoryear{{Wan} et~al.,}{{Wan}
  et~al.}{2020}]{Wan:2020Natur.583..768W}
{Wan} Z.,  et~al., 2020, \mn@doi [\nat] {10.1038/s41586-020-2483-6}, \href
  {https://ui.adsabs.harvard.edu/abs/2020Natur.583..768W} {583, 768}

\bibitem[\protect\citeauthoryear{{Welch} et~al.,}{{Welch}
  et~al.}{2022a}]{Welch:2022}
{Welch} B.,  et~al., 2022a, \mn@doi [\nat] {10.1038/s41586-022-04449-y}, \href
  {https://ui.adsabs.harvard.edu/abs/2022Natur.603..815W} {603, 815}

\bibitem[\protect\citeauthoryear{{Welch} et~al.,}{{Welch}
  et~al.}{2022b}]{Welch:2022ApJ...940L...1W}
{Welch} B.,  et~al., 2022b, \mn@doi [\apjl] {10.3847/2041-8213/ac9d39}, \href
  {https://ui.adsabs.harvard.edu/abs/2022ApJ...940L...1W} {940, L1}

\bibitem[\protect\citeauthoryear{{Yajima}, {Nagamine}, {Zhu}, {Khochfar}  \&
  {Dalla Vecchia}}{{Yajima} et~al.}{2017}]{Yajima:2017ApJ...846...30Y}
{Yajima} H.,  {Nagamine} K.,  {Zhu} Q.,  {Khochfar} S.,   {Dalla Vecchia} C.,
  2017, \mn@doi [\apj] {10.3847/1538-4357/aa82b5}, \href
  {https://ui.adsabs.harvard.edu/abs/2017ApJ...846...30Y} {846, 30}

\makeatother
\end{thebibliography}

\appendix
\section{BSC Profiles}
\label{sec:Profiling BSCs}

\begin{figure}
	\includegraphics[width=\columnwidth]{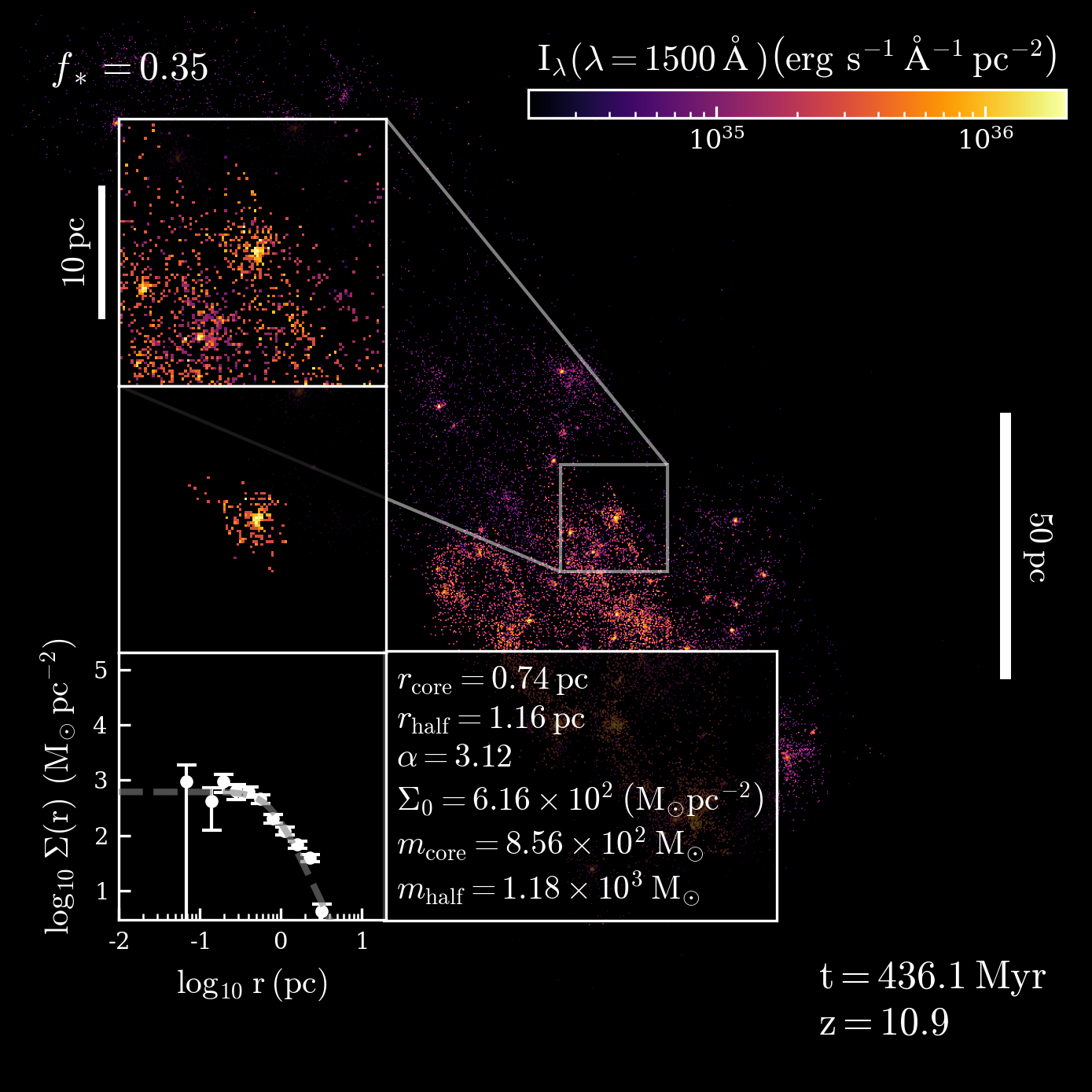}
    \caption{Demonstration of how the BSC profiles are computed. The specific surface brightness ($I_{\lambda}$ at 1500~\AA) of the $f_{*} = 0.35$ galaxy is depicted. The inset panels (\textit{top to bottom}) show a magnified view of the galaxy centered on a selected BSC, the background stars and other clusters subtracted from the BSC, and the projected density profile of this extracted BSC along with the fitted parameters.}
    \label{fig:Profile Demo} 
\end{figure}

To quantify the extent of the BSCs, we use a halo finder. The BSCs are first isolated from the field stars and other surrounding clusters which affects the derived values for $r_{\mathrm{core}}$ and $\alpha$, before the surface density is profiled. Fig. \ref{fig:Profile Demo} shows this operation on a single BSC in the low-SFE galaxy.

\section{Surface Brightness Relationships}\label{app:B}

In Fig.~\ref{fig:Light Projection}, we show the specific surface brightness $I_{\nu}$ at 1500~\AA\ in Jansky arcsec$^{-2}$ and in mag arcsec$^{-2}$ ($\mu_{AB}$), corrected for cosmological redshift dimming using the following relationships:
\begin{equation}
     \begin{split}
     I_\nu &= 1.56 \times 10^{-6} \frac{\rm{Jy}}{\rm{arcsec^2}}
     \left(\frac{I_\lambda}{10^{36}~\text{ergs~s}^{-1}~\text{\AA}^{-1}~\text{pc}^{-2}}\right)
     \left(\frac{1+z}{10}\right)^{-4},\\
    \mu_{AB} &= \left[ 23.4-2.5 \log_{10} \left(\frac{I_\lambda}{10^{36}~\text{ergs~s}^{-1}~\text{\AA}^{-1}~\text{pc}^{-2}}\right) \right. \\
    &\left. + 10 \log_{10}\left(\frac{1+z}{10}\right)\right]~\frac{\rm mag}{\rm{arcsec}^2},
    \end{split}
\label{Eq Flux}
\end{equation}
where we have used the relationship $\mu_{AB} = -2.5 \log_{10} I_{\nu} + 8.90$ to convert from Jansky to AB magnitude, and the surface brightness definition at $z\ll 1$: $I_\nu= f_\nu/\theta^2$ (at~10 pc, which we then correct for $z$), where the specific flux in Jansky is:
\begin{equation}
    \frac{f_{\nu}}{\rm Jy} = 7.5 \times 10^{10} \left( \frac{\lambda}{\text{1500~\AA}}\right)^2 
    \frac{f_\lambda}{\mrm{erg} \mrm{s}^{-1}~\text{\AA}^{-1}.  \mrm{cm}^{-2}}
\end{equation}
We have obtained the specific surface brightness $I_{\lambda}$ at 1500~\AA\ in Eq.~(\ref{Eq Flux}) from the stellar mass and age of the stellar population in each pixel using instantaneous burst stellar population models in \cite{1999ApJS..123....3L}, assuming a standard Salpeter IMF.

\bsp	
\label{lastpage}
\end{document}